# Encoding call-by-push-value in the $\pi$-calculus

Benjamin Bennetzen, Nikolaj Rossander Kristensen, and Peter Buus Steffensen

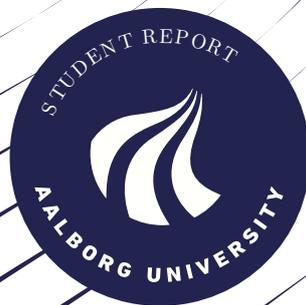

STUDENT REPORT

AALBORG UNIVERSITY

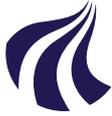

**Title:**
Encoding call-by-push-value in the π-calculus

**Theme:**
Theoretical Computer Science, Computational Calculi, Process Calculi, Coq

**Project Period:**
Spring Semester 2025

**Project Group:**
cs-25-sv-10-02

**Participants:**
Benjamin Bennetzen
Nikolaj Rossander Kristensen
Peter Buus Steffensen

**Supervisor:**
Hans Hüttel

**Date:**
2025-06-12

**Copies:** 1

**Page Numbers:** 56

**Abstract:**
In this report we define an encoding of Levy's call-by-push-value λ-calculus (CBPV) in the π-calculus, and prove that our encoding is both sound and complete. We present informal (by-hand) proofs of soundness, completeness, and all required lemmas. The encoding is specialized to the internal π-calculus (πI-calculus) to circumvent certain challenges associated with using de Bruijn index in a formalization, and it also helps with bisimulation as early-, late- and open-bisimulation coincide in this setting, furthermore bisimulation is a congruence. Additionally, we argue that our encoding also satisfies the five criteria for good encodings proposed by Gorla in [1], as well as show similarities between Milner's original encodings of call-by-value and call-by-name in the π-calculus and our encoding.

This paper includes encodings from CBPV in the πI-calculus, asynchronous polyadic π and the local π-calculus. We begin a formalization of the proof in Coq for the soundness and completeness of the encoding in the πI-calculus. Not all lemmas used in the formalization are themselves formally proven. However, we argue that the non-proven lemmas are reasonable, as they are proven by hand, or amount to Coq formalities that are straightforward given informal arguments.

# Summary

This report introduces and explores an encoding of Levy's call-by-push-value (CBPV) $\lambda$-calculus in the $\pi$-calculus, presenting detailed proofs demonstrating both the soundness and completeness of the encoding. CBPV is a computational calculus that clearly separates values from computations, effectively encompassing the traditional paradigms of call-by-name (CBN) and call-by-value (CBV), both of which have significantly influenced functional programming languages such as Haskell, OCaml, and F#. By introducing a unified encoding for these paradigms, the report contributes notably to the ongoing discourse around computational paradigms and their expressive power in concurrent computational models.

A crucial part of the contribution involves demonstrating that the encoding satisfies Gorla's five properties for effective encodings. These criteria include operational correspondence, divergence reflection, success sensitiveness, compositionality, and name invariance. We examine how our unified encoding relates to and compares with Milner's foundational encodings of CBV and CBN in the $\pi$-calculus, providing insightful analysis into their similarities as well as their differences.

We further demonstrate the adaptability of our encoding by extending our encoding to prominent variations of the $\pi$-calculus, including the internal, asynchronous, polyadic, and local $\pi$-calculi. Particular emphasis is given to the internal $\pi$-calculus. Specifically, in the internal $\pi$-calculus, early, late, and open bisimulations coincide. Forthermore bisimulation is a congruence. This significantly simplifies the choice of equivalence and congruence relation between encoded processes.

Moreover, the report includes the initial formalization efforts conducted in Coq, aiming to provide a machine-checked verification of the soundness and and completeness proofs for the encoding. Although this formalization remains incomplete at the time of the report, significant progress has been achieved. Key proofs have been captured in Coq, while remaining lemmas are provided with informal arguments supporting why these unproven lemmas are reasonable.

We conclude by identifying several promising areas for future research to build upon our results. Foremost among these is the need to fully complete the Coq formalization effort, as completing these formal proofs would provide a more robust and verifiable guarantee of correctness. Beyond this immediate goal, we also propose deeper explorations into theoretical correspondences with Milner's original CBV and CBN encodings. Such analyses could enhance our understanding of computational paradigms, particularly how they might be effectively unified or parallelized in practice.

Additionally, extending the existing CBPV encoding to incorporate more complex and advanced constructs from Levy's original CBPV framework would also be interesting to investigate further. This would potentially offer a richer and more versatile encoding, capable of capturing more nuanced computational behaviours inherent in more sophisticated functional languages. We also see potential value in refining Gorla's criteria specifically tailored for CBPV and similar computational frameworks, potentially leading to more specialized yet powerful standards for assessing encoding quality.

Other research directions include an in-depth investigation into potential encodings of CBPV within the asynchronous and local variants of the $\pi$-calculus, which could yield theoretical insights into asynchronous computation and locality constraints within concurrent systems. Furthermore, examining correspondences between type systems in CBPV and those in various $\pi$-calculus formulations is highlighted as a promising research path. Establishing such corre-spondences could illuminate deeper structural parallels and facilitate the integration of strong typing discipline into concurrent computation frameworks.

Overall, this report represents a comprehensive theoretical insight into understanding computational paradigms, contributing to the literature and laying a solid foundation for ongoing and future research endeavors.

# Contents

# Appendix

# 1 Introduction

The $\lambda$-calculus is one of the most well studied calculi, and it is the base of all functional programming languages. The $\lambda$-calculus has a large variety of different extensions that represent distinct reduction paradigms. Two of the most useful are the call-by-name (CBN) and the call-by-value (CBV) paradigms introduced by Plotkin in [2]. Additionally, Plotkin shows how to simulate each of the two paradigms in the other one. These two paradigms have been very important for the development of functional programming languages, with CBV being used in ML style languages such as OCaml and F#, and CBN (or call-by-need) being used in Miranda style languages such as Haskell and Clean. Amongst the many variations of the $\lambda$-calculus, there exist some that encompass both of these aforementioned reduction paradigms. One such variation, is the call-by-push-value (CBPV) $\lambda$-calculus, introduced by Levy in [3]. CBPV makes a distinction between values and computations, in which one can encode CBN as a value reduction and CBV as a computation reduction, thereby subsuming both paradigms. This means that programming languages based on CBPV do not need to choose between being CBV or CBN. In [4], Rizkallah et. al continue the work of Levy, by introducing a formal equational theorem for CBPV. Another variation is the Bang calculus, introduced by Ehrhard and Guerrieri in [5], which is similar to CBPV, but is untyped. The parametric $\lambda$-calculus as introduced by Paolini in [6], also encapsulates both CBV and CBN. Moggi created what he describes as "a simple programming language" (we did not find it simple) which could be used to represent the computational aspects of different paradigms [7]. Somewhat similar is Takahashi's work about parallel reductions in the $\lambda$-calculus [8].

Like the $\lambda$-calculus, the $\pi$-calculus has been extensively studied, resulting in numerous variations. These include the higher-order $\pi$-calculus introduced by Sangiorgi in [9], allowing processes to send not only names but other processes. The polyadic $\pi$-calculus from Milner et. al in [10] allows multiple names to be sent simultaneously over a single channel. The internal $\pi$-calculus by Sangiorgi in [11] binds all outputs, thereby restricting channel interactions. Furthermore, the asynchronous $\pi$-calculus, presented by Boudol in [12], removes continuations after sending, and the local $\pi$-calculus introduced by Merro and Sangiorgi in [13] limits names received via communication to only be used in outputs.

An interesting way to gain insights into different calculi is by encoding them in each other. This can be used to show the expressive powers of calculi in relation to each other. In the $\pi$-calculus, one of the most famous results is presented in Milner's "Functions as Processes" [14], where the $\lambda$-calculus, both CBN and CBV, were encoded in the $\pi$-calculus. This result showed a way to evaluate functions using concurrent processes, a nice result for parallelizing functional programming languages. However, Milner needed separate encodings for CBN and CBV. An interesting question is if it is possible to encode both CBV and CBN with a single unified encoding. Sangiorgi did something similar to this with his work in [16], where he showed a way to encode both CBN and CBV to a higher order $\pi$-calculus, through an intermediate encoding, and then encoded this higher order $\pi$-calculus in the standard $\pi$-calculus. However, this required extra encoding steps, which might be avoided. Furthermore, his unified encoding was from a process calculi, where it might be more useful if the unified encoding was from a $\lambda$-calculus, meaning it is closer in behaviour to CBV and CBN.

Further work has explored encoding different $\lambda$-calculi into restricted variations of the $\pi$-calculus, such as [11], which encodes CBN into the internal $\pi$-calculus. The interesting aspect of these encodings is to determine what is sufficient to fully encode the $\lambda$-calculus. These encodings prove that it is possible to encode the $\lambda$-calculus in variations of the $\pi$-calculus, which are much more restricted and limited.

Moreover, various criteria for assessing the quality of encodings between calculi have emerged. Notably in [1], Gorla introduced five key properties that encodings between process calculi should satisfy to be considered effective simulations.

Our contribution to this line of research is an encoding of CBPV in the $\pi$-calculus, demonstrating that it unifies CBN and CBV within a single encoding. We prove that our encoding satisfies the properties of encodings presented in [1], and argue that it has a meaningful correspondence with Milner's encodings in [14]. Furthermore, we show that it is possible to encode CBPV in the $\pi I$-calculus, whilst satisfying the same properties presented in [1]. This is interesting, as the $\pi I$-calculus is a calculus where every output is bound, meaning that outputting free names is not necessarily needed to encode CBPV. Additionally, we argue that the $\pi I$-calculus is easier to formalize using de Bruijn indices compared to standard the $\pi$-calculus. The central parts of the proof of soundness and completeness of the $\pi I$-calculus encoding have been formalized in Coq.

The report is structured as follows: In Chapter 2 we motivate and introduce the call-by-push-value $\lambda$-calculus, as well as the $\pi$-calculus. In Chapter 3 we define an encoding of CBPV in the $\pi$-calculus, and define an operational correspondence, which also defines what it means for the encoding to be both sound and complete. This is followed by an analysis of whether the use of Milner's direct encodings of CBN and CBV in the $\pi$-calculus commute with Levy's encodings of CBN and CBV in CBPV, when used in conjunction with our encoding of CBPV in the $\pi$-calculus. The chapter concludes with an adaptation of Gorla's five properties of encodings – to analyze the efficacy of our encoding. In Chapter 4 we describe a partial formalization for the proof of soundness and completeness of the encoding, and argue for the correctness of any lemmas used in the formalization that are not formally proven. Lastly Chapter 5 summarizes the results of the report and discuses any further work.

# 2 λ-calculi and the π-calculus

In this chapter, we introduce and examine several calculi that form the basis for understanding computational models: The call-by-value (CBV) and call-by-name (CBN) variants of the $\lambda$-calculus, the call-by-push-value (CBPV) $\lambda$-calculus and the $\pi$-calculus.

We begin by reviewing the call-by-value and call-by-name $\lambda$-calculi, highlighting their differences and the limitations inherent to each. This motivates the introduction of CBPV, which aims to unify and generalize the strengths of these paradigms by clearly distinguishing between values and computations.

Following the introduction of CBPV, we introduce the $\pi$-calculus, a formalism designed for modeling concurrent computation through message-passing and process interactions.

## 2.1 Call-by-value and call-by-name

Here will give an introduction to the CBV and CBN paradigms of the $\lambda$-calculus. This will also serve as motivation as to why one would want to use a calculus that encompasses both of these paradigms.

The syntax and semantics for the CBN $\lambda$-calculus can be seen in Definition 1, and the syntax and semantics of the CBV $\lambda$-calculus can be seen in Definition 2. The semantics are the same as presented by Plotkin in [2].

Syntactically the two calculi both have variables, abstractions, and applications. They differ only by the fact that the CBN variant only has terms, while the CBV variant has variables and abstractions under a syntactic category for values.

**Definition 1.** (CBN $\lambda$-calculus)

Syntax:

$$M, N \coloneqq x \mid MN \mid \lambda x.M$$

Reduction rules:

$$\text{(Subject-reduction)} \frac{M \to M'}{M\ N \to M'N} \qquad \text{(Application-N)} \frac{}{(\lambda x.M)N \to M\{N/x\}}$$

**Definition 2.** (CBV $\lambda$-calculus)

Syntax:

$$M, N \coloneqq V \mid MN$$
$$V \coloneqq x \mid \lambda x.M$$

Reduction rules:

$$(\text{Subject-reduction}) \ \frac{M \rightarrow M'}{M \ N \rightarrow M' \ N} \quad (\text{Object-reduction}) \ \frac{N \rightarrow N'}{V \ N \rightarrow V \ N'}$$

$$(\text{Application-V}) \ \frac{}{\lambda x.M \ V \rightarrow M\{^V/_x\}}$$

The main difference between CBV and CBN appears in the case where the argument of an application is evaluated. Although their expressive power is the same, their termination capabilities are not. In CBV the argument is evaluated before we perform the substitution, where as with CBN it uses a lazy evaluation, that only evaluates the expression when it is needed. This results in CBN and CBV having different termination for some expressions, as seen in Example 1.

**Example 1.** (Termination difference between CBV and CBN) In this example we show how using the CBN semantics one can reduce the $\lambda$-expression shown below, while using the CBV semantics the expression will never terminate.

$$(\lambda x.y)((\lambda x.xx)(\lambda x.xx))$$

Using CBN semantics one can apply the (Application-N) rule to get the following reduction:

$$(\lambda x.y)((\lambda x.xx)(\lambda x.xx)) \rightarrow y$$

Here we say that the process reduces to the **terminal value**, or final value, of $y$.

However, using CBV semantics one has to use the (Subject-reduction) rule, meaning the expression $(\lambda x.xx)(\lambda x.xx)$ has to be reduced. This expression will never terminate, as one will continually have to apply the (Application-V) rule, leading to the infinite reduction sequence below.

$$(\lambda x.xx)(\lambda x.xx) \rightarrow (\lambda x.xx)(\lambda x.xx) \rightarrow ...$$

The lacy evaluation of CBN enables the termination of the expression in Example 1. CBN however, sometimes need to evaluate the same term many times, which in some sense is very inefficient. It would therefore be preferable to have both evaluation strategies in one calculus. Levy has done exactly that with his CBPV, which unifies CBN and CBV within a unified framework. We say that CBPV subsumes both CBN and CBV, meaning that it faithfully can represent both CBV and CBN without obscuring their constructs or losing their original meaning.

Although bidirectional translations between CBV and CBN exist, as well as translations into other calculi such as the linear $\lambda$-calculus or Moggi's computational $\lambda$-calculus [3], Levy argues for the importance of a unified language that naturally subsumes both evaluation strategies. He identifies two primary reasons for this: Firstly, Levy points out that separate treatments of CBV and CBN evaluation strategies "make each language appear arbitrary," whereas "a unified language might be more canonical." Secondly, from a practical standpoint, Levy notes that without a unified language, "each time we create a new style of semantics – e.g., Scott semantics, operational semantics, game semantics, continuation semantics, etc. – we always need to do it twice, once for each paradigm." Thus, a unified approach could significantly reduce redundancy and eliminate the necessity of choosing between CBV and CBN strategies. Furthermore, this unified approach provides a semantic foundation for two of the biggest branches of functional languages. All languages following the functional paradigm are in some form or another based on the $\lambda$-calculus, of which the ML branch of languages is based on CBV and the Miranda branch of languages are based on CBN or the more refined call-by-need. Using CBPV provides a common semantic foundation for both the ML branch and the Miranda branch of functional languages.

Following this motivation, we will now introduce CBPV in greater detail.

## 2.2 Call-by-push-value

In this section we introduce a reduced version of CBPV that is introduced in [3] and [17]. We have tried to reduce the calculus presented in [3] to such an extent that only the constructs needed for computational effects are present, as these are the constructs that we want to represent in the $\pi$-calculus.

### 2.2.1 Syntax

The formation rules representing the syntax for CBPV can be seen in Definition 3. We use $M \gg= \lambda x.N$ instead of $M$ to $x$ in $N$ as it was first introduced. The reason being, that this uses a more modern notation for the bind operation, inspired by monads from functional programming languages. In Levy's newest paper on CBPV, [18], he has also renamed the **produce** construct with the more suitable name **return**, further motivating this notation, as we now have the two prototypical operations for monads, namely bind and return. The monad formed in this version of CBPV is very similar to that of the identity monad. Values are simply wrapped and unwrapped within a context indicating that the value is part of a computation. We have only included the constructs needed to encode CBV and CBN as they were presented in [14], as the encodings presented by Milner is what we will compare our encoding with.

**Definition 3.** (CBPV Syntax)

$$M, N \coloneqq V$$
$$\mid \lambda x.M$$
$$\mid M\ V$$
$$\mid \textbf{force } V$$
$$\mid \textbf{return } V$$
$$\mid M \gg= \lambda x.N$$

$$V \coloneqq x \mid \textbf{thunk } M$$

We let $M$ and $N$ range over **Expr**, the set of all possible expressions, and we let $V$ range over **Val**, the set of all possible values. We also let **Var** denote the countably infinite set containing all variables, where $x, y, z, ... \in$ **Var**.

The syntax consists of computations, denoted as $M$ and $N$, and values, denoted as $V$. A value can either be a variable $x$, or a suspended computation, written as **thunk** $M$. A **thunk** represents a computation that has been suspended and will only be executed when invoked by **force** $M$. It is here that we get the behaviour of CBN, as we are able to delay the computation and thereby substitute with **thunk** terms, and not just variables or abstractions like in CBV. We also have abstraction, $\lambda x.M$, application, $M\ V$, and the *bind* term mentioned earlier.

### 2.2.2 Free and bound variables

An interesting thing about the CBPV is that, like the $\pi$-calculus, it contains two binders. As usual, a variable is only free if it is within the scope of an abstraction or a binding with the same name, as per Definition 4. And a variable is only bound if it is used in an abstraction or binding, as seen in Definition 5.

**Definition 4.** (Free Variables) The free variables of a terms in **Expr** and values in **Val** are defined inductively as follows.

$$\text{fv}(\lambda x.M) = \text{fv}(M) - \{x\}$$
$$\text{fv}(M\ V) = \text{fv}(M) \cup \text{fv}(V)$$
$$\text{fv}(\textbf{force}\ V) = \text{fv}(V)$$
$$\text{fv}(\textbf{return}\ V) = \text{fv}(V)$$
$$\text{fv}(M \ggeq \lambda x.N) = \text{fv}(M) \cup \text{fv}(N) - \{x\}$$
$$\text{fv}(x) = \{x\}$$
$$\text{fv}(\textbf{thunk}\ M) = \text{fv}(M)$$

**Definition 5.** (Bound Variables) The bound variables of a terms in **Expr** and values in **Val** are defined inductively as follows.

$$\text{bv}(\lambda x.M) = \{x\} \cup \text{bv}(M)$$
$$\text{bv}(M\ V) = \text{bv}(M) \cup \text{bv}(V)$$
$$\text{bv}(\textbf{force}\ V) = \text{bv}(V)$$
$$\text{bv}(\textbf{return}\ V) = \text{bv}(V)$$
$$\text{bv}(M \ggeq \lambda x.N) = \text{bv}(M) \cup \{x\} \cup \text{bv}(N)$$
$$\text{bv}(x) = \{\}$$
$$\text{bv}(\textbf{thunk}\ M) = \text{bv}(M)$$

We use Definition 6 to determine whether a variable name appears anywhere in a process. This helps ensure that any name we choose as *fresh* does not conflict with any name already present in the process.

**Definition 6.** (Variables) The variable names of terms in **Expr** and values in **Val** are defined as follows.

$$\text{v}(M) = \text{bv}(M) \cup \text{fn}(M)$$

### 2.2.3 Substitution

We define the substitution of free variables to be capture avoiding as seen in Definition 7.

**Definition 7.** (Substitution)

$$(\lambda x.M)\{^{M'}/_y\} = \begin{cases} \lambda x.M & x = y \\ \lambda x.M\{^{M'}/_y\} & x \neq y \wedge x \neq M' \\ (\lambda z.M\{^z/_x\})\{^{M'}/_y\} & x \neq y \wedge x = M' \wedge z \notin \text{fv}(M) \cup \text{fv}(M') \cup \{x, y\} \end{cases}$$

$$(MV)\left\{{M'}/{y}\right\} = M\left\{{M'}/{y}\right\}V\left\{{M'}/{y}\right\}$$

$$(\textbf{force }V)\left\{{M'}/{y}\right\} = \textbf{force }\left(V\left\{{M'}/{y}\right\}\right)$$

$$(\textbf{return }V)\left\{{M'}/{y}\right\} = \textbf{return }\left(V\left\{{M'}/{y}\right\}\right)$$

$$(M \gg= \lambda x.N)\left\{{M'}/{y}\right\} = \left(M\left\{{M'}/{y}\right\}\right) \gg= (\lambda x.N)\left\{{M'}/{y}\right\}$$

$$x\left\{{M'}/{y}\right\} = \begin{cases} M' & x = y \\ x & x \neq y \end{cases}$$

$$(\textbf{thunk }M)\left\{{M'}/{y}\right\} = \textbf{thunk }\left(M\left\{{M'}/{y}\right\}\right)$$

### 2.2.4 Small-step semantics

For the small-step semantics we have taken inspiration from the approach of Rizkallah et. al in [17]. The semantics of Rizkallah et. al has been altered slightly, as we omit the use of their unrolling function $\leadsto$, which would unroll a term such as "letrec x = λy.y in $N$". We do not need this, as we do not have mutual recursive aliasing, which would require the unroll operation. Instead, the already unrolled term can be found in the conclusion of the rule, instead of unrolling it in the premise of the rule.

**Definition 8.** (Small step semantics)

$$(\text{Binding-base}) \frac{}{(\textbf{return }V) \gg= \lambda x.N \to N\left\{{V}/{x}\right\}}$$

$$(\text{Binding-evolve}) \frac{M \to M'}{M \gg= \lambda x.N \to M' \gg= \lambda x.N}$$

$$(\text{force-thunk}) \frac{}{\textbf{force }(\textbf{thunk }M) \to M}$$

$$(\text{Application-base}) \frac{}{(\lambda x.M)\ V \to M\left\{{V}/{x}\right\}} \qquad (\text{Application-evolve}) \frac{M \to M'}{M\ V \to M'\ V}$$

We define $\mapsto$ as the transitive and reflexive closure of $\to$.

*Remark.* The syntax of CBPV allows us to create erroneous terms, that are "good CBPV terms that are nevertheless stuck" [17].

**Example 2.** (Erroneous terms) If $x$ is free, then $(\textbf{force }x)$ is an erroneous term as we are trying to force computation on a variable, for which there exists no further computations.

$$\textbf{force }x \gg= \lambda y.M$$

Likewise it would also be erroneous to use an abstraction on the left-hand side of a bind operator.

$$\lambda x.x \gg= \lambda y.M$$

However, such stuck terms can be avoided with the use of a type system.

### 2.2.5 Typing

In this section we introduce a type system for CBPV. We let **Types** denote the set of all possible types constructable from the formation rules below.

**Definition 9.** (Types)

$$A \coloneqq U\underline{B} \mid T$$
$$\underline{B} \coloneqq FA \mid A \to \underline{B}$$

The type system makes a clear distinction between values and computations, representing values with some type $A$, and computations with some type $\underline{B}$. Historically, computation types have also been underlined for clarity. A value type $A$, can either be a type $U\underline{B}$, which is a **thunk** of some computation $\underline{B}$, or a type $T$, which is a value belonging to some set of base types $\mathcal{B}$. A computation type $\underline{B}$ can either be a type $FA$, representing computation which returns a value of type $A$, or a type $A \to \underline{B}$ representing an abstraction that takes an argument of type $A$, and returns an expression of type $\underline{B}$.

Before presenting the typing rules for the type system, we formally define typing contexts as seen in Definition 10.

**Definition 10.** (Typing Context) A typing context $\Gamma$ is a partial function from variables to types (**Var** $\rightharpoonup$ **Types**). The notation $\Gamma, x : A$ adds $x$ to the typing context, meaning that $x$ is then assigned to the type $A$, shadowing any previous assignments of $x$ in $\Gamma$.

The typing rules are the same as presented by Levy in [18]. The typing relation $\vdash^{\mathrm{v}}$ takes a type context $\Gamma$, and a value $V$, and returns a type of form $A$. The typing relation $\vdash^{\mathrm{c}}$ takes a type context $\Gamma$ and a computation $B$, and returns a type of form $\underline{B}$.

**Definition 11.** (Typing rules)

$$(\text{T-Force}) \; \frac{\Gamma \vdash^{\mathrm{v}} V : U\underline{B}}{\Gamma \vdash^{\mathrm{c}} \textbf{force } V : \underline{B}} \qquad\qquad (\text{T-Thunk}) \; \frac{\Gamma \vdash^{\mathrm{c}} M : \underline{B}}{\Gamma \vdash^{\mathrm{v}} \textbf{thunk } M : U\underline{B}}$$

$$(\text{T-Bind}) \; \frac{\Gamma \vdash^{\mathrm{c}} M : FA \quad \Gamma, x : A \vdash^{\mathrm{c}} N : \underline{B}}{\Gamma \vdash^{\mathrm{c}} M \gg= \lambda x.N : \underline{B}} \qquad (\text{T-Return}) \; \frac{\Gamma \vdash^{\mathrm{v}} V : A}{\Gamma \vdash^{\mathrm{c}} \textbf{return } V : FA}$$

$$(\text{T-Application}) \; \frac{\Gamma \vdash^{\mathrm{c}} M : A \to \underline{B} \quad \Gamma \vdash^{\mathrm{v}} V : A}{\Gamma \vdash^{\mathrm{c}} M \; V : \underline{B}} \qquad (\text{T-Abstraction}) \; \frac{\Gamma, x : A \vdash^{\mathrm{c}} M : \underline{B}}{\Gamma \vdash^{\mathrm{c}} \lambda x.M : A \to \underline{B}}$$

$$(\text{T-Variable}) \; \frac{(x : A) \in \Gamma}{\Gamma \vdash^{\mathrm{v}} x : A}$$

We define well-typed expressions as terms that are typed with these rules. We define terminal terms as terms that are either a value, an abstraction or return. The typing rules ensure **type** **safety**, meaning that well-typed expressions will never get stuck, that is, reach a state in which there is no further reduction, yet the expression is a binding, force or an application. **Type safety** can be described as a type system for which both **type preservation** and **progress** hold. [19]

- **Type preservation** is the property that types are preserved through reductions. More formally, if $M : T$ and $M \to M'$, then $M' : T$. [19]

- **Progress** is the property that a well-typed expression is either a terminal term or it can take another reduction. More formally, if $M : T$ then $M$ is a value, return or abstraction, or there exists some $M'$ such that $M \to M'$. [19]

Earlier, we mentioned some erroneous expressions: expression that are syntactically correct but are neither terminal terms nor reducible. In Example 3 and Example 4 we have highlighted two such expressions and how the type system ensures type safety.

**Example 3.** ((**force** $x$) is possibly erroneous) In this example, we show how the possibly erroneous expression (**force** $x$) is not erroneous if it is well-typed.

By inspection of the typing rules, one can see that only the rule (T-Force) can be used to type a **force** expression. It states that what we are trying to **force** should be of type $U\underline{B}$, that being a **thunk**. By further inspection of the rules, one can see that only when the **thunk** operator is used in the (T-Thunk) rule will something be typed as $U\underline{B}$.

This means that, for well-typed expressions, **force** will only be used on **thunks**, and the erroneous cases where one tries to **force** on other values will be ill-typed. Consequently a **force** $x$ can only be well-typed, if the typing environment $\Gamma$ includes the judgement $x : U\underline{B}$, which implies a **thunk** could be substituted with $x$.

**Example 4.** ($\lambda x.x \gg= \lambda y.M$ is ill-typed) In this example we show how the syntactically correct, but erroneous expression $\lambda x.x \gg= \lambda y.M$, is ill-typed. If one were to try and construct a sequence of derivations proving that the expression is well-typed, one would quickly come to a contradiction. Specifically, that for a use of the bind operator to be well-typed, then $\lambda x.x$ needs to have the type $FA$. However, by inspection of the rules it is evident that such an expression can only have the type $A \to \underline{B}$.

Instead, what one should probably do in this case is to transform the expression into (**return thunk** $\lambda x.$ **return** $x$) $\gg= \lambda y.M$, this expression is well-typed as shown by the following sequence of derivations:

$$
\text{(T-Bind)} \cfrac{
\text{(T-Return)} \cfrac{
\text{(T-Thunk)} \cfrac{
\text{(T-Abstraction)} \cfrac{
\text{(T-Return)} \cfrac{
\text{(T-Variable)} \cfrac{x : A \in \Gamma, x : A}{\Gamma, x : A \vdash^{\mathrm{v}} x : A}
}{\Gamma, x : A \vdash^{\mathrm{c}} \mathbf{return}\ x : FA}
}{\Gamma \vdash^{\mathrm{c}} \lambda x.\ \mathbf{return}\ x : A \to FA}
}{\Gamma \vdash^{\mathrm{v}} \mathbf{thunk}\ \lambda x.\ \mathbf{return}\ x : U(A \to FA)}
}{\Gamma \vdash^{\mathrm{c}} \mathbf{return\ thunk}\ \lambda x.\ \mathbf{return}\ x : F(U(A \to FA))} \quad \Gamma, y : U(A \to FA) \vdash^{\mathrm{c}} M : \underline{B}
}{\Gamma \vdash^{\mathrm{c}} (\mathbf{return\ thunk}\ \lambda x.\ \mathbf{return}\ x) \gg= \lambda y.M : \underline{B}}
$$

### 2.2.6 Encoding call-by-value and call-by-name to call-by-push-value

We will now present how to encode CBV and CBN in CBPV, as shown by Levy in [20], but with our reduced syntax. It should be noted that in [20], the CBV and CBN he encodes to CBPV are typed with boolean, sum and functions type, where ours are untyped.

The encoding of CBV can be seen in Definition 12. With variables we use **return** to lift the variable into a computation, and similarly we use **return**, in conjunction with **thunk**, to turn an abstraction into a value which can be lifted into a computation. When encoding application we have to enforce a specific evaluation order, first the subject then the object, and lastly substitution. This is done through the use of binds, as per the semantics for CBPV the left operand of a bind has to reduce to a **return** $V$ before continuing. This enforces the evaluation order previously described. The use of the **force** is needed as abstractions are encoded as **thunks**.

**Definition 12.** (Encoding of CBV)

$$\llbracket x \rrbracket^v = \textbf{return } x$$

$$\llbracket \lambda x.M \rrbracket^v = \textbf{return thunk } \lambda x.\llbracket M \rrbracket^v$$

$$\llbracket M \ N \rrbracket^v = (\llbracket M \rrbracket^v \gg= (\lambda f.\llbracket N \rrbracket^v \gg= \lambda x.\textbf{force } f \ x))$$

As can be seen in Definition 13 it is more straightforward to enforce the lazy evaluation order of CBN. Variables are encoded as a **force** $x$, forcing evaluation of expressions only when they are strictly needed. Abstractions just further encode sub-expressions. Applications can be encoded by using a **thunk** on the object, thereby delaying evaluation of it; and allowing substitution as soon as the subject has reduced to an abstraction.

**Definition 13.** (Encoding of CBN)

$$\llbracket x \rrbracket^n = \textbf{force } x$$

$$\llbracket \lambda x.M \rrbracket^n = \lambda x.\llbracket M \rrbracket^n$$

$$\llbracket M \ N \rrbracket^n = \llbracket M \rrbracket^n \ (\textbf{thunk } \llbracket N \rrbracket^n)$$

Examples of using these encodings can be found in Appendix E.

## 2.3 The $\pi$-calculus

We will now introduce the monadic $\pi$-calculus and the proof techniques which will be used later to argue for the soundness and completeness of our encoding.

### 2.3.1 Syntax

The set of processes **Proc** denotes the set of all possible processes constructable from the formation rules in Definition 14, where $a, b, c, x, y, z... \in \textbf{Name}$, the set containing all names.

**Definition 14.** (Syntax of $\pi$-calculus)

$$P, Q := \bar{a}x.P \mid a(x).P \mid (\nu x)P \mid P|Q \mid !P \mid 0$$

The syntax consists of six constructors. $(\overline{a}x.P)$ is the output prefix, which reads as a process can send the name $x$ along the channel $a$. $(a(x).P)$ is the input prefix, which means a process can receive a name along a channel $a$ and then substitute it for the bound name $x$ in the following process $P$. We may also refer to the channel along which names are sent as the subject, and the names sent and received as the object. $(\nu x)P$ creates a new binding of $x$ in $P$. $P \mid Q$ is the parallel composition of the processes $P$ and $Q$. $!P$ is the replication of the process $P$, intuitively meaning that there are as many $P$'s running in parallel as needed. Lastly $0$ is the null process, the process at which no further actions can be taken. In most instances we will not write the null constructor explicitly, and as such $\overline{a}x$ should be read as $\overline{a}x.0$. We define free and bound names in the $\pi$-calculus in the usual way, with $\mathrm{bn}(P)$ to mean the set of bound names in $P$, $\mathrm{fn}(P)$ to mean the set of free names in $P$ and $n(P)$ to mean the union of all free and bound names in $P$.

We formally define the notion of subjects and objets in the usual way, as shown in the inductive definition of the four functions $\mathrm{sub_{in}}$, $\mathrm{sub_{out}}$, $\mathrm{obj_{in}}$, and $\mathrm{obj_{out}}$, seen in Definition 15. $\mathrm{sub_{in}}$ is a function $\mathbf{Proc} \rightarrow 2^{\mathbf{Name}}$, which takes process and collects all names that appear as subjects in an input prefix. The three latter functions function similarly. These four functions will prove to be useful, as some of our lemmas make restrictions on when certain names can appear as subjects, or objects, on inputs and outputs.

**Definition 15.** (Subjects and objects of a process) We inductively define the input subjects in a process as the function $\mathrm{sub_{in}}$, the input objects in a process as the function $\mathrm{obj_{in}}$, the output subjects in a process as the function $\mathrm{sub_{out}}$, and the output objects in a process as the function $\mathrm{obj_{out}}$.

$$\mathrm{sub_{in}}(\overline{a}x.P) = \mathrm{sub_{in}}(P) \qquad\qquad \mathrm{sub_{out}}(\overline{a}x.P) = \{a\} \cup \mathrm{sub_{out}}(P)$$
$$\mathrm{sub_{in}}(a(x).P) = \{a\} \cup \mathrm{sub_{in}}(P) \qquad\qquad \mathrm{sub_{out}}(a(x).P) = \mathrm{sub_{out}}(P)$$
$$\mathrm{sub_{in}}((\nu x)P) = \mathrm{sub_{in}}(P) \setminus \{x\} \qquad\qquad \mathrm{sub_{out}}((\nu x)P) = \mathrm{sub_{out}}(P) \setminus \{x\}$$
$$\mathrm{sub_{in}}(P|Q) = \mathrm{sub_{in}}(P) \cup \mathrm{sub_{in}}(Q) \qquad\qquad \mathrm{sub_{out}}(P|Q) = \mathrm{sub_{out}}(P) \cup \mathrm{sub_{out}}(Q)$$
$$\mathrm{sub_{in}}(!P) = \mathrm{sub_{in}}(P) \qquad\qquad \mathrm{sub_{out}}(!P) = \mathrm{sub_{out}}(P)$$
$$\mathrm{sub_{in}}(0) = \emptyset \qquad\qquad \mathrm{sub_{out}}(0) = \emptyset$$

$$\mathrm{obj_{in}}(\overline{a}x.P) = \mathrm{obj_{in}}(P) \qquad\qquad \mathrm{obj_{out}}(\overline{a}x.P) = \{x\} \cup \mathrm{obj_{out}}(P)$$
$$\mathrm{obj_{in}}(a(x).P) = \{x\} \cup \mathrm{obj_{in}}(P) \qquad\qquad \mathrm{obj_{out}}(a(x).P) = \mathrm{obj_{out}}(P)$$
$$\mathrm{obj_{in}}((\nu x)P) = \mathrm{obj_{in}}(P) \qquad\qquad \mathrm{obj_{out}}((\nu x)P) = \mathrm{obj_{out}}(P)$$
$$\mathrm{obj_{in}}(P|Q) = \mathrm{obj_{in}}(P) \cup \mathrm{obj_{in}}(Q) \qquad\qquad \mathrm{obj_{out}}(P|Q) = \mathrm{obj_{out}}(P) \cup \mathrm{obj_{out}}(Q)$$
$$\mathrm{obj_{in}}(!P) = \mathrm{obj_{in}}(P) \qquad\qquad \mathrm{obj_{out}}(!P) = \mathrm{obj_{out}}(P)$$
$$\mathrm{obj_{in}}(0) = \emptyset \qquad\qquad \mathrm{obj_{out}}(0) = \emptyset$$

### 2.3.2 Substitution and $\alpha$-equivalence

In the $\pi$-calculus, we must ensure that performing a substitution does not lead to name capture, as this would alter the meaning of a process. In cases where name capture would occur, we perform an $\alpha$-conversion to avoid it. This requires finding a fresh name — a name not used in the process — such that we can safely rename the variables that would otherwise be captured. In Definition 16 we write $z \notin n(P) \cup \{u, v\}$ to mean that we find the fresh name of $z$.

We define substitution, context and $\alpha$-equivalence as outlined in Definition 16, Definition 17 and Definition 18, which corresponds to the definitions given in [21].

**Definition 16.** (Substitution) We define substitution of $v$ with $u$, written as $\{{}^u/_v\}$ inductively on the structure of processes:

$$0\{{}^u/_v\} = 0$$

$$((\nu x)P)\{{}^u/_v\} = \begin{cases} (\nu x)P & x = v \\ (\nu z)(P\{{}^z/_x\})\{{}^u/_v\} & x = u \wedge z \notin n(P) \cup \{u, v\} \\ (\nu x)P\{{}^u/_v\} & \text{otherwise} \end{cases}$$

$$(x(y).P)\{{}^u/_v\} = \begin{cases} x(y).P & x \neq v \wedge y = v \\ u(y).P\{{}^u/_v\} & x = v \wedge y \neq v \\ x(z).(P\{{}^z/_y\})\{{}^u/_v\} & y = u \wedge x \neq v \wedge z \notin n(P) \cup \{u, v\} \\ u(z).(P\{{}^z/_y\})\{{}^u/_v\} & y = u \wedge x = v \wedge z \notin n(P) \cup \{u, v\} \\ u(y).P & x = v \wedge y = v \\ x(y).P\{{}^u/_v\} & \text{otherwise} \end{cases}$$

$$(\overline{x}y.P)\{{}^u/_v\} = \begin{cases} \overline{u}y.P\{{}^u/_v\} & x = v \wedge y \neq v \\ \overline{x}u.P\{{}^u/_v\} & x \neq v \wedge y = v \\ \overline{u}u.P\{{}^u/_v\} & x = v \wedge u = v \\ \overline{x}y.P\{{}^u/_v\} & \text{otherwise} \end{cases}$$

$$(P \mid P')\{{}^u/_v\} = (P\{{}^u/_v\}) \mid (P'\{{}^u/_v\})$$

$$(!P)\{{}^u/_v\} = !(P\{{}^u/_v\})$$

We also need to define a context in order to argue that two processes exhibit the same behaviour under bisimulation, see Section 2.3.4.

**Definition 17.** (Context) A process context $C$, is defined with the following syntax:

$$C := \overline{a}x.C \mid a(x).C \mid !C \mid (\nu x)C \mid P|C \mid C|P \mid [\cdot]$$

$C[P]$ is the expression where the hole in $C$ is filled with the process $P$, and denotes the closed term obtained by substituting $P$ for the $[\cdot]$ occurring in $C$ [22].

**Definition 18.** ($\alpha$-equivalence) Let $C[P]$ be a process where $P$ has a binding as its outermost constructor. Then the process $C[Q]$ can be obtained from the process $C[P]$, by the change of the outermost bound name in $P$, called an $\alpha$-conversion.

- If $P = (\nu x)P'$ then $Q = (\nu y)P'\{{}^y/_x\}$ where $y \notin (\text{fn}(P'))$
- If $P = a(x).P'$ then $Q = a(y).P'\{{}^y/_x\}$ where $y \notin (\text{fn}(P'))$

We say that two processes $P$ and $Q$ are $\alpha$-equivalent, written $P \equiv_\alpha Q$, if there exists a sequence of changes of bound names starting at $P$ and ending at $Q$.

The definition of $\alpha$-equivalence is used in the bisimilarity laws in Proposition 1. We use it to argue that two processes with different bound names have the same behaviour.

### 2.3.3 Early labeled transition semantics

We will be using the early labeled transition semantics as presented by Hirschkoff in [23]. The transition system has the actions in the set **Action** seen on Definition 19. The derivation rules on Definition 20, describe the transition relation **Proc** × **Action** × **Proc**.

**Definition 19.** (Actions)

$$\mu ::= a(x) \mid \overline{a}x \mid \overline{a}(x) \mid \tau$$

**Definition 20.** (Early labeled transition semantics of the $\pi$-calculus)

$$(\text{In}) \frac{}{a(x).P \xrightarrow{a(y)} P\{^y/_x\}} \qquad\qquad (\text{Out}) \frac{}{\overline{a}x.P \xrightarrow{\overline{a}x} P}$$

$$(\text{Com}) \frac{Q \xrightarrow{a(x)} Q' \quad P \xrightarrow{\overline{a}x} P'}{P \mid Q \xrightarrow{\tau} P' \mid Q'} \qquad (\text{Par}) \frac{P \xrightarrow{\mu} P'}{P \mid Q \xrightarrow{\mu} P' \mid Q} \text{bn}(\mu) \cap \text{fn}(Q) = \emptyset$$

$$(\text{Res}) \frac{P \xrightarrow{\mu} P'}{(\nu x)P \xrightarrow{\mu} (\nu x)P'} x \notin \text{n}(\mu) \qquad (\text{Rep}) \frac{P \mid !P \xrightarrow{\mu} P'}{!P \xrightarrow{\mu} P'}$$

$$(\text{Open}) \frac{P \xrightarrow{\overline{a}x} P'}{(\nu x)P \xrightarrow{\overline{a}(x)} P'} x \neq a \qquad (\text{Close}) \frac{P \xrightarrow{a(x)} P' \quad Q \xrightarrow{\overline{a}(x)} Q'}{P \mid Q \xrightarrow{\tau} (\nu x)(P' \mid Q')} x \notin \text{fn}(P)$$

We have omitted the symmetric versions of the (Com), (Par) and (Close) rules. We will informally call it having "barbs", when a process can do an input or output action.

### 2.3.4 Early bisimulation and congruence

We will write $\Rightarrow$ as the transitive and reflexive closure of $\xrightarrow{\tau}$. We will write $P \xRightarrow{\mu} P'$ to mean $P \Rightarrow Q \xrightarrow{\mu} Q' \Rightarrow P'$. We define early strong bisimulation and early weak bisimulation of processes as seen in Definition 21 and Definition 22. We note that for the case of bound output, we use alpha-conversion instead of substitutions as in [24], for notational conciseness.

**Definition 21.** (Early strong bisimulation) A relation $S$ on processes is an early strong simulation if $PSQ$ implies

1. If $P \xrightarrow{a(x)} P'$, then for each name $b$, a process $Q'$ exists s.t. $Q \xrightarrow{a(x)} Q'$ and $P'\{^b/_x\}SQ'\{^b/_x\}$.
2. If $P \xrightarrow{\alpha} P'$ for $\alpha = \overline{a}b$ or $\alpha = \tau$, then $Q'$ exists s.t. $Q \xrightarrow{\alpha} Q'$ and $P'SQ'$.
3. If $P \xrightarrow{\overline{a}(b)} P'$ then there exists a $Q'$ and $Q''$ s.t. $Q \equiv_\alpha Q'$ and $Q' \xrightarrow{\overline{a}(b)} Q''$ and $P'SQ''$

$S$ is an early strong bisimulation if $S$ and $S^{-1}$ are early strong simulations. Two processes $P$ and $Q$ are early strong bisimilar, written $P \sim_E Q$ if $PSQ$, for some early strong bisimulation $S$ [24]. $\sim_E$ is the largest early strong bisimulation.

For early strong bisimulation, $P$ and $Q$ must always be able to match the exact action that the other process takes. For early weak bisimulation, internal communications ($\tau$) can be 'answered' by doing any number of internal communications, this included zero. Input and output can also be answered after any number of $\tau$-transition has been performed.

**Definition 22.** (Early weak bisimulation) A relation $S$ on processes is an early weak simulation if $PSQ$ implies

1. If $P \xrightarrow{a(x)} P'$, then for each name $b$, a process $Q'$ exists s.t. $Q \xRightarrow{a(x)} Q'$ and $P'\{^b/_x\}SQ'\{^b/_x\}$.
2. If $P \xrightarrow{\overline{a}b} P'$ then a $Q'$ exists s.t. $Q \xRightarrow{\overline{a}b} Q'$ and $P'SQ'$.
3. If $P \xrightarrow{\tau} P'$ then a $Q'$ exists s.t. $Q \Rightarrow Q'$ and $P'SQ'$.
4. If $P \xrightarrow{\overline{a}(b)} P'$ then there exists a $Q'$ and $Q''$ s.t. $Q \equiv_\alpha Q'$ and $Q' \xRightarrow{\overline{a}(b)} Q''$ and $P'SQ''$

$S$ is an early weak bisimulation if $S$ and $S^{-1}$ are early weak simulations. Two processes $P$ and $Q$ are early weak bisimilar, written $P \approx_E Q$ if $PSQ$, for some early weak bisimulation $S$ [24]. $\approx_E$ is the largest early weak bisimulation.

**Definition 23.** (Early weak congruent) Two processes $P$ and $Q$ are early weak congruent, written $P \mathrel{\dot{\approx}}_E Q$, if for every substitution $\sigma$ it holds that $P\sigma \approx_E Q\sigma$. [24]

We will be using a number of bisimilarity laws throughout this report, both to clarify what is interesting in processes, by garbage collecting redundant parts, as well as to rewrite processes using up-to-bisimilarity, to be able to apply various lemmas. The bisimilarity laws presented in Proposition 1 are simple laws, that all are pretty straightforward why they hold. Many of these laws are similar to structural congruence laws. We refer to [25] and [26] for proof, as each can be proven with a simple bisimilarity proof.

**Proposition 1.** (Bisimilarity laws)

$$P \mid 0 \approx_E P \qquad P \mid Q \approx_E Q \mid P \qquad P \mid (Q \mid R) \approx_E (P \mid Q) \mid R$$

$$(\nu x)0 \approx_E 0 \qquad (\nu x)(\nu y)P \approx_E (\nu y)(\nu x)P$$

$$(\nu x)(P \mid Q) \approx_E P \mid (\nu x)Q \quad \text{if } x \notin \mathrm{fn}(P)$$

$$!P \approx_E P \mid !P \qquad !0 \approx_E 0 \qquad !!P \approx_E !P \qquad !(P \mid Q) \approx_E !P \mid !Q$$

$$(\nu x)(x(a).P) \approx_E 0 \qquad (\nu x)(\overline{x}a.P) \approx_E 0 \qquad (\nu x)(!x(a).P) \approx_E 0$$

$$\frac{P \equiv_\alpha Q}{P \approx_E Q}$$

With this, we can now introduce our encoding of CBPV to the $\pi$-calculus.

# 3 Encoding Call-by-push-value in the $\pi$-calculus

This chapter presents an encoding of the Call-by-push-value (CBPV) $\lambda$-calculus in the $\pi$-calculus, positioning it within the established tradition of encoding various forms of $\lambda$-calculi using process calculi.

Historically, different encodings of $\lambda$-calculi in the $\pi$-calculus have highlighted computational equivalences and differences between calculi, clarifying their operational semantics and expressive capabilities. The motivation behind these encodings lies in their capacity to reveal aspects of computational behaviour, especially in sequential computations represented as concurrent processes. Our contribution in this area is an encoding of CBPV in the $\pi$-calculus, thereby encoding a $\lambda$-calculus which allows for both lazy and eager evaluation in the $\pi$-calculus. Furthermore, CBPV lends itself very well to extensions with various monads, thereby laying a very basic groundwork for the study of monads modelled as concurrent processes. The encoding's operational correspondence is established through proofs of soundness and completeness.

Additionally, we analyze the relationships between Milner's direct encodings, which encode Call-by-name (CBN) and Call-by-value (CBV) $\lambda$-calculi directly in the $\pi$-calculus, and Levy's encoding, which encode CBN and CBV in CBPV, combined with our encoding of CBPV in the $\pi$-calculus. This comparative analysis provides insight into how these distinct approaches relate and differ.

Lastly, to assess the quality of our encoding, we adapt Gorla's five criteria to evaluate our encoding. These criteria provide clear standards to measure the effectiveness and clarity of our encoding.

## 3.1 Encoding variations of the $\lambda$-calculus in the $\pi$-calculus

Encoding the $\lambda$-calculus, and variations thereof, in the $\pi$-calculus is not a new concept. It has been an area of research, ever since Milner made his encoding of the CBN and CBV paradigms of the $\lambda$-calculus in the $\pi$-calculus in his seminal paper "Functions as Processes" [14].

Before we delve into encodings of the $\lambda$-calculus, we would like to motivate why Milner's encoding was so important, and why the continued study in the area is valuable. According to Sangiorgi, Milner's original encoding serves as an exercise to show the expressiveness of the $\pi$-calculus. Furthermore, the encodings Milner made show that the sequential behaviour of the $\lambda$-calculus could function as concurrent processes. This gives one a foundation for the semantics of how sequentially written functions can run concurrently [27]. Indeed this could potentially be useful for programming languages specifically designed to be executed on a GPU, such as Futhark [28], Dex [29], Accelerate [30] and Bend [31]. However, there exists many programming constructs which are not easily expressed in the $\lambda$-calculus. This motivates making extensions to the $\lambda$-calculus that contain these constructs and encodes this newly extended $\lambda$-calculus in the $\pi$-calculus.

Numerous different extensions of the $\lambda$-calculus has been encoded to various different $\pi$-calculi. One approach is to encode the $\lambda$-calculus into a more expressive $\pi$-calculus. Such as in [9], where Sangiorgi has encoded the $\lambda$-calculus to the Higher-Order $\pi$-calculus (HO$\pi$), and combined with his compilation of HO$\pi$ in the $\pi$-calculus, one can gain the exact same encoding as the one presented by Milner. Gérard Boudol has in [32] created an extension of the λ-calculus called name-passing-λ, which he encoded to his blue π-calculus, that being an extension of the π-calculus that is more reminiscent of the λ-calculus.

The encoding of λ-calculus to π-calculus is still a relevant field of study, and some more recent results in encoding the λ-calculus that is worth mentioning is [33], where Sangiorgi encodes the call-by-need λ-calculus in the polyadic local asynchronous π-calculus. The call-by-need calculus is closer to how more current programming languages implements call-by-name, where the argument is only evaluated once, and not multiple times. However, his encoding does not satisfy all the properties he wanted with regards to β-reductions in all contexts, showing an example of an area that could be explored. Another feature commonly seen in programming languages is that of references, a concept which is very unnatural to the purely functional nature of the λ-calculus. This stands in rather stark contrast to the π-calculus, where references are innate. Another recent result is where Prebet has extended the λ-calculus with references, and encodes it in the asynchronous internal π-calculus [34]. In our initial encoding of CBPV we have drawn inspiration from Milners encoding in [14].

## 3.2 Encoding CBPV in the π-calculus

We will now look at our encoding of CBPV in the π-calculus, which can be seen in Definition 24. We define $\mathcal{V}$ to be the function from names in the π-calculus to the corresponding variables in CBPV. We define $\mathcal{V}(x, y)$ to mean $\mathcal{V}(x), \mathcal{V}(y)$.

**Definition 24.** (Encoding of CBPV to π-calculus) For two distinct names $u, r \notin n(M, V)$ :

$$\llbracket \lambda x.M \rrbracket_u^r = u(s).s(u).s(r).s(x).\llbracket M \rrbracket_u^r \qquad \mathcal{V}(s) \notin \mathrm{fv}(M)$$

$$\llbracket M\ V \rrbracket_u^r = (\nu p)(\nu q)(\llbracket M \rrbracket_p^q \mid (\nu s)\overline{p}s.\overline{s}u.\overline{s}r.\llbracket V \rrbracket_s^r) \quad \mathcal{V}(p, q) \notin \mathrm{fv}(M, V)$$

$$\llbracket \mathbf{force}\ V \rrbracket_u^r = (\nu p)(\llbracket V \rrbracket_p^r \mid p(y).(\nu s)\overline{y}s.\overline{s}u.\overline{s}r) \qquad \mathcal{V}(p) \notin \mathrm{fv}(V)$$

$$\llbracket \mathbf{return}\ V \rrbracket_u^r = \llbracket V \rrbracket_r^r$$

$$\llbracket M \gg= \lambda x.N \rrbracket_u^r = (\nu z)((\nu u)\llbracket M \rrbracket_u^z \mid z(x).\llbracket N \rrbracket_u^r) \qquad \mathcal{V}(z) \notin \mathrm{fv}(M, N)$$

$$\llbracket V \rrbracket_u^r = (\nu y)(\overline{u}y.\llbracket y := V \rrbracket) \qquad \mathcal{V}(y) \notin \mathrm{fv}(V)$$

$$\llbracket y := x \rrbracket = !y(w).\overline{x}w$$

$$\llbracket y := \mathbf{thunk}\ M \rrbracket = !y(s).s(u).s(r).\llbracket M \rrbracket_u^r \qquad \mathcal{V}(s) \notin \mathrm{fv}(M)$$

In many ways the encoding takes inspiration from Milner's encoding of the CBN and CBV λ-calculi in the π-calculus seen in [14]. Many constructs are encoded similarly, an example being abstractions having an input prefix receiving a channel for the additional arguments, followed by input prefixes for the argument to the abstraction. Like Milner we also use the notion of handles $\llbracket \rrbracket_u^r$, however, instead of using one handle, our encoding has two handles; the top handle being the return handle $\llbracket \rrbracket^r$, and the bottom $\llbracket \rrbracket_u$ being the argument handle. The reasoning behind the "return" handle is that we need to know where to send our value when we have reduced down to a **return** V, the same way that the argument handle is used to announce that the process has reduced down to a value. If it just used the same handle, then if we had $\llbracket V \gg= \lambda x.M \rrbracket_u$ it would act the same as $\llbracket \mathbf{return}\ V \gg= \lambda x.M \rrbracket_u$, which should not be the case.

It is important to emphasize that each step in the encoding of any process is deterministic. However, since the resulting calculus of the encoding is the π-calculus, the reduction sequence becomes non-deterministic. This stands in contrast to the deterministic reduction strategy of CBPV, where the order of reductions is strictly defined. In the $\pi$-calculus, multiple reduction paths may exist, and different reductions can be applied at different times, leading to variation in the sequence of computational steps. Despite this non-determinism in reduction, an important property is preserved in our encoding: all reduction paths ultimately lead to the same result (up to multiple unfoldings of replications), regardless of the order in which reductions are applied. In fact, at all times only a single reduction path is possible in our encoding (again, up to multiple uses of the (Rep)-rule from Definition 20).

Within our encoding we have also taken the liberty of making a shorthand $[\![y := V]\!]$, which is a notation for granting access to the value $V$ through the name $y$. Below we give a more in depth description of each of the clauses in the encoding.

**(Abstraction)**: Similarly to Milner's encoding in [14], abstractions are encoded by having inputs. However, we need to have more inputs since we need to receive 2 handles instead of 1. Therefore the first object is the name on which further objects are received $u(s)$. This will ensure that the following objects are received in the correct order. The second object is where further arguments will be received in future applications. The third object is the return handle. The last object we receive is a pointer to the value that should be substituted. This is followed by the encoding of $M$ in which the handle $u, r$ has been replaced with the received names.

**(Application)**: We encode an application by making a private name $p$ on which the encoding of $M$ will receive a private channel $s$. $M$ is expected to reduce to an abstraction, and as such we send the object $u$ on the channel $s$ indicating on which channel it should receive further arguments after the application. Afterwards, it will receive the return handle $r$, and at last, it will be set up to communicate with the encoding of $V$, which will be able to send access to itself along the channel $s$. We also need a restriction on the return handle $q$, because otherwise, if we for an example had the encoding of $[\![((\mathbf{return}\ V_1)\ V_2) \gg= \lambda x.M]\!]_u^r$, then this could send $V_1$ to the binding, even though it was not fully reduced to only a Return.

**(Force)**: We encode **force** such that it is be able to unlock a **thunk**, since the **thunk** will need to be informed of both the return handle and the argument handle. To ensure that the information is received both at the same location, and in the correct order, the force first sends a new restricted name $(\nu s)\overline{y}s$, that will be the channel the rest will be sent over. Because of the encoding of values, this restricted name will either be able to be redirected to a **thunk**, or be sent directly to one.

**(Return)**: We encode the return of a value such that its argument handle is replaced with its return handle $r$.

**(Bind)**: A binding is encoded by making a private channel $a$ where the communication between the two operands happens. $M$ is expected to reduce to a return. This means that the encoding of $M$ will give access to this value along the handle $a$, which is is also the channel where the right operand will receives its argument. We restrict the handle $u$, to stop an outer application to happen, for example in the $\lambda$-expression $(\lambda x.M \gg= \lambda z.N)\ V$, where $\lambda x.M$ might otherwise receive arguments from $V$.

**(Value)**: These are encoded by creating a new private channel $y$ which we share on the channel $u$ such that the receiver has access to the encoded value.

**(Pointer to variable)**: A pointer to a variable is encoded as a forwarder, a notion introduced by Honda and Yoshida in [35]. This means that through the pointer $y$ one can make outputs on channel $x$.

**(Pointer to thunk)**: The pointer to thunk is encoded to take an input on the channel that has the name of the pointer, and the object is a channel on which it first receives the argument handle $u$, and afterwards the return handle $r$.

**Example 5.** (Encoding of a small example) $(\lambda x.\ \textbf{force}\ x)(\textbf{thunk}\ x)$

Here, we demonstrate how a small example behaves under our encoding. When bisimilarity laws $\approx_E$ are applied, it is because the scope of a restricted name appears only as an input, as in $(\nu y_1)(!y_1.N)$. Since the channel $y_1$ can never be accessed externally, no scope extrusion can occur, resulting in both the process and the restriction being redundant. They can therefore be removed for greater readability. Given our encoding, this is the only possible reduction sequence for $[\![M]\!]$ and $[\![V]\!]$ in the example (up to multiple unfoldings of the replication).

Expected reduction: $(\lambda x.\ \textbf{force}\ x)(\textbf{thunk}\ x) \to (\textbf{force}\ (\textbf{thunk}\ x)) \to x$

What follows is a simplified version of the full example, with only some of the bisimilarity laws being used to avoid clutter. For the full step-by-step encoding and reduction of the example, we refer to Appendix F.

We begin with the encoding of an application (1.1), where the left-hand side of the parallel $M$ is first encoded as an abstraction, and then the sub-term is encoded as **force** $V'$, with the value $V'$ encoded as $x$, resulting in (1.2). After several reduction steps, we reach (1.3). In (1.4) we use bisimilarity laws to remove inaccessible restrictions and subprocesses.

The right-hand side $V$ is encoded as $[\![V]\!]_s^r$ in (1.2), then encoded as a **thunk**, and finally represented as the value $x$ in (1.5). Through a sequence of reduction steps, we arrive at (1.6), and by applying bisimilarity laws, we clean up to obtain (1.7).

$$= (\nu p)(\nu q)\big([\![M]\!]_p^q \mid (\nu s)\overline{p}s.\overline{p}u.\overline{p}r.[\![V]\!]_s^r\big) \tag{1.1}$$

$$= (\nu p)(\nu q)(p(s).s(u).s(r).s(x).(\nu p_1)$$
$$\quad \big((\nu y)(\overline{p_1}y.!y(w).\overline{x}w) \mid p_1(y).(\nu s')\overline{y}s'.\overline{s'}u.\overline{s'}r)$$
$$\quad \mid (\nu s)\overline{p}s.\overline{p}u.\overline{p}r.(\nu y_1)(\overline{p}y_1.[\![y_1 := V]\!])) \tag{1.2}$$

$$\Rightarrow (\nu y_1)((\nu y)(!(y(w).\overline{y_1}w) \mid (\nu s)(\overline{y_1}s \mid \overline{s}u.\overline{s}r) \mid [\![y_1 := V]\!]) \tag{1.3}$$

$$\approx_E (\nu y_1)((\nu s)(\overline{y_1}s \mid \overline{s}u.\overline{s}r) \mid [\![y_1 := V]\!]) \tag{1.4}$$

$$= (\nu y_1)((\nu s)(\overline{y_1}s \mid \overline{s}u.\overline{s}r) \mid !y_1(s).s(u).s(r).(\nu y)(\overline{u}y.!y(w).\overline{x}w)) \tag{1.5}$$

$$\Rightarrow (\nu y_1)(\nu s)(!y_1(s).s(u).s(r).(\nu y)(\overline{u}y.!y(w).\overline{x}w) \mid (\nu y)(\overline{u}y.!y(w).\overline{x}w)) \tag{1.6}$$

$$\approx_E (\nu y)(\overline{u}y.!y(w).\overline{x}w) \tag{1.7}$$

This is exactly the encoding of $[\![x]\!]$.

An interesting thing to notice is, that when we have an encoding of an application $M\ V$ or a binding $M \gg= \lambda x.N$, it is only $M$ that is not behind prefix. This means that the only reductions available are those that involve the encoding of $M$, which is also what we would expect when reading the (Application-evolve)- and (Binding-evolve)-rules. We note that in the encoding $[\![\textbf{force}\ V]\!]_u^r$, $V$ is not behind prefix, as values are encoded in a way that does not disturb the rest of the program.

However, for this encoding to be meaningful, the reductions that the encoding can do must match the reductions that the original term could for all possible terms. This property is expressed in Theorem 1. We use $\mapsto$ to be the reflexive and transitive closure of $\to$ in CBPV.

**Theorem 1.** (Operational correspondence)
Let $M, N$ be expressions in the $\lambda$-calculus, let $r, u$ be distinct names, where $\mathcal{V}(r, u) \notin \text{fv}(M) \cup \text{fv}(N)$ and let $P, P'$ be processes in the $\pi$-calculus. We then have:

1. (Sound) if $M \to N$ then $\exists P. [\![M]\!]_u^r \Rightarrow P$ and $P \approx_E [\![N]\!]_u^r$.
2. (Complete) $\forall P.$ If $[\![M]\!]_u^r \Rightarrow P$, then $\exists N, P'.P \Rightarrow P'$ and $P' \approx_E [\![N]\!]_u^r$, and $M \mapsto N$.

   Furthermore, if $P \neq P'$ then $\forall a, b.P \overset{\overline{a}b}{\nrightarrow}$ and $P \overset{\overline{a}(b)}{\nrightarrow}$ and $P \overset{a(b)}{\nrightarrow}$.

Theorem 1 expresses that the encoding must be both sound (1) and complete (2) for it to be operationally correspondent. We also specify that all intermediate processes have no outgoing actions, but is only able to do internal reductions, until it reaches a process that is early bisimilar with an encoding of a term. This is not normally part of a completeness theorem, but it is useful for our proof of the completeness theorem, as well as a nice property of the encoding. In the definition of these we use $\approx_E$ as we must allow for additional internal communications in the $\pi$-calculus encoding to match a single action in the $\lambda$-calculus. We will now go through an example to show how the encoding and its reductions behave.

**Example 6.** (Encoding **return** and bind)

We will now encode the expression "**return** $y \ggg \lambda x.x$", and reduce it in the $\pi$-calculus. We expect this to reduce to a process early weak bisimilar to the encoding of $y$.

$$
\begin{aligned}
[\![\textbf{return } y \ggg \lambda x.x]\!]_u^r &= (\nu a)((\nu u)[\![\textbf{return } y]\!]_a^a \mid a(x).[\![x]\!]_u^r) \\
&= (\nu a)((\nu u)[\![y]\!]_a^a \mid a(x).((\nu a_2)\overline{u}a_2.[\![a_2 := x]\!])) \\
&= (\nu a)(\nu a_3)((\nu u)\overline{a}a_3.[\![a_3 := y]\!] \mid a(x).((\nu a_2)\overline{u}a_2.[\![a_2 := x]\!])) \\
&= (\nu a)(\nu a_3)((\nu u)\overline{a}a_3.!a_3(w).\overline{y}w \mid a(x).((\nu a_2)\overline{u}a_2.!a_2(w).\overline{x}w)) \\
&\to (\nu a)(\nu a_3)((\nu u)!a_3(w).\overline{y}w \mid (\nu a_2)\overline{u}a_2.!a_2(w).\overline{a_3}w) \\
&\approx_E (\nu a_2)(\nu a_3)((\nu u)!a_3(w).\overline{y}w \mid \overline{u}a_2.!a_2(w).\overline{a_3}w)
\end{aligned}
$$

This is the only possible reduction, as we now need another parallel process that can receive on the channel $u$ for us to reduce further.

This is early bisimilar to the behaviour of $[\![y]\!]_u^r$:

$$
[\![y]\!]_u^r = (\nu a)(\overline{u}a.(!a(w).\overline{y}w))
$$

Informally this is because they both output a bound name on $u$, and everything that is sent to that bound name will be sent via the channel $y$. For the encoding of "**return** $y \ggg \lambda x.x$" this just has an extra step, where the value, to be returned, on the channel $y$ is first send to the parallel process of $!a_3(w).\overline{y}w$ and then sent on the channel $y$.

We will now go to the proof of Theorem 1.

## 3.3 Soundness and completeness

We will now give an outline for the proof that our encoding is sound and complete. For the full proof, see Appendix A.

Before we can prove Theorem 1, we need Lemma 1. Lemma 1 is necessary with regards to substitution of values, as we in the $\pi$-calculus cannot substitute whole processes, but instead rely on pointers, whereas in CBPV we can substitute with values. The proof of Lemma 1 can be found in Appendix A.

**Lemma 1.** (Substitution-Lemma)

$$\llbracket M\{^V/_x\} \rrbracket_u^r \approx_E (\nu y)(\llbracket M \rrbracket_u^r \{^y/_x\} \mid \llbracket y := V \rrbracket)$$

We will throughout the proof of Theorem 1 take note of how many reductions in our encoding is equivalent to a reduction in CBPV, as this will be used later for a theorem about divergence reflection. The proof of Theorem 1 has two parts, soundness (1) and completeness (2). First we use induction on $\rightarrow$ to prove (1), and show that for all possible reductions in CBPV, the encoding can match them. For (2), we use structural induction on the term $M$, to show that no matter how the term is constructed, its behaviour is the same, when encoded in the $\pi$-calculus.

We will here show the proof of (1) for the (Application-base)-rule, and (2) when $M$ is an application.

*Proof.*
**Soundness of an application**:

We now look at the (Application-base)-rule.

$$\text{(Application-base)} \;\dfrac{}{(\lambda x.M)\ V \rightarrow M\{^V/_x\}}$$

For (1) to hold, it must be the case that $\llbracket (\lambda x.M)\ V \rrbracket_u^r$ can reduce to a process that is early bisimilar with $\llbracket M\{^V/_x\} \rrbracket_u^r$.

We begin with the encoding:

$$\llbracket (\lambda x.M)\ V \rrbracket_u^r = (\nu p)(\nu q)(p(s).s(u).s(r).s(x).\llbracket M \rrbracket_u^r \mid (\nu s)\overline{p}s.\overline{s}u.\overline{s}r.((\nu q')\overline{s}q'.\llbracket q' := V \rrbracket))$$

We will now reduce this:

$$(\nu p)(\nu q)(p(s).s(u).s(r).s(x).\llbracket M \rrbracket_u^r \mid (\nu s)\overline{p}s.\overline{s}u.\overline{s}r.((\nu q')\overline{s}q'.\llbracket q' := V \rrbracket))$$
$$\xrightarrow{\tau} (\nu p)(\nu q)(\nu s)(s(u).s(r).s(x).\llbracket M \rrbracket_u^r \mid \overline{s}u.\overline{s}r.((\nu q')\overline{s}q'.\llbracket q' := V \rrbracket))$$
$$\xrightarrow{\tau} (\nu p)(\nu q)(\nu s)(s(r).s(x).\llbracket M \rrbracket_u^r \mid \overline{s}r.((\nu q')\overline{s}q'.\llbracket q' := V \rrbracket))$$
$$\xrightarrow{\tau} (\nu p)(\nu q)(\nu s)(s(x).\llbracket M \rrbracket_u^r \mid ((\nu q')\overline{s}q'.\llbracket q' := V \rrbracket))$$
$$\xrightarrow{\tau} (\nu p)(\nu q)(\nu s)(\nu q')\big(\llbracket M \rrbracket_u^r \{^{q'}/_x\} \mid \llbracket q' := V \rrbracket\big)$$

We will now show the following, using up-to-bisimulation, meaning we can apply our bisimulation laws:

$$(\nu p)(\nu q)(\nu q')(\nu s)\big(\llbracket M \rrbracket_u^r \{^{q'}/_x\} \mid \llbracket q := V \rrbracket\big) \approx_E \llbracket M\{^V/_x\} \rrbracket_u^r$$

First of all, we can ignore the restriction $(\nu p)$, $(\nu q)$ and $(\nu s)$ as there is not any instances of $p$,$q$ or $s$ anymore. Therefore, they can easily be removed using our bisimulation laws. We then have:

$$(\nu q')\big(\llbracket M \rrbracket_u^r \{^{q'}/_x\} \mid \llbracket q' := V \rrbracket\big) \approx_E \llbracket M\{^V/_x\} \rrbracket_u^r$$

This can be concluded using Lemma 1. With this, we showed we can match the reduction in CBPV with 4 reductions in the $\pi$-calculus, and therefore proved (1), when it is concluded using the (Application-base)-rule. We also notice, that because we use Lemma 1, we "remove" a link, that might increase the number of reductions to match future reductions by 1, since if $V = y$, we have that $[\![q' := y]\!] = !q'(a).\overline{y}a$ because if:

$$[\![M]\!]^r_u \{ {}^y/_x \} \overset{\overline{y}a}{\to}$$

Then without using Lemma 1 we would have:

$$(\nu q')\big([\![M]\!]^r_u \big\{ {}^{q'}/_x \big\} \mid !q'(a).\overline{y}a\big) \overset{\tau}{\to} (\nu q')\big([\![M]\!]^r_u \big\{ {}^{q'}/_x \big\} \mid !q'(a).\overline{y}a \mid \overline{y}a\big) \overset{\overline{y}a}{\to}$$

Here we see the extra $\tau$-reduction needed for future reductions.

**Completeness of an application**:

We look at the encoding of an application, where

$$M = M' \; V'$$

We will be using structural induction on $M$, and our inductive hypothesis will be the following: If $[\![M']\!]^b_a \overset{\tau}{\to} Q$, then $Q \Rightarrow Q'$ and $Q' \approx_E [\![N]\!]^b_a$, without being able to communicate with its environment until it reaches Q'. We also use induction on $\Rightarrow$, which means we will assume that forall Q, where $[\![M \; V]\!]^r_u \Rightarrow Q$ and $Q \approx_E [\![N']\!]^r_u$, then if $N' \neq M$, (2) holds for $N'$. This essentially mean we will prove (2) until we reach a process that is early weak bisimilar with a new term.

We have that the encoding of this will be:

$$[\![M' \; V']\!]^r_u = (\nu a)(\nu b)\big([\![M']\!]^b_a \mid (\nu s)\overline{a}s.\overline{s}u.\overline{s}r.[\![V']\!]^r_u\big)$$
$$= (\nu a)(\nu b)\big([\![M']\!]^b_a \mid (\nu s)\overline{a}s.\overline{s}u.\overline{s}r.(\nu y).\overline{s}y.[\![y := V']\!]\big)$$

For this to reduce, there are two cases; either $[\![M']\!]^b_a \overset{\tau}{\to}$, or $[\![M']\!]^b_a$ communicates with the output $\overline{a}s$. In case of $[\![M']\!]^b_a \overset{\tau}{\to}$, where we have internal communication, we have from the inductive hypothesis know, that it will further reduce to a process that is early weak bisimilar to the encoding of a CBPV expression, without having any outgoing outputs or inputs, until it reaches a process weak bisimilar with an encoding. Therefore, without loss of generality, we can ignore the internal communications in $[\![M']\!]^b_a$, as it will eventually end up as a process that is early weak bisimilar to an encoding of a CBPV expression again.

In the other case, if it is able to reduce, it must be the case that $[\![M']\!]^b_a \overset{a(s)}{\to} Q$. By inspection of the encoding rules, for this to be the case, $M'$ must be of the form $\lambda x.M''$, as in all other encoding rules, either their sub-expressions are restricted with regards to the argument handle, or behind output prefixes. We therefore have:

$$(\nu a)(\nu b)\big([\![\lambda x.M'']\!]^b_a \mid (\nu s)\overline{a}s.\overline{s}u.\overline{s}r.(\nu y).\overline{s}y.[\![y := V']\!]\big)$$
$$= (\nu a)(\nu b)\big(a(s).s(u).s(r).s(x).[\![M'']\!]^r_u \mid (\nu s)\overline{a}s.\overline{s}u.\overline{s}r.(\nu y).\overline{s}y.[\![y := V']\!]\big)$$
$$\to (\nu a)(\nu b)(\nu s)\big(s(u).s(r).s(x).[\![M'']\!]^r_u \mid \overline{s}u.\overline{s}r.(\nu y).\overline{s}y.[\![y := V']\!]\big)$$

Now we need to show that there exists a $P'$ and a $N$, where we can reduce to $P'$, and $P' \approx_E [\![N]\!]^r_u$, and we have that $M \to N$. We begin by further reducing, until we reach a suitable process. We notice that in all intermediate processes, the process as a whole has no barbs, meaning that it can only do internal communications, and not output or input actions. This is the case since all communication happens on the channel $s$, that is restricted:

$$(\nu a)(\nu b)(\nu s)\big(s(u).s(r).s(x).[\![M']\!]_u^r \mid \bar{s}u.\bar{s}r.(\nu y).\bar{s}y.[\![y := V']\!]\big)$$

$$\Rightarrow (\nu a)(\nu b)(\nu s)(\nu y)\big([\![M']\!]_u^r\{^y/_x\} \mid [\![y := V']\!]\big)$$

This will be our $P'$. We then see that if $N = M'\big\{^{V'}/_x\big\}$, we have that $(\lambda x.M')V' \to M'\big\{^{V'}/_x\big\}$, and we can then show that $P' \approx_E [\![M'\big\{^{V'}/_x\big\}]\!]_u^r$. We note that for each intermediate process until here, we can choose the same $P'$ and $N$, and they will not have had any outgoing behaviour, as the channels for communication have been restricted until now. From Lemma 1 and our bisimilarity laws we have the following:

$$(\nu a)(\nu b)(\nu s)(\nu y)\big([\![M']\!]_u^r\{^y/_x\} \mid [\![y := V']\!]\big) \approx_E (\nu y)\big([\![M']\!]_u^r\{^y/_x\} \mid [\![y := V']\!]\big)$$

$$\approx_E [\![M'\big\{^{V'}/_x\big\}]\!]_u^r$$

We have reached a process that is early weak bisimilar to an encoding of a $\lambda$-expression, so by the inductive hypothesis, this will hold for future reductions. With this, we can say that (2) holds when $M$ is an application. We of course also noticed that this is the same reduction as the reduction in the proof of (1) for application.

<div style="text-align: right">□</div>

Having showcased how to prove Theorem 1, we will now look at some of the properties of our encoding.

## 3.4 Correspondence with direct encodings

In [14], Milner encodes both CBV and CBN in the $\pi$-calculus. We now look at how similar our encodings are to his two encodings. Specifically, if you encode either a variable or an abstraction directly in the $\pi$-calculus, or if you first encode it to CBPV, and then use our encoding, you get very similar processes as a result. The different paths to encode CBV and CBN in the $\pi$-calculus can be seen on Figure 1. It would be very nice if this diagram was commutative, meaning regardless of the path through the diagram, you ended up with the same process. However, this is not exactly the case, but we will argue that you end up with processes that are very similar in behaviour.

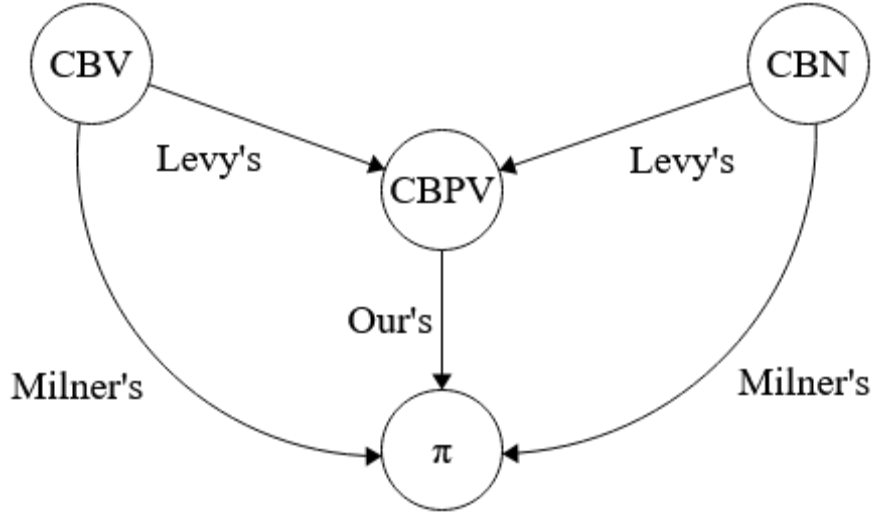

Figure 1: Graph that shows the different ways to encode CBV and CBN in the $\pi$-calculus. Milner's encodings are defined in Definition 26 and Definition 25.

**Definition 25.** (Milner's encoding of CBN)

$$\mathcal{L}[\![M\ N]\!]_u = (\nu v)(\mathcal{L}[\![M]\!]_v \mid (\nu x)\overline{v}x..\overline{v}u.\mathcal{L}[\![x := N]\!]) \qquad \mathcal{V}(x) \notin \mathrm{fv}(N), \mathcal{V}(v) \notin \mathrm{fv}(M)$$

$$\mathcal{L}[\![\lambda x.M]\!]_u = u(x).u(v).\mathcal{L}[\![M]\!]_v \qquad\qquad\qquad\qquad \mathcal{V}(v) \notin \mathrm{fv}(M)$$

$$\mathcal{L}[\![x]\!]_u = \overline{x}u$$

$$\mathcal{L}[\![x := M]\!] = !x(w).\mathcal{L}[\![M]\!]_w$$

**Definition 26.** (Milner's encoding of CBV)

$$\mathcal{V}[\![M\ N]\!]_p = (\nu qr)(q(y).(\nu v)\overline{y}v.r(z).\overline{v}z.\overline{v}p \mid \mathcal{V}[\![M]\!]_q \mid \mathcal{V}[\![N]\!]_r) \qquad \mathcal{V}(q,r) \notin \mathrm{fv}(N) \cup \mathrm{fv}(M)$$

$$\mathcal{V}[\![V]\!]_p = (\nu y)\overline{p}y.\mathcal{V}[\![y := V]\!] \qquad\qquad\qquad\qquad \mathcal{V}(y) \notin \mathrm{fv}(V)$$

$$\mathcal{V}[\![y := x]\!] = !y(w).\overline{x}w$$

$$\mathcal{V}[\![y := \lambda x.M]\!] = !y(w).w(x).w(p).\mathcal{V}[\![M]\!]_p$$

In Milner's encoding of the CBN $\lambda$-calculus, an abstraction is encoded as:

$$\mathcal{L}[\![\lambda x.M]\!]_u = u(x).u(v).\mathcal{L}[\![M]\!]_v$$

This is almost what our encoding of an abstraction in CBPV is, with the exception of receiving the additional "return" handle, and ensuring a secure channel for communication. This also matches neatly with Levy's encoding of CBN in CBPV, where $[\![\lambda x.M]\!]^n = \lambda x.[\![M]\!]^n$. However, variables are encoded differently, where we have in Milner's CBN encoding that $\mathcal{L}[\![x]\!]_u = \overline{x}u$. However, if we encode CBN to CBPV using Definition 13 by Levy, and then encode it in the $\pi$-calculus using our encoding from Definition 24, we actually get something similar:

$$[\![\ [\![x]\!]^n\ ]\!]^r_u = [\![\textbf{force}\ x]\!]^r_u = (\nu p)(\overline{p}y.!y(a).\overline{x}a \mid p(y).(\nu s)\overline{y}s.\overline{s}u.\overline{s}r)$$

This can reduce down to the following:

$$(\nu p)((\nu y)\overline{p}y.!y(a).\overline{x}a \mid p(y).(\nu s)\overline{y}s.\overline{s}u.\overline{s}r) \Rightarrow (\nu p)(\nu y)(\nu s)(!y(a).\overline{x}a \mid \overline{s}u.\overline{s}r \mid \overline{x}s)$$
$$\approx_E (\nu s)(\overline{x}s \mid \overline{s}u.\overline{s}r)$$

Here we also see that our encoding outputs on the channel $x$, however, what we output is a little different, as it is a channel on which we can receive the two handles. This stems from the fact that we have two handles in our encoding, but there is only one in Milner's encoding. However, if we instead of using a monadic encoding, used a polyadic encoding $\mathcal{P}[\![\ ]\!]$, then this could be avoided. In Definition 27 we show a full translation to a polyadic $\pi$-calculus encoding. If we for example encoded **force** $x$ in the polyadic $\pi$-calculus, we would have:

$$\mathcal{P}[\![\textbf{force } x]\!]_u^r = (\nu p)((\nu y)\overline{p}y.!y(u,r).\overline{x}[u,r] \mid p(y).\overline{y}[u,r])$$
$$\Rightarrow (\nu p)((\nu y)\overline{p}y.!y(u,r).\overline{x}[u,r] \mid \overline{x}[u,r])$$
$$\approx_E \overline{x}[u,r]$$

As seen here, our encoding would then closely match Milner's encodings. In fact we could get an even more similar encoding, if we used a bit of foresight when encoding CBPV. We write $e[\![\ ]\!]$ for this 'smart' encoding, in the following way:

$$e[\![\textbf{force } x]\!]_u^r = \overline{x}[u,r]$$
$$e[\![\textbf{force thunk } M]\!]_u^r = [\![M]\!]_u^r$$

However, we felt that this was overspecifying our encoding, instead of relying on the syntax of CBPV. If we now look at CBV, Milner's encoding of a variable is:

$$\mathcal{V}[\![x]\!]_p = (\nu y)\overline{p}y.!y(w).\overline{x}w$$

This is also the encoding of a variable in our CBPV encoding. If we however first encode the CBV to CBPV using Definition 12 by Levy, we get that $[\![x]\!]^v = \textbf{return } x$. If we encode this, we get that $[\![\textbf{return } x]\!]_u^r = [\![x]\!]_r^r$. So our encoding is similar to the encoding of variables in the CBV, with the exception that instead of $u$, we use the other handle $r$. If we inspect the encoding of an abstraction in the CBV, we have that in Milner's encoding this is:

$$\mathcal{V}[\![\lambda x.M]\!]_p = (\nu y)\overline{p}y.!y(w).w(x).w(p)[\![M]\!]_p$$

This is very different from our encoding of abstractions, but this comes from the fact that when encoding abstractions from CBV to CBPV we have that $[\![\lambda x.M]\!]^v = \textbf{return thunk } \lambda x.[\![M]\!]^v$. If we further encode this with our encoding, we get:

$$[\![\textbf{return thunk } \lambda x.[\![M]\!]^v]\!]_u^r = (\nu y)\big(\overline{r}y.!y(s).s(u).s(r).\big(u(s).s(u).s(r).s(x).[\![\ [\![M]\!]^v\ ]\!]_u^r\big)\big)$$

This is in essence the same as the encoding of an abstraction in CBV, with it outputting a pointer, and then receiving a lot of set-up on that pointer. If we used a polyadic encoding, we would be even closer, as we would not need the intermediate channels of communication. We would then have:

$$\mathcal{P}[\![\textbf{return thunk } \lambda x.[\![M]\!]^v]\!]_u^r = (\nu y)\big(\overline{r}y.!y(u',r').u'(u,r,x).\big([\![\ [\![M]\!]^v\ ]\!]_u^r\big)\big)$$

This is still different from Milner's encoding, because instead of receiving a secure link, and then $x$ and the handle $p$, we first receive the handles $u$ and $r$, which in some sense act as our secure channel here, where we then receive on that secure channel $x$ and our handles for the rest of the encoding.

**Definition 27.** (Polyadic encoding of CBPV in the $\pi$-calculus)

$$\mathcal{P}[\![\lambda x.M]\!]_u^r = u(v,r).u(x).\mathcal{P}[\![M]\!]_v^r \qquad\qquad v \notin \text{fn}(M)$$

$$\mathcal{P}[\![M\ V]\!]_u^r = (\nu pq)\big(\mathcal{P}[\![M]\!]_p^q \mid \overline{p}[u,r].\mathcal{P}[\![V]\!]_p^r\big) \qquad p \notin \text{fn}(M,V)$$

$$\mathcal{P}[\![\textbf{force}\ V]\!]_u^r = (\nu p)\big(\mathcal{P}[\![V]\!]_p^r \mid p(y).\overline{y}[u,r]\big) \qquad\qquad p \notin \text{fn}(V)$$

$$\mathcal{P}[\![\textbf{return}\ V]\!]_u^r = \mathcal{P}[\![V]\!]_r^r$$

$$\mathcal{P}[\![M \gg= \lambda x.N]\!]_u^r = (\nu z)((\nu u)\mathcal{P}[\![M]\!]_u^z \mid z(x).\mathcal{P}[\![N]\!]_u^r) \qquad z \notin \text{fn}(M,N)$$

$$\mathcal{P}[\![V]\!]_u^r = (\nu y)\big(\overline{u}y.\mathcal{P}[\![y := V]\!]\big) \qquad\qquad y \notin \text{fn}(V)$$

$$\mathcal{P}[\![y := x]\!] = !y(u,r).\overline{x}[u,r]$$

$$\mathcal{P}[\![y := \textbf{thunk}\ M]\!] = !y(u,r).\mathcal{P}[\![M]\!]_u^r \qquad\qquad s \notin \text{fn}(M)$$

We hope to have shown that if instead of just encoding either the CBV or CBN directly in the $\pi$-calculus using Definition 25 and Definition 26, if you use our encoding, as an intermediate language, you would, in the end, still get a process with a behaviour closely matching that of the direct encoding. We have attempted to define the exact correspondence between the encodings in Definition 28, Definition 29 and Conjecture 1.

**Definition 28.** (Correspondence between CBN and double encoding)
For all distinct names $u, r$ and two processes $P_{\text{CBN}}, P_{\text{CBPV}}$, we say that $\ominus_u^r$ is the largest relation where $P_{\text{CBN}} \ominus_u^r P_{\text{CBPV}}$ implies the following:

(1). If $\exists x.P_{\text{CBN}} \xoverset{\overline{x}u}{\Rightarrow} Q_{\text{CBN}}$, then $\exists Q_{\text{CBPV}}.P_{\text{CBPV}} \xoverset{\overline{x}[u,r]}{\Rightarrow} Q_{\text{CBPV}}$ and $Q_{\text{CBN}} \ominus_u^r Q_{\text{CBPV}}$

(2). If $\exists w, x, p.P_{\text{CBN}} \xoverset{u(w)}{\Rightarrow} Q'_{\text{CBN}} \xoverset{w(x)}{\Rightarrow} Q''_{\text{CBN}} \xoverset{w(p)}{\Rightarrow} Q_{\text{CBN}}$, then $\exists p'.\exists Q', Q''.P_{\text{CBPV}} \xoverset{u(p',r)}{\Rightarrow} Q' \xoverset{u(x)}{\Rightarrow} Q''$ and $Q_{\text{CBN}} \ominus_u^r Q''$

Similarly, we have for CBV the following correspondence between encodings:

**Definition 29.** (Correspondence between CBV and double encoding)
For distinct names $u, r$ and two processes $P_{\text{CBV}}, P_{\text{CBPV}}$, we say that $\odot_u^r$ is the largest relation where $P_{\text{CBV}} \odot_u^r P_{\text{CBPV}}$ implies the following:

(1). If $\exists y.P_{\text{CBV}} \xoverset{\overline{u}(y)}{\Rightarrow} Q_{\text{CBV}}$, then $\exists r.\exists Q_{\text{CBPV}}.P_{\text{CBPV}} \xoverset{\overline{r}(y)}{\Rightarrow} Q_{\text{CBPV}}$ and $Q_{\text{CBV}} \odot_u^r Q_{\text{CBPV}}$

(2). If $\exists y.P_{\text{CBV}} \xoverset{y(x)}{\Rightarrow} Q'_{\text{CBV}} \xoverset{y(u)}{\Rightarrow} Q_{\text{CBV}}$, then $\exists u', r', r.\exists Q', Q'', Q'''.P_{\text{CBPV}} \xoverset{y(u',r')}{\Rightarrow} Q' \xoverset{u'(u,r)}{\Rightarrow} Q'' \xoverset{u'(x)}{\Rightarrow} Q'''$ and $Q_{\text{CBV}} \odot_u^r Q'''$

With these, we can define a result we believe our encoding satisfies:

**Conjecture 1.** (Encoding stays inside correspondence)
For all terms $M$, and handles $u, r \notin \text{fn}(M)$, we have:

(1). $\mathcal{L}[\![M]\!]_u \ominus_u^r [\![[\![M]\!]^n]\!]_u^r$

(2). $\mathcal{V}[\![M]\!]_u \odot_u^r [\![[\![M]\!]^v]\!]_u^r$

What Conjecture 1 essentially says is that our encodings have a way to match the behaviour of Milner's encodings. This could most likely be proven using a co-inductive proof. It might also be possible to derive it from the encodings of variables and abstractions, as well as the soundness and completeness property. However we leave the proof of this correspondence as future work.

Now that we have mentioned the similarities, we should also note that Milner has encoded a slightly different CBV calculus, as his CBV allows parallel reductions, meaning he allows the object of the application to reduce before the subject has reduced to a value. This is also evident in his encoding of CBV, as his encoding of application does not have the object behind prefix. However, the CBV we have been showing properties of, has a strict reduction order. Since convergence is deterministic in Milner's CBV, the expressive power of the calculi are still the same. Essentially, all terms will eventually reduce to the same term, regardless of which of the two CBV you use.

## 3.5 Unified properties of encodings

In [1], Gorla presents 5 properties to evaluate the quality of encodings between process calculi. These are general enough that it is also interesting to see if our encoding satisfies them, despite it being an encoding of a computational calculus, and not a process calculus. The properties are defined as in Definition 30 for our encoding. We let $M \overset{\omega}{\mapsto}$ denote divergence, i.e. that $M$ has an infinite reduction sequence, and let $P \overset{\omega}{\to}$ denote the same property for processes.

We also need to define a "success" property. We define this in the CBPV as reducing to a **return**. In the $\pi$-calculus, our "success" property should match this by being equivalent to the encoding of a **return**. In our encoding this can be satisfied when we are able to send a pointer on the handle $r$. By abuse of notation, we let $C[P_1, ..., P_n]$ denote a context where there is multiple holes in $C$, and each hole is filled with a process $P$, where $P \in \{P_1...P_n\}$.

**Definition 30.** (Properties of encodings) An encoding from CBPV in the $\pi$-calculus should satisfy the following:

(1). (Compositionality) An encoding $[\![\cdot]\!]_u^r$ is compositional if, for every term constructor *op* of terms in CBPV, there exists a context $C$ and a set of names $N = \{n_1, ...n_m\}$, such that $[\![op(M_1, ..., M_k)]\!]_u^r = C\Big[[\![M_1]\!]_{n_1}^{n_2}; ...[\![M_k]\!]_{n_{m-1}}^{n_m}\Big]$.

(2). (Name invariance) An encoding $[\![\cdot]\!]_u^r$ is name invariant, if, for every term $M$, and every substitution $\sigma$ in CBPV, we have that:

$$[\![M\sigma]\!]_u^r \begin{cases} = [\![M]\!]_u^r \sigma' & \text{if } \sigma \text{ is injective} \\ \approx_E [\![M]\!]_u^r \sigma' & \text{otherwise} \end{cases}$$

Where $\sigma'$ is a substitution in the $\pi$-calculus, and $\sigma'(\textbf{Val}^{-1}(a)) = \textbf{Val}^{-1}(\sigma(a))$.

(3). (Operational correspondence) An encoding $[\![\cdot]\!]_u^r$ satisfies the property of operational correspondent if it is:

- *Sound* : For all $M \Rightarrow N$, $[\![M]\!]_u^r \Rightarrow P$ and $P \approx_E [\![N]\!]_u^r$.
- *Complete* : For all $P$ where $[\![M]\!]_u^r \Rightarrow P$, there exists an $N$ so that $M \mapsto N'$ and $P \Rightarrow P'$ and $P' \approx_E [\![N]\!]_u^r$

(4). (Divergence reflection) An encoding $[\![\cdot]\!]_u^r$ reflects divergence, if for every $M$ where $[\![M]\!]_u^r \overset{\omega}{\to}$, it holds that $M \overset{\omega}{\mapsto}$.

(5). (Success sensitiveness) An encoding $[\![\cdot]\!]_u^r$ is success sensitive if, for every $M$, it holds that $M \mapsto \textbf{return } V$ if and only if $[\![M]\!]_u^r \Rightarrow P$ and $P \overset{\overline{r}(y)}{\to} P'$.

Our encoding satisfies property 1, as each encoding can be described as a context around subencodings. We also believe it satisfies the property of 2, under the assumption that the handles $u$ and $r$ do not appear in the substitution. If they did, we could just choose some other handles. This is because we use equivalent names in the encoding of a term, as in the term itself, and all other names are restricted. We have proven that our encoding also satisfies property 3, even satisfying a stronger completeness requirement. However, as stated in [1], this is not really an impressive feat, as an encoding that just replaces everything with 0, would also satisfy property 1-3. On the other hand, property 4 and 5 require our encoding to actually show that our encoding matches divergence, and reaches an acceptance state, if the encoded term does. We will now prove that our encoding also satisfies property 4 and 5.

*Proof.*

**Property 4**: To prove that our encoding satisfies property 4, we will prove the contra-positive form, which states that:

$$a \to b \text{ iff } \neg b \to \neg a$$

In our context this means that we will prove:

$$\neg \left( M \overset{\omega}{\mapsto} \right) \text{ implies } \neg \left( [\![ M ]\!]_u^r \overset{\omega}{\to} \right)$$

If $M$ does not diverge, it means that in $n$ reductions, it reduces to a terminal value. We then need to show that $[\![ M ]\!]_u^r$ reduces to a final process within some bounded number of reductions. We saw from the completeness proof, that no infinite reductions are introduced, or in other words, each process would terminate to a process that is early weak bisimilar with an encoding of a term, or a final process, with finite many reductions. We have in the proof for soundness (and completeness) shown that we can mimic each transition with a finite amount of steps in the $\pi$-calculus, without introducing divergence. We should note, that for each time we use Lemma 1, to simplify a substitution, we actually "hide" a possible extra reduction for all future reductions (from an introduced link). However, since at the outermost level it at most takes 4 reductions to match a reduction in CBPV, and then future reduction might use 1 reduction more, we can still have a upper bound on the number of reductions made in the $\pi$-calculus to match a single reduction in CBPV, which is $4 + n$. We also realize that if the term is **force** $x$, then there is two possible reductions in the $\pi$-calculus, even though $n = 0$. Furthermore, these two reductions can also have been incremented by each previous reduction because of introduced links, before we reach the final term **force** $x$. We can therefore bound the number of reductions made in the $\pi$-calculus to reach a final process by $(4 + n) * n + (n + 2)$. Since we have shown that we have an upper bound on the number of reductions, we therefore have proven the contra-positive, and the original must also be true, that:

$$[\![ M ]\!]_u^r \overset{\omega}{\to} \text{ implies } M \overset{\omega}{\mapsto}$$

This in some sense also tells us something about how "efficient" our encoding is, since it can match any computation in CBPV with $O(n^2)$ reductions in the $\pi$-calculus.

**Property 5**: If property 5 is to hold, it must be the case that we can have an output on the channel $r$, if and only if we end up with a process that is early weak bisimilar with $[\![ \textbf{return } V ]\!]_u^r$. By inspection of the encoding, this is also the case, as in all other cases, either there is restrictions that stops a subprocesses from outputting on the $r$ handle, or such an output is behind prefix. And from the soundness and completeness proof, we have shown that any encoding reduces down to a process that is early weak bisimilar with the encoding of a **return**, if and only if the encoded term also would reduce to a **return**. Therefore, our encoding also satisfies property 5.

$\square$

We could give a more specialized version of property 5, by saying something about the final $P'$ where $P \overset{\overline{r}y}{\to} P'$, as it could be represented in the following way, for some context $C$ with multiple holes:

$$P' = C[\llbracket y := a_1 \rrbracket, ..., \llbracket a_n := V \rrbracket]$$

We could also specify other correctness properties, such as **force** $x$ having a special action, (this would be that $\llbracket M \rrbracket_u^r \Rightarrow P$ and $P \overset{\overline{x}(u,r)}{\to}$), as this is something that the encoding of CBV in CBPV could reduce down to. Similarly, if CBN reduces down to an abstraction, then in CBPV we would reduce down to a **return thunk** $M$, which would also have special "barbs" we could recognize, besides just the pointer send on $r$.

# 4 Formalization

In this chapter we will be formalizing CBPV and the $\pi$-calculus, which we later specialize to the internal $\pi$-calculus, $\pi I$-calculus, in Coq. As we are not the first to formalize the CBPV or the $\pi$-calculus in a proof assistant, we will be drawing inspiration/solutions from the literature. For formalizing the $\pi$-calculus, our main inspiration was an already existing formalization by Hirschkoff in [36], where he used de Bruijn indices and the locally nameless approach to formalize the $\pi$-calculus in Coq.

We start the chapter with an introduction to de Bruijn indices. Following this we go through the steps of formalizing the body of knowledge introduced in the previous chapters, going trough the necessary adjustments needed to make it easily implementable in Coq. When appropriate, we have opted to use equations and figures instead of code to present the formalization, as this approach provides a clearer overview. The entirety of the Coq formalization, with inline comments referencing the report's definitions, can be found at [37].

## 4.1 De Bruijn indices and the locally nameless approach

When formalizing the proofs in Coq we need to be able to represent binders and their bound occurrences. To tackle this, one can use higher order abstract syntax, nominal logic or de Bruijn indices to represent names. As our previous work has been focused on the use of de Bruijn indices, we continue using an approach that is very similar, namely the locally nameless approach. For a detailed discussion of our reasons for our choice of using de Bruijn indices we refer to our earlier work [21].

The central idea to de Bruijn indices is that all names are represented with natural numbers, such that each number "refers to the nesting level of its binder" [21]. This was first introduced by de Bruijn for the $\lambda$-calculus in [38], which our introductory example will be based on. Our realization of the $\lambda$-calculus with de Bruijn indices in Coq follows Stark as presented in [39]. The main advantage of this is that $\alpha$-equivalence comes for free, in the sense that it will equate to syntactic equality. This also means that $\alpha$-conversion is not needed, which implies that there is no need to algorithmically find fresh names. However, de Bruijn indices also have the drawbacks of being less readable to humans, but more importantly there is a lot of bookkeeping to ensure that the bound occurrences always refer to the correct binder. The latter becomes especially evident in the $\pi$-calculus as scope-extrusion allows restrictions to swap places.

We define the syntax for nameless representation of the $\lambda$-calculus using de Bruijn indices as seen in Definition 31. We let $\mathbf{Expr_{DB}}$ range over the set of possible terms.

**Definition 31.** ($\lambda$-calculus syntax with de Bruijn index)
$$M, N \Coloneqq n \mid MN \mid \lambda M \quad \text{where } n \in \mathbb{N}_0$$

**Example 7.** (De Bruijn index using the nameless representation)

Using a named representation one can write:
$$\lambda f.f((\lambda x.fx)yzy)$$

Transforming the term to a nameless representation, the variable $f$ is first referenced as 0, and then later as 1 as another binder has been introduced. However, the free variable $y$ is also represented as a 1, because $y$ is only used under a single binder, and as such it is still free. The free variable $z$ gets an index of 2, as it is also free, but refers to another free name than $y$.

$$\lambda 0((\lambda 10)121)$$

Keeping track of free variables can be a bit tedious, so it is often convenient to write $(y + n)$, where $y \in \mathbb{N}_0$. With this approach, the variable $y$ remains free regardless of how many binders are added. In general, we add to the index of a variable the number of binders it is nested under, ensuring it stays free. This method also makes it easier to read and write expressions with free variables—when we write a variable as $(y + 1)$, we signal that it skips 1 binder to be just beyond any binder, effectively marking the point where the free variables begin. The number $n$ then serves to uniquely identify the free variable, allowing us to distinguish between different ones or recognize when two occurrences refer to the same variable. The example would then be:

$$\lambda 0((\lambda 10)(y + 1)(z + 1)(y + 1))$$

As reasoned above, this is why the variable originally indexed as 2 becomes $(z + 1)$ and not $(z + 2)$, as the $+1$ accounts for the single binder we have skipped, placing us just past the binding scope and correctly preserving its status as a free variable.

In the case of the $\lambda$-calculus, a substitution is a function $\mathbb{N}_0 \to \mathbf{Expr_{DB}}$, though this definition will vary depending on the calculus. Renamings are a subclass of substitutions, only returning names, meaning a renaming is a function $\mathbb{N}_0 \to \mathbb{N}_0$. We use $\sigma$ and $\varepsilon$ to denote substitutions and renamings respectively.

We define three auxiliary functions; id, shift, and extend.

**Definition 32.** (id: $\mathbb{N} \to \mathbb{N}$) We define id as the renaming:

$$\mathbf{id}(n) = n$$

**Definition 33.** (shift $\uparrow$: $\mathbb{N} \to \mathbb{N}$) We define shift as the renaming:

$$\uparrow n = n + 1$$

**Definition 34.** (extend) We define extend as a function $\mathbf{Expr_{DB}} \to (\mathbb{N}_0 \to \mathbf{Expr_{DB}}) \to \mathbb{N}_0 \to \mathbf{Expr_{DB}}$, meaning an extension of a substitution.

$$(M \triangleright \sigma)(n) = \begin{cases} M & n = 0 \\ \sigma(n - 1) & n > 0 \end{cases}$$

Lifting as defined in Definition 35 is needed to define a substitution that is capture-avoiding, and is specifically used when passing over a binder. It achieves capture-avoiding behaviour by using extend to ensure that any 0's remain unchanged, and composing the substitution with an instantiation (described in the following paragraph) of the shift renaming, such that any terms substituted into the surrounding term also account for another occurrence of a binder.

**Definition 35.** (Lifting)

$$\Uparrow \sigma = 0 \triangleright (\{\uparrow\} \circ \sigma)$$

With the previous auxiliary functions defined, instantiations of substitution can be defined rather easily as seen on Definition 36. It is simply a matter of propagating the substitution down through any sub-terms. When a binder occurs we need to use lift, and when a variable occurs we apply the substitution.

**Definition 36.** (Instantiation of a substitution)

$$n\{\sigma\} = \sigma(n)$$
$$(MN)\{\sigma\} = (M\{\sigma\})(N\{\sigma\})$$
$$(\lambda M)\{\sigma\} = \lambda M\{\Uparrow \sigma\}$$

$\beta$-reduction is defined as in Definition 37. Following the definition we have Example 8 where we try to show how all these definitions tie together.

**Definition 37.** (Performing a $\beta$-reduction)

$$(\lambda M)N \rightarrow M\{N \triangleright \mathrm{id}\}$$

**Example 8.** (Substitution in the $\lambda$-calculus using de Bruijn indices)

For this example we will be using the following $\beta$-reduction:

$$(\lambda x.xyz(\lambda y.x))y \rightarrow yyz(\lambda z.y)$$

Rewriting it to de Bruijn we get:

$$(\lambda 012(\lambda 1))0 \rightarrow 001(\lambda 1)$$

We now show the reduction using Definition 32 to Definition 37. From (2.6) to (2.8) we attach primes, (e.g. $0'$) to numbers to show where they originate.

$$(\lambda 012(\lambda 1))0 \rightarrow (((01)2)(\lambda 1))\{0 \triangleright \mathrm{id}\} \tag{2.1}$$
$$= (((01)\{0 \triangleright \mathrm{id}\}2\{0 \triangleright \mathrm{id}\})(\lambda 1)\{0 \triangleright \mathrm{id}\}) \tag{2.2}$$
$$= (((0\{0 \triangleright \mathrm{id}\}1\{0 \triangleright \mathrm{id}\})2\{0 \triangleright \mathrm{id}\})(\lambda 1)\{0 \triangleright \mathrm{id}\}) \tag{2.3}$$
$$= (((00)1)(\lambda 1)\{0 \triangleright \mathrm{id}\}) \tag{2.4}$$
$$= (((00)1)(\lambda 1 \Uparrow \{0 \triangleright \mathrm{id}\})) \tag{2.5}$$
$$= ((00)1)(\lambda 1'\{0' \triangleright (\uparrow \circ (0 \triangleright \mathrm{id}))\} \tag{2.6}$$
$$\{0' \triangleright (\uparrow \circ (0 \triangleright \mathrm{id}))\}(1') = \begin{cases} 0' & 1' = 0 \\ \{\uparrow \circ (0 \triangleright \mathrm{id})\}(1'-1) & 1' > 0 \end{cases}$$
$$= ((00)1)(\lambda 0''\{\uparrow \circ (0 \triangleright \mathrm{id})\}) \tag{2.7}$$
$$\{\uparrow \circ (0 \triangleright \mathrm{id})\}(0'') = \begin{cases} \uparrow 0 & 0'' = 0 \\ \uparrow \mathrm{id}(0''-1) & 0'' > 0 \end{cases}$$
$$= ((00)1)(\lambda \uparrow 0) \tag{2.8}$$
$$= ((00)1)(\lambda 1) \tag{2.9}$$

For (2.1) we remove the left most $\lambda$ and insert the extend function. In (2.2) and (2.3) we continuously use the $(MN)\{\sigma\}$ rule, from Definition 36, to apply the extend function. For (2.4) we perform all of the extend, where 0 is replaced with 0, indexes higher than 0 are decreased by 1 and id is used. In (2.5) we pass over a $\lambda$, and given Definition 36, we must lift $\sigma$, such that the indexes of 1 will be substituted with 1, to account for the binder. Lifting $\sigma$, given by Definition 35, gives us (2.6). We use extend from Definition 34 to first reduce to (2.7), where we actually count the index to the right of the $\lambda$ down by one, given by $(1' - 1)$, before we shift it from 0 to 1 and get (2.9), which is the result we expected.

With this we conclude the presentation of de Bruijn indices. However, while this comes close to what is used in our formalization, there is still a slight alteration that needs to be made, namely the nameless representation. We will be using the locally nameless approach as presented by Charguéraud in [40], where you divide free and bound variables into two distinct syntactic categories, where bound variables use de Bruijn indices and free variables use names. Continuing Example 7 we would have:

$$\text{Nameless:} \qquad \lambda 0((\lambda 10)(y + 1)(z + 1)(y + 1))$$

$$\text{Locally nameless:} \ \lambda 0((\lambda 10)yzy)$$

Our reasoning for this approach instead of using de Bruijn indices is that it allows us to say a lot about which binders outside of an encoded term is referenced inside the encoded term, without ever having to unfold the encoding. However, there is also the drawback that it is now possible to make nonsensical terms, by this we mean terms where a de Bruijn index is used, but there is no binder that it is referring to. To combat this we will in the following sections define what it means for a term, and a process, to be well-formed with respect to some number. Another thing of note is that this slightly changes how substitutions and renamings work, but it is simply a case of ignoring free variables.

## 4.2 Formalizing call-by-push-value

Now that all of the prerequisite theory has been covered we begin the formalization in Coq with CBPV. We start with the rewriting of the syntax for CBPV from Definition 3 to use de Bruijn indices, where $x$ is a free variable represented by a name and $n$ is a bound variable represented by an integer.

**Definition 38.** (CBPV with de Bruijn indices)

$$M, N ::= V \mid \lambda M \mid M \ V \mid \textbf{force } V \mid \textbf{return } V \mid M \gg= N$$

$$V ::= x \mid n \mid \textbf{thunk } M$$

As the locally nameless approach allows for terms that are not well-formed, we inductively define what it means for a term $M$ to be well-formed with respect to some number $n$ using the relation $\vDash$ as seen in Definition 39. To be well-formed with respect to some number states something about how far bound occurrences within a term may refer to binders outside of it. A small example could be the term $\lambda\lambda M$. If $0 \vDash M$, we know that there are no bound occurrences inside of $M$ that refer to the two binders outside of $M$. On the other hand if $1 \vDash M$, we know that there are definitely no bound occurrences inside of $M$ that refer to the outermost binder, however there may be some that refer to the binder just outside of $M$.

**Definition 39.** (Well-formed terms)

$$(\text{wf-abs}) \ \frac{n + 1 \vDash M}{n \vDash \lambda M} \qquad\qquad (\text{wf-bind}) \ \frac{\begin{array}{c} n \vDash M \\ n + 1 \vDash N \end{array}}{n \vDash M \gg= N} \qquad\qquad (\text{wf-force}) \ \frac{n \vDash V}{n \vDash \textbf{force } V}$$

$$(\text{wf-app}) \frac{n \vDash M \quad n \vDash V}{n \vDash MV} \qquad (\text{wf-return}) \frac{n \vDash V}{n \vDash \textbf{return } V} \qquad (\text{wf-thunk}) \frac{n \vDash M}{n \vDash \textbf{thunk } M}$$

$$(\text{wf-bv}) \frac{m < n}{n \vDash m} \qquad (\text{wf-fv}) \frac{}{n \vDash x}$$

In Definition 40, we define substitution. It is important to note that, unlike in the $\lambda$-calculus, substitutions in CBPV apply to values rather than terms. Specifically, we have $\sigma : \mathbb{N} \to \textbf{Val}$. We continue using id, shift, extend and lifting from Section 4.1.

**Definition 40.** (Instantiation of substitutions in CBPV)

$$(\lambda M)\{\sigma\} = \lambda M\{\Uparrow \sigma\}$$
$$(M\ N)\{\sigma\} = M\{\sigma\}\ N\{\sigma\}$$
$$(\textbf{force } V)\{\sigma\} = \textbf{force } V\{\sigma\}$$
$$(\textbf{return } V)\{\sigma\} = \textbf{return } V\{\sigma\}$$
$$(M \ggg N)\{\sigma\} = M\{\sigma\} \ggg N\{\Uparrow \sigma\}$$
$$(\textbf{thunk } M)\{\sigma\} = \textbf{thunk } M\{\sigma\}$$
$$(n)\{\sigma\} = \sigma(n)$$
$$(x)\{\sigma\} = x$$

The implementation of this is not quite as straightforward, as Definition 40 is not structurally recursive, which is a requirement in Coq. This is due to instantiations being defined with the use of lifting, and lifting being defined with the use of instantiations. In [39], Stark outlines the previous work of Adams and how he in [41] describes how to define instantiation of substitutions through instantiation of renamings. In the outline this process is divided into five steps, however, for our purposes only the first four are needed.

First we define the lifting of renamings. Here `>>>` is notation for forward composition, and shift has been replaced by the successor constructor $S$.

```
1  Definition extend_rn (s : nat) (rn : nat -> nat) (n : nat) :=
2    match n with
3    | 0 => s
4    | S n => rn n
5    end.
6
7  Definition lift_rn (rn : nat -> nat) := extend_rn 0 (rn >>> S).
```

Second we define instantiation of renamings , using the definition of lifting on renamings.

```
1  Fixpoint int_rn (s : term) (rn : nat -> nat) :=
2    match s with
3    | Val v => Val (int_rn_value v rn)
4    | Abs s => Abs (int_rn s (lift_rn rn))
5    | App s v => App (int_rn s rn) (int_rn_value v rn)
```

```
6       | Force v => Force (int_rn_value v rn)
7       | Ret v => Ret (int_rn_value v rn)
8       | Bind s t => Bind (int_rn s rn) (int_rn t (lift_rn rn))
9       end
10
11  with int_rn_value (v : value) (rn : nat -> nat) :=
12      match v with
13      | Var (BV n) => Var (BV (rn n))
14      | Var (FV n) => Var (FV n)
15      | Thunk s => Thunk (int_rn s rn)
16      end.
17
18  Notation "s << rn >>" := (int_rn s rn) (at level 90, left associativity).
19  Notation "v <.< rn >.>" := (int_rn_value v rn) (at level 90, left
        associativity).
```

Thirdly we can define the composition of a substitution with an instantiation of a renaming, which we can then use to define the lifting of a substitution.

```
1   Definition extend_subst_lam (v : value) (subst : nat -> value) (n : nat) :=
2       match n with
3       | 0 => v
4       | S n' => subst n'
5       end.
6
7   Notation "v {}> subst" := (extend_subst_lam v subst) (at level 81, left
        associativity).
8
9   Definition compose_subst_int_rn
10      (subst : nat -> value)
11      (rn : nat -> nat)
12      (n : nat)
13      := int_rn_value (subst n) rn.
14
15  Definition lift_subst (subst : nat -> value) :=
16      extend_subst_lam (Var (BV 0)) (compose_subst_int_rn subst S).
17
18  Notation "^ subst" := (lift_subst subst) (at level 81, left associativity).
```

Fourthly we can use the definition of lifting of substitutions to define the instantiation of a substitution.

```
1   Fixpoint int_subst (s : term) (subst : nat -> value) :=
2       match s with
3       | Val v => Val (int_subst_value v subst)
4       | Abs s => Abs (int_subst s (lift_subst subst))
5       | App s v => App (int_subst s subst) (int_subst_value v subst)
6       | Force v => Force (int_subst_value v subst)
```

```
7        | Ret v => Ret (int_subst_value v subst)
8        | Bind s t => Bind (int_subst s subst) (int_subst t (lift_subst subst))
9        end
10
11  with int_subst_value (v : value) (subst : nat -> value) :=
12    match v with
13    | Var (BV n) => subst n
14    | Var (FV n) => Var (FV n)
15    | Thunk s => Thunk (int_subst s subst)
16    end.
17
18  Notation "s {{ subst }}" := (int_subst s subst) (at level 90, left
     associativity).
19  Notation "v {.{ subst }.}" := (int_subst_value v subst) (at level 90, left
     associativity).
```

With instantiation of substitutions defined+ we can now define the reduction relation as seen in Definition 41. A small implementation detail of this is that when making substitutions in Coq we have to use the `BV >>> Var` type constructors to turn a natural number into a term. When we are following Definition 38 we can just use id as natural numbers are part of the syntax.

**Definition 41.** (CBPV reduction relation using de Bruijn indices)

$$\text{(binding-base)} \; \frac{}{\textbf{return } V \gg= M \longrightarrow M\{V \rhd \text{id}\}}$$

$$\text{(binding-evolve)} \; \frac{N \to N'}{N \gg= M \to N' \gg= M}$$

$$\text{(force-thunk)} \; \frac{}{\textbf{force } (\textbf{thunk } M) \to M}$$

$$\text{(Application-base)} \; \frac{}{(\lambda M)V \to M\{V \rhd \text{id}\}} \quad \text{(Application-evolve)} \; \frac{M \to M'}{MV \to M'V}$$

With this, we can now move on to our work with formalizing the $\pi$-calculus.

## 4.3 Formalizing the $\pi$-calculus

In the following section we formalize the $\pi$-calculus. We start by expanding on the theory of de Bruijn indices, as the standard implementation is not quite enough for us.

### 4.3.1 Challenges of formalizing the $\pi$-calculus with de Bruijn index

In the $\pi$-calculus, substitution is limited to replacing names with other names, and not names with terms as in $\lambda$-calculus or values as in CBPV, which means that renamings and substitutions effectively coincide. As a result, within the context of $\pi$-calculus, we will refer to renamings as substitutions.

Due to scope extrusion, binders in the $\pi$-calculus can swap places. This behaviour complicates de Bruijn indices, as the previously introduced substitutions can not replicate this behaviour. We therefore introduce the swap substitution as seen in Definition 42. Example 9 illustrates the complication further.

**Definition 42.** (swap)

$$\updownarrow (n) = \begin{cases} 1 & n = 0 \\ 0 & n = 1 \\ n & \text{otherwise} \end{cases}$$

**Example 9.** An example of a process where scope extrusion occurs could be the following:

$$(\nu x)(\nu y)(\bar{a}y.\bar{a}x.\bar{a}y) \mid a(u).\overline{u}v \xrightarrow{\tau} (\nu y)((\nu x)(\bar{a}x.\bar{a}y) \mid \overline{y}v)$$

In the process one can see that the binders $(\nu x)$ and $(\nu y)$ swap places, and this further complicates the nameless representation seen below:

$$(\nu)(\nu)\big(\overline{(a+2)}0.\overline{(a+2)}1.\overline{(a+2)}0\big) \mid a().\overline{0}(v+1) \xrightarrow{\tau} (\nu)\big((\nu)\big(\overline{(a+2)}0.\overline{(a+2)}1\big) \mid \overline{0}(v+1)\big)$$

Due to the binders swapping places, one has to adjust the bound occurrences in the sub-process $\overline{(a+2)}1.\overline{(a+2)}0$, such that 0 and 1 also swap places. In the general case a restriction may jump over any number of other restrictions as seen below:

$$(\nu x_n)...(\nu x_1)(\bar{a}x_1.....\bar{a}x_n.\bar{a}x_1) \mid a(u).\overline{u}v \xrightarrow{\tau} (\nu x_1)((\nu x_n)...(\nu x_2)(\bar{a}x_2.....\bar{a}x_n.\bar{a}x_1) \mid \overline{x_1}v)$$

In the nameless representation, things become slightly more complex. To the right of the parallel composition operator, we do not need to do anything, because when the input prefix is removed, the scope is extruded, so the 0 refers to the correct binder and $v+1$ remains free. To the left of the parallel composition every number above 0 has to be shifted down and 0 has to be replaced by $n-1$. This can be done by using a sequence of swaps, one for every time two restrictions are swapped.

$$\prod_{n-1}^{0}(\nu)\big(\overline{(a+n)}0.....\overline{(a+n)}(n-1).\bar{a}0\big) \mid a().\overline{0}(v+1) \xrightarrow{\tau}$$

$$(\nu)\left(\prod_{n-2}^{0}(\nu)(\bar{a}0.....\bar{a}(n-2).\bar{a}(n-1)) \mid \overline{0}v+1\right)$$

We show this pictorially in Figure 2.

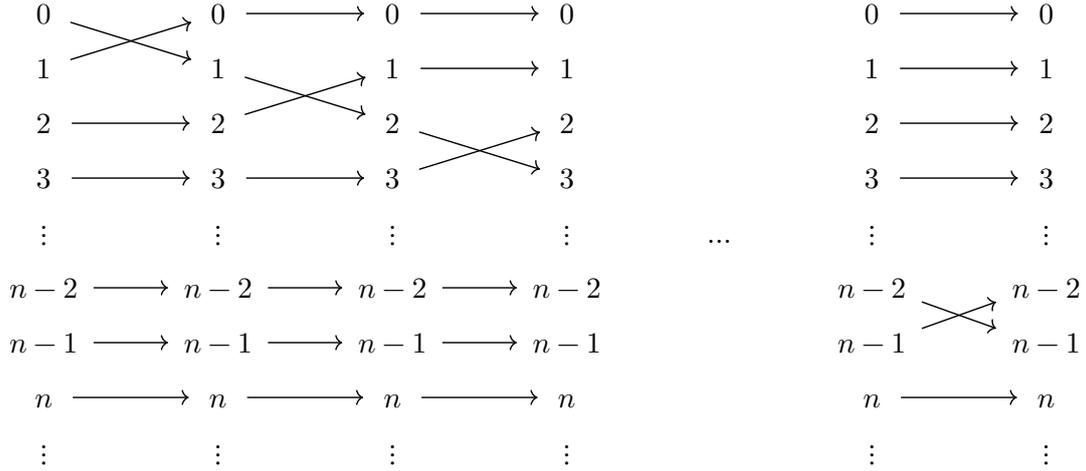

Figure 2: Sequencing swaps.

In this report we have introduced the $\pi$-calculus with the early labeled transition semantics. However there is a problem when formalizing this variation of the $\pi$-calculus. We realized an issue when trying to translate the semantic rules into a version using de Bruijn indices. The semantics for the (In) rule would depend on the type of communication between two processes. When communication occurs using a free output one uses the (Com) rule, which means that one has to instantiate an $(0 \triangleright \text{id})$ substitution on the sub-process under the input prefix, as there would be one less binder surrounding the sub-process. However, when communication occurs using a bound output, one uses the (Close) rule, this means one has to do nothing on the sub-process under prefix as a restriction has taken the place of the input prefix. The problem arises from the fact that there are both free and bound outputs, and as such one could potentially fix this by splitting the input action up into two, one for when communication using the (Com) rule, and one for the (Close) rule. One could also limit themselves to only use either free or bound outputs. Here the internal $\pi$-calculus ($\pi I$-calculus) by Sangiorgi becomes very useful as no free outputs exist in this subset of the $\pi$-calculus [11]. As will be discussed in the following section there are numerous advantages to using the $\pi I$-calculus. As such moving forward in the formalization we will be using the $\pi I$-calculus.

### 4.3.2 Transitioning to the $\pi I$-calculus

The $\pi I$-calculus is a variation on the $\pi$-calculus where every output is bound [11]. This creates a symmetry where input and output, instead of communicating, are agreeing on which name to use. This is possible since $\alpha$-conversion of bound names does not change the behaviour of processes. Combined with de Bruijn indices, this notion becomes even simpler as it is no longer a question of whether two process, through $\alpha$-conversion, can find a name which they agree on, but simply do they agree on the index. By using the $\pi I$-calculus many of the somewhat arbitrary choices that one makes also disappear, such as when choosing a bisimulation relation. This is is because in the $\pi I$-calculus early-, late- and open-bisimulation all coincide, making the choice of bisimulation almost unambiguous. Furthermore bisimulation is a congruence, further simplifying ones choices. In the $\pi I$-calculus substitution is not used. As such, the differences between the early and late labelled transition semantics also disappear. Since all outputs are bound, one can reason better about processes, meaning that one can make use of arithmetic laws, that will not necessarily hold when free outputs are permitted. Sangiorgi also used the $\pi I$-calculus in the context of encoding the $\lambda$-calculus, which we can draw inspiration from in our own encoding [11]. Lastly, since it is possible to encode CBPV in $\pi I$-calculus, it means that bound output is enough, and that it is not necessary to have free outputs.

The syntax for the $\pi I$-calculus is defined in Definition 43.

**Definition 43.** ($\pi I$-calculus syntax)

$$P, Q \ ::= \ a(x).P \ \mid \ \overline{a}(x).P \ \mid \ (\nu x)P \ \mid \ !P \ \mid \ x \to y \ \mid \ P|Q \ \mid \ 0$$

First and foremost, the central difference between the syntax of the $\pi$-calculus and the $\pi I$-calculus is that output is replaced with the $\overline{a}(x).P$ construct instead. This essentially acts as shorthand for $(\nu x)\overline{a}x.P$. The consequence of this is that output is also a binder. Another noticeable difference from the syntax of the $\pi$-calculus is that we include a new construct $x \to y$, which we will refer to as a link. This is required because, to encode CBPV in the $\pi I$-calculus, replication is not enough, as we also need recursion [11]. For similar reasons the $\pi I$-calculus version presented by Sangiorgi uses constant application rather than replication. The reasoning behind needing recursion, comes from the fact, that the link we had in the $\pi$-calculus, $!y(a).\overline{x}a$, cannot work in the $\pi I$-calculus, as all outputs have to be bound. Therefore, we need a new form of link. However, seeing as we have been using replication throughout this project, we found it simpler to embed the needed constants directly into the syntax, as links would be the only new construct needed for our encoding. We also suspect that only introducing links, and not recursion in full, should make formalization easier, as one would not have to keep track of user defined constants.

With this, we can now define the actions of the $\pi I$-calculus in Definition 44.

**Definition 44.** (Actions in $\pi I$-calculus)

$$\mu \ ::= \ a(x) \ \mid \ \overline{a}(x) \ \mid \ \tau$$

From these, we can then define the transition system for $\pi I$-calculus shown in Definition 45. We have chosen to include a simple rule for link, that defines its behaviour. The idea behind a link is that when you send a name to a link, a pointer to the received name gets send out. One thing that is noticeable in the $\pi I$-calculus, that is not present in in the standard $\pi$-calculus, is the use of $\alpha$-conversion. However, since we will be formalizing the $\pi I$-calculus with de Bruijn indices we get this for free. We have excluded the symmetric versions of the (Par) and (Com) rules.

**Definition 45.** (The $\pi I$-calculus transition system)

$$\text{(Out)} \ \frac{}{\overline{a}(x).P \xrightarrow{\overline{a}(x)} P} \qquad\qquad \text{(In)} \ \frac{}{a(x).P \xrightarrow{a(x)} P}$$

$$\text{(Par)} \ \frac{P \xrightarrow{\mu} P'}{P \mid Q \xrightarrow{\mu} P' \mid Q} \text{bn}(\mu) \cap \text{fn}(Q) = \emptyset \qquad \text{(Com)} \ \frac{Q \xrightarrow{a(x)} Q' \quad P \xrightarrow{\overline{a}(x)} P'}{P \mid Q \xrightarrow{\tau} (\nu x)(P' \mid Q')}$$

$$\text{(Res)} \ \frac{P \xrightarrow{\mu} P'}{(\nu x)P \xrightarrow{\mu} (\nu x)P'} x \notin \text{n}(\mu) \qquad \text{(Link)} \ \frac{!x(a).\overline{y}(b).(b \to a) \xrightarrow{x(a)} P'}{x \to y \xrightarrow{x(a)} P'}$$

$$\text{(Rep)} \ \frac{P \mid !P \xrightarrow{\mu} P'}{!P \xrightarrow{\mu} P'} \qquad\qquad \text{(Alpha)} \ \frac{P \equiv_\alpha Q \quad Q \xrightarrow{\mu} P'}{P \xrightarrow{\mu} P'}$$

With this, we can now encode CBPV in the $\pi I$-calculus. We have chosen to write it in the polyadic $\pi I$-calculus, to keep the encoding notationally concise, and within the margins of a page. The monadic version can be found in Appendix D.4. Since the polyadic $\pi I$-calculus can easily be encoded in the monadic $\pi I$-calculus, this can be done without loss of generality [11]. However, because of advice we received from a personal correspondence with Daniel Hirschkoff, we decided to stick to the monadic $\pi$-calculus in the formalization.

**Example 10.** (From polyadic- to monadic-$\pi I$-calculus)

To encode polyadic to monadic we just have to rewrite it, using the following idea for tuples larger than 1:

$$\overline{e}(a, b, c).\mathcal{P}[\![M]\!] = \overline{e}(x).\overline{x}(a).\overline{x}(b).\overline{x}(c).\mathcal{M}[\![M]\!] \quad x \notin \text{fn}(M)$$
$$e(a, b, c).\mathcal{P}[\![M]\!] = e(x).x(a).x(b).x(c).\mathcal{M}[\![M]\!] \quad x \notin \text{fn}(M)$$

The idea is to establish a connection using the channel $e$, and send the data on the linking channel $x$. However, this encoding from polyadic to monadic assumes that communication on a channel $e$ always has the same arity [1], meaning that we never encounter a process such as:

$$\overline{e}(a, b).P \mid e(a, b, c).Q$$

In the polyadic $\pi I$-calculus this would not be able to communicate. However, if we just used the straightforward translation to monadic, we would get the following:

$$\overline{e}(c).\overline{c}(a).\overline{c}b.P \mid e(c).e(a).e(b).e(c.).Q \Rightarrow P \mid e(c.).Q$$

Now $P$ is not behind prefix anymore, despite its polyadic counterpart still being behind a prefix. This is something we have to be aware of.

With this, we we will now define our encoding from CBPV in the $\pi I$-calculus in Definition 46.

**Definition 46.** (Encoding of CBPV in $\pi I$-calculus)

$$\mathcal{J}[\![\lambda x.M]\!]_u^r = \overline{u}(e).e(u, r, x).\mathcal{J}[\![M]\!]_u^r \qquad\qquad e \notin \text{fn}(M)$$
$$\mathcal{J}[\![M\ V]\!]_u^r = (\nu abcd)\big(a(e).c(x).\overline{e}(u', r', x').$$
$$(u' \to u \mid r' \to r \mid x' \to x)$$
$$\mid \mathcal{J}[\![M]\!]_a^b \mid \mathcal{J}[\![V]\!]_c^d\big) \qquad\qquad a, b, c, d \notin \text{fn}(M) \cup \text{fn}(V)$$
$$\mathcal{J}[\![\mathbf{force}\ V]\!]_u^r = (\nu ab)\big(\mathcal{J}[\![V]\!]_a^b \mid a(y).\overline{y}(u', r').(u' \to u \mid r' \to r)\big) \qquad a, b \notin \text{fn}(V)$$
$$\mathcal{J}[\![\mathbf{return}\ V]\!]_u^r = \mathcal{J}[\![V]\!]_r^r$$
$$\mathcal{J}[\![M \gg= \lambda x.N]\!]_u^r = (\nu ab)\big(\mathcal{J}[\![M]\!]_a^b \mid b(x).\mathcal{J}[\![N]\!]_u^r\big) \qquad\qquad a, b \notin \text{fn}(M) \cup \text{fn}(N)$$
$$\mathcal{J}[\![V]\!]_u^r = (\overline{u}(y).\mathcal{J}[\![y := V]\!]) \qquad\qquad y \notin \text{fn}(V)$$
$$\mathcal{J}[\![y := x]\!] = !y(u, r).\overline{x}(u', r').(u' \to u \mid r' \to r)$$
$$\mathcal{J}[\![y := \mathbf{thunk}\ M]\!] = !y(u, r).\mathcal{J}[\![M]\!]_u^r$$

We have for clarity chosen to write the encoding of $\mathcal{J}[\![y := x]\!]$ as a single unfolding of a link, to show what will eventually be sent along this link is always the two handles used to "force" a thunk. This also stops an unintended behaviour, where an application could communicate on the link. But in essence, the encoding of a variable is similar to the encoding in the $\pi$-calculus, where it is a link used to send a "force" along.

However, if we want to transform this into a monadic encoding, we run into the aforementioned problem of arities not matching from Example 10. In the case we have the CBPV expression:

$\mathcal{I}[\![(\mathbf{thunk}\ M)\ V]\!]_u^r =$

$(\nu abcd)\big(a(e).c(x).\overline{e}(u',r',x').(u' \to u \mid r' \to r \mid x' \to x) \mid \mathcal{I}[\![\mathbf{thunk}\ M]\!]_a^b \mid \mathcal{I}[\![V]\!]_c^d\big) =$

$(\nu abcd)\big(a(e).c(x).\overline{e}(u',r',x').(u' \to u \mid r' \to r \mid x' \to x) \mid \overline{a}(e).!e(u',r').\mathcal{I}[\![M]\!]_{u'}^{r'} \mid \overline{c}(x).\mathcal{I}[\![x := V]\!]\big)$

This can reduce down to a process that encounters this issue:

$(\nu abcd)\big(a(e).c(x).\overline{e}(u',r',x').(u' \to u \mid r' \to r \mid x' \to x) \mid \overline{a}(e).!e(u',r').\mathcal{I}[\![M]\!]_{u'}^{r'} \mid \overline{c}(x).\mathcal{I}[\![x := V]\!]\big)$

$\Rightarrow (\nu abcd)(\overline{e}(u',r',x').(u' \to u \mid r' \to r \mid x' \to x) \mid !e(u',r').\mathcal{I}[\![M]\!]_{u'}^{r'} \mid \mathcal{I}[\![x := V]\!]$

Here we encounter the exact issue that there is an output on the channel $e$ that has a different arity than the input on the channel $e$. Now this CBPV expression is obviously always ill-typed, so we could restrict our encoding to only work on well-typed CBPV expressions, and then this would not be an issue. However, one could also add an extra check to the end of the translation of the encoding of force and thunk. Specifically, we have the following:

$$\mathcal{M}[\![y := \mathbf{thunk}\ m]\!] = !y(s).s(u).s(r).\overline{s}(s).[\![M]\!]_u^r$$

$$\mathcal{M}[\![\mathbf{force}\ V]\!]_u^r = (\nu ab)\big([\![V]\!]_a^b \mid a(y).\overline{y}(s).\overline{s}(u').\overline{s}(r').s(s).(u' \to u \mid r' \to r)\big)$$

In all other cases, the standard translation from monadic to polyadic $\pi I$-calculus can be used. Now there is an extra check-communication, that stops the aforementioned wrong communication, without enabling other wrong communications. The monadic encoding can be found in the appendix at Appendix D.4

The proof of soundness and completeness follows much of the same structure as the one for the $\pi$-calculus, for the full proof see Appendix B. This encoding also satisfies property 1-5 of [1], which is also argued in Appendix B.

### 4.3.3 Formalizing $\pi I$-calculus with de Bruijn Indices

We will now look at how we formalized the $\pi I$-calculus. We define the syntax for the monadic $\pi I$-calculus using de Bruijn indices as in Definition 47. Actions are defined in Definition 48.

**Definition 47.** ($\pi I$-calculus with de Bruijn indices syntax)

$$P, Q ::= \mathrm{ch}().P \mid \overline{\mathrm{ch}}().P \mid (\nu)P \mid !P \mid \mathrm{ch}_1 \to \mathrm{ch}_2 \mid P|Q \mid 0$$

$$\mathrm{ch} ::= x \mid n$$

**Definition 48.** (Actions with de Bruijn indices)

$$\mu ::= n() \mid \overline{n}() \mid \tau$$

As with CBPV we also define what is it means for a process to be well-formed with respect to some number. This is done in Definition 49.

**Definition 49.** (Well-formed process)

$$(\text{wf-fn}) \frac{}{n \vDash x} \qquad (\text{wf-bn}) \frac{m < n}{n \vDash m} \qquad (\text{wf-nil}) \frac{}{n \vDash 0}$$

$$(\text{wf-out}) \frac{n \vDash a \quad n+1 \vDash P}{n \vDash \overline{a}().P} \qquad (\text{wf-in}) \frac{n \vDash a \quad n+1 \vDash P}{n \vDash a().P} \qquad (\text{wf-res}) \frac{n+1 \vDash P}{n \vDash (\nu)P}$$

$$(\text{wf-link}) \frac{n \vDash a \quad n \vDash b}{n \vDash a \rightarrow b} \qquad (\text{wf-rep}) \frac{n \vDash P}{n \vDash \,!P} \qquad (\text{wf-par}) \frac{n \vDash a \quad n \vDash b}{n \vDash a \mid b}$$

The definitions for instantiations of substitutions follow the same principles as the instantiations previously defined, as seen in Definition 50 and Definition 51.

**Definition 50.** (Instantiation of substitutions in the $\pi I$-calculus)

$$(\text{ch}().P)\{\sigma\} = (\text{ch}\{\sigma\})().P\{\Uparrow \sigma\}$$
$$(\overline{\text{ch}}().P)\{\sigma\} = \overline{(\text{ch }\{\sigma\})}().P\{\Uparrow \sigma\}$$
$$((\nu)P)\{\sigma\} = (\nu)P\{\Uparrow \sigma\}$$
$$(!P)\{\sigma\} = \,!P\{\sigma\}$$
$$(P \mid Q)\{\sigma\} = P\{\sigma\} \mid Q\{\sigma\}$$
$$(\text{ch}_1 \rightarrow \text{ch}_2)\{\sigma\} = \text{ch}_1\{\sigma\} \rightarrow \text{ch}_2\{\sigma\}$$
$$0\{\sigma\} = 0$$
$$x\{\sigma\} = x$$
$$n\{\sigma\} = \sigma(n)$$

**Definition 51.** (Instantiation of substitutions on actions)

$$(\text{ch}())\{\sigma\} = (\text{ch }\{\sigma\})()$$
$$(\overline{\text{ch}}())\{\sigma\} = \overline{\text{ch}\{\sigma\}}()$$
$$\tau\{\sigma\} = \tau$$
$$x\{\sigma\} = x$$
$$n\{\sigma\} = \sigma(n)$$

We define the labelled transition system as seen in Definition 52, where we have omitted the symmetric versions of (Par), (Par-tau) and (Com). To transform the transition system in Definition 45 we continue to use the same strategies as before, but we also draw inspiration from [42], in which Perera and Cheney present a late transition semantics for the $\pi$-calculus using de Bruijn indices. We have most notably taken inspiration from their use of the swap substitution. Another advantage of using $\pi I$-calculus is that much of the bookkeeping of indices disappears, due to the fact that only bound outputs are permitted. This means that unlike Perera and Cheney, we do not have to use extend to account for free outputs, as they can never occur.

**Definition 52.** (The $\pi I$-calculus transition system using de Bruijn index)

$$\text{(Out)} \;\frac{}{\overline{n}().P \stackrel{\overline{n}()}{\to} P} \qquad\qquad \text{(In)} \;\frac{}{n().P \stackrel{n()}{\to} P}$$

$$\text{(Par-tau)} \;\frac{P \stackrel{\tau}{\to} P'}{P \mid Q \stackrel{\tau}{\to} P' \mid Q} \qquad \text{(Par)} \;\frac{P \stackrel{\alpha}{\to} P'}{P \mid Q \stackrel{\alpha}{\to} P' \mid Q\{\uparrow\}} \; \alpha \in \{\overline{n}(), n()\}$$

$$\text{(Res-tau)} \;\frac{P \stackrel{\tau}{\to} P'}{(\nu)P \stackrel{\tau}{\to} (\nu)P'} \qquad \text{(Res)} \;\frac{P \stackrel{\alpha\{\uparrow\}}{\to} P'}{(\nu)P \stackrel{\alpha}{\to} (\nu)P'\{\updownarrow\}} \; \alpha \in \{\overline{n}(), n()\}$$

$$\text{(Com)} \;\frac{P \stackrel{n()}{\to} P' \qquad Q \stackrel{\overline{n}()}{\to} Q'}{P \mid Q \stackrel{\tau}{\to} (\nu)(P' \mid Q')} \qquad \text{(Link)} \;\frac{(!n().\overline{m+1}().0 \to 1) \stackrel{n()}{\to} p'}{n \to m \stackrel{n()}{\to} P'}$$

$$\text{REP} \;\frac{P \mid !P \stackrel{\alpha}{\to} Q}{!P \stackrel{\alpha}{\to} Q}$$

As previously mentioned the notion that processes in the $\pi I$-calculus do not communicate, but rather agree on names becomes simpler when used in conjunction with de Bruijn indices. A small example could be the process $a(x).x(u) \mid \overline{a}(y).\overline{y}(v)$, where the processes can use the (Alpha) rule and $\alpha$-conversion to agree on the name $x$, thereby transitioning to the process $(\nu x)(x(u) \mid \overline{x}(v))$. However when using de Bruijn indices, or the locally nameless approach as we do, the process would be $a().0() \mid \overline{a}().\overline{0}()$. No $\alpha$-conversion is needed and the process can transition to $(\nu)\big(0() \mid \overline{0}()\big)$.

Additionally weak bisimulation has also been formalized, as seen in Definition 53. The definition contains three clauses. First is the is standard definition of weak bisimulation, second is an up-to structural congruence clause, and thirdly an up-to context clause. Calling the second structural congruence is a legacy from earlier stages of formalizing, but they are the same as the bisimilarity laws presented in Proposition 1, except with a different name. The reasoning behind including the second and third clause in the definition, is that these are behaviours that have already been proven, such as in [26]. Clause (1) also differs from the definition of early weak bisimulation, again this is due to using the $\pi I$-calculus where early-, late- and open-bisimulation coincide. We note that $\stackrel{\tau}{\Rightarrow}$ is defined as $\Rightarrow$.

**Definition 53.** (Weak bisimulation) A binary relation $R \subset \textbf{Proc} \times \textbf{Action} \times \textbf{Proc}$ is a weak bisimulation if the following holds.

$$\Big(\forall a, P'.P \stackrel{a}{\to} P' \to \exists Q'.Q \stackrel{a}{\Rightarrow} Q' \land P'RQ'\Big) \land \Big(\forall a, Q'.Q \stackrel{a}{\to} Q' \to \exists P'.P \stackrel{a}{\Rightarrow} P' \land P'RQ'\Big)$$

Given a bisimulation relation $R$, two processes $P$ and $Q$ are weak bisimilar, written $P \approx Q$, if one of the following holds
1. $PRQ \to P \approx Q$
2. $P \equiv Q \to P \approx Q$
3. $\forall C.P \approx Q \to C[P] \approx C[Q]$

$\approx$ is the largest relation closed under these rules.

## 4.4 Formalization of the encoding

In this section we present how we have formalized our encoding form CBPV to $\pi I$-calculus. The encoding has been implemented as a function, and for the most part is a straightforward translation. As outlined in Example 11, the main difficulty is that a term has fewer binders than the process corresponding to the encoded term. This means that if a bound variable in a term has to refer to the corresponding binder in the encoded term, we have to account for the addition of any new binders introduced during the encoding. To resolve this issue, we need to continually keep track of the lack of correspondence in the indices. We do so by creating a list of pairs, where the first element of each pair is the index in CBPV, and the second element is the resulting index in the $\pi I$-calculus. A consequence of this is the need to include a new parameter `refs` in the formalization of the encoding.

**Example 11.** When encoding a CBPV term additional binders are introduced. This is not a problem when using channel names as it is easy to find the binding occurrence for a variable, but for de Bruijn indices we have to account for the varying number of binders. If we look at a simple term, and naively try to encode it:

$$\mathcal{M}[\![\lambda x.x]\!]^r_u = \overline{u}(e).e(s).s(u).s(r).s(x).\mathcal{J}[\![x]\!]^r_u$$
$$= \overline{u}(e).e(s).s(u).s(r).s(x).(\overline{u}(y).y \to x)$$

We will now try to translate this to de Bruijn indices. We will write **db** as the function from $x$ to a de Bruijn index:

$$\mathcal{J}[\![\lambda.0]\!]^r_u = \overline{u}().0().0().1().2().\mathcal{J}[\![x]\!]^1_2$$
$$= \overline{u}().0().0().1().2().(\overline{2}().0 \to \mathbf{db}(x))$$

We now need to figure out what $\mathbf{db}(x)$ should be substituted with. A naive idea might be to just substitute the same value as in the CBPV, where $x$ got substituted with 0:

$$\mathcal{J}[\![\lambda.0]\!]^r_u = \overline{u}().0().0().1().2().\mathcal{J}[\![0]\!]^1_2$$
$$= \overline{u}().0().0().1().2().(\overline{2}().0 \to 0)$$

However the $x$ in the original encoding now refers to the input of the replication, and not the $s(x)$ where the $x$ is bound.

To maintain the list, and increment the indices in it, we define the auxiliary function `incRefs`.

```
1  Fixpoint incRefs (n : nat) (m : nat) (refs : list (nat * nat)) :=
2    match refs with
3    | [] => []
4    | (n', m') :: ps => (n + n', m + m') :: (incRefs n m ps)
5    end.
```

We also need to be able to find the correct index that corresponds to the relevant binder when encoding a variable. This is the motivation for our recursive definition `findRef`, that finds the corresponding index in `refs`. Take notice of the case in which the result is simply 42. If that happens, it means a variable in CBPV is in the syntactic category for bound variables, yet its binder is not in the `refs` list. This should only happen if the encoded term is not well-formed with respect to 0, and if this happens in proof scripts it is a good indicator to start looking for a falsehood in one of the premises. One might want to remove this case, but that makes the function non-total. Other approaches could have been to use an option type or composing functions. The option type would require us to perform the encoding within an option monad or pollute the encoding with constants checks for a failed computation. Composition of functions was not chosen either, because while composition is easy, decomposing functions is much harder than decomposing a list.

```
1  Fixpoint findRef (n : nat) (refs : list (nat * nat)) :=
2    match refs with
3    | [] => 42 (* this case should never happen *)
4    | (x, y) :: ps =>
5        if n =? x
6        then y
7        else findRef n ps
8    end.
```

Using these auxiliary functions one can define the encoding. The definition follows the encoding presented in Definition 46, with the exception that reference indices are implicitly incremented. The `encode` definition is there to represent the initial call to encode a term, which ensures a couple of small, yet quite important details. Firstly it ensures that the handles $u$ and $r$ are different and that the references list is empty. Secondly it also ensures that $u$ and $r$ are bound, justifying them being in the syntactic category for bound names.

```
1  Definition pointer (p : proc) :=
2    Rep (In (BN 0) (In (BN 0) (In (BN 1) p))).
3
4  Fixpoint encode (s : term) (u r : nat) (refs : list (nat * nat)) :=
5    match s with
6    | Abs s => Out (BN u) (In (BN 0) (In (BN 1)
7        (In (BN 2) (encode s 2 1 ((0,0) :: incRefs 1 4 refs)))))
8    | App s v => (Res ^^ 4) (Par
9        (In (BN 3) (In (BN 2) (Out (BN 1) (Out (BN 2) (Out (BN 3) (Par
10          (Link (BN 2) (BN (9 + u)))
11          (Par
12            (Link (BN 1) (BN (9 + r)))
13            (Link (BN 0) (BN 3))
14          )
15        ))))))
16        (Par
17          (encode s 3 2 (incRefs 0 4 refs))
18          (Out (BN 1) (encode_value v (incRefs 0 4 refs)))
19        )
20      )
21    | Force v => (Res ^^ 2) (Par
22        (Out (BN 1) (encode_value v (incRefs 0 2 refs)))
23        (In (BN 1) (Out (BN 0) (Out (BN 0) (Out (BN 1) (Par
24          (Link (BN 1) (BN (6 + u)))
25          (Link (BN 0) (BN (6 + r)))
```

```
26          )))))
27        )
28      | Ret v => Out (BN r) (encode_value v refs)
29      | Bind s t => (Res ^^ 2) (Par
30          (encode s 1 0 (incRefs 0 2 refs))
31          (In (BN 0) (encode t (3 + u) (3 + r) ((0,0) :: incRefs 1 3 refs)))
32        )
33      | Val v => Out (BN u) (encode_value v refs)
34      end
35
36  with encode_value (v : value) (refs : list (nat * nat)) :=
37    pointer(
38      match v with
39      | Var n =>
40        match n with
41        | (BV m) => Link (BN 0) (BN (findRef m refs + 4))
42        | (FV m) => Link (BN 0) (FN m)
43        end
44      | Thunk s => encode s 1 0 (incRefs 0 4 refs)
45      end
46    ).
47
48  Notation "$ v ; refs $" := (encode_value v refs) (at level 90, left
      associativity).
49  Notation "$ s ; u ; r ; refs $" := (encode s u r refs) (at level 90, left
      associativity).
50
51  Definition encode' (s : term) := Res (Res ($ s ; 1 ; 0 ; [] $)).
```

## 4.5 Soundness and Completeness

In this section we present the current state of our formalization of the soundness and completeness proofs for the encoding. We will discuss how they differ from their hand-written variations and outline the overarching ideas behind the proof scripts. We will also address some missing details, that one will likely need to be adding if they were to continue working on the formalization.

Soundness from clause (1) in Theorem 1, has a straightforward translation with only two differences. Firstly, since we are using $\pi I$-calculus we can simply use weak bisimulation, no need for early weak bisimulation. Secondly we add the premise that $s$ must be well-formed with respect to 0. This is done to ensure that there are no bound variables in $s$ that do not have a binder that they are referencing.

```
1  Theorem sound:
2    forall s,
3      wf_term 0 s ->
4      forall t,
5        s --> t ->
```

```
6        forall u r,
7          exists P,
8            (encode s u r []) =()> P /\ (P ~~ (encode t u r [])).
```

In the proof script for soundness the first step is to do induction on the reduction hypothesis `s --> t`. In the base cases (Force-thunk), (Binding-base) and (Application), we did as many reduction as possible on the process, and then it was often a case of using structural congruence, and our formalized versions of Lemma 5 and Lemma 6 clause (2) and (4) (these can be found in Appendix B), to prove the bisimilarity clause. In the inductive cases (Application-evolve) and (Binding-evolve), it was mostly a case of destructuring the inductive hypothesis into two hypotheses, one stating the reduction for the evolving step, and the other stating bisimilarity. One could then use these hypotheses to show reduction and bisimilarity for the whole process.

Completeness from clause (2) in Theorem 1 differs quite a bit, as the completeness clause was strengthened after the formalization had begun. Firstly we see that `s` either has a single reduction or no reduction, which is different from $\Rightarrow$ used in Theorem 1 which is the reflexive and transitive closure of $\rightarrow$. Secondly there is no clause stating that the intermediary stage process `P` should only be able to make internal communications, and have no barbs. Thirdly the transition from the encoding of `s` is a single step silent action, and not a other weak silent action. The completeness used in Coq is much weaker, as it only holds for a single reduction of a encoded term, where the completeness in Theorem 1 would say for all processes we can reduce to.

```
1 Theorem complete:
2   forall s P u r,
3     wf_term 0 s ->
4     (encode s u r []) -( a_tau )> P ->
5       exists P' t,
6         P =()> P' /\
7         P' ~~ encode t u r [] /\
8         (s --> t \/ s = t).
```

The proof script started by doing induction on `s`, where it turned out to be important not to weaken the induction hypothesis by using the `intros` tactic on any further variables in the goal, before doing the induction. The `Val`, `Abs` and `Return` cases were quite simple as the encoded terms could not make a reduction, and as such a simple inversion on the `(encode s u r [])` -`( a_tau )> P` hypothesis sufficed to prove these cases. One should also expect this as neither of these terms can make a reduction. The odd one out is the `Force` case, as the encoding of a **force** term can make a $\tau$-transition, even in the case that **force** is used on a variable. We use inversions on `(encode s u r [])` -`( a_tau )> P` hypothesis to find this transition, and destructured the value that **force** was applied to, upon reaching the transition. In the case where the value is a variable we destructured further into bound and free variables. For a bound variable we proved the falsehood of the premise `wf_term 0 s`. For free variables we proved that any further transitions would still be weak bisimilar to `s`. In the case where the value is a **thunk** (`Force Thunk s'`) we prove that `P` would transition into something that is weak bisimilar with the encoding of `s'`.

In these proof scripts we make use of lemmas that have yet to be formally proven, something that we will discuss in the following section. It is likely that these lemmas will make it necessary to add further premises. One premise that we think is likely needed is one stating that $u$ and *r* should be different. However, by looking at the definition of `encode'`, one can see that this is an easy premise to enforce.

## 4.6 Unformalized lemmas

In the proofs presented in Section 4.5, we have relied on a combination of lemmas introduced in this paper and a few additional ones. While some of these lemmas have been formally proven in Coq, others—though essential—have not (yet) been proven in Coq. For this section, we have collected the unproven lemmas and will argue for their validity: although they are tedious or cumbersome to formalize, their correctness is either evident from the by-hand proofs, presented in the report, or easily justified by informal reasoning.

It should be noted that many of these lemmas are missing two minor adjustments, that are likely to make them not hold. We mention these adjustments here to spare reader of the reiteration of these mistakes. Firstly, there should be a premise stating that the handle *u* and *r* should be different for everything except the encoding of a value, something one can easily enforce be simply making the initial call to encode with two different numbers. Secondly some lemmas remove restrictions surrounding a call to an encoding. This means that the references inside this call no longer line up, and as such these lemmas would need to incorporate decrementing the references. Both these adjustments should only incur slight changes to what has already been proven, and would mostly impact the as of yet unproven lemmas.

### 4.6.1 Bisimulation

We have many lemmas that concern bisimulation. This is because bisimulation has both been a tedious and difficult property to prove. The reason being that most bisimulation proofs are spent proving that most transitions are not possible, an aspect which is not seen much in by-hand proofs, as it is most often the case that it is immediately obvious which transitions are possible for a given process. A possible solutions to this may have been to look further into creating our own tactics, as proving the falsehood of premises did to a certain extent follow the same structure, which was to perform long sequences of inversions on the correct premises until a contradiction would arise. However, the sequences would slightly change depending on the process, and our Coq knowledge is still not quite at the stage where we could implement such logic into our tactics.

The first two lemmas state that weak bisimulation is both symmetric and transitive. The first follows trivially from the definition of weak bisimulation, as the second clause of its definition is a symmetry clause. The second we can prove by constructing weak bisimulation relation $R$, where $p, r \in R$. By assumption there must exist bisimulation relations $R_1$ and $R_2$ s.t. $p, q \in R_1$ and $q, r \in R_2$, one can then construct the bisimulation relation $R = \{(p, r) \mid (p, q) \in R_1, (q, r) \in R_2\}$.

```
1  Lemma wb_sym:
2    forall p q,
3      p ~~ q -> q ~~ p.
4
5  Lemma wb_trans:
6    forall p q r,
7      p ~~ q -> q ~~ r -> p ~~ r.
```

`wb_par` follows from the fact that in $\pi I$-calculus bisimulation is naturally a congruence.

```
1  Lemma wb_par:
2    forall p q r s,
3      p ~~ q -> r ~~ s -> Par p r ~~ Par q s.
```

`res_rep_in_0` follows trivially as neither of the processes can make an action. This lemma also matches one of our bisimiliarity laws.

```
1  Lemma res_rep_in_0:
2    forall p,
3      Res (Rep (In (BN 0) p)) ~~ Nil.
```

`app_val_bisim` is a bisimilarity between two processes, that we did not complete the proof of. It can intuitively be understood as follows. If the left-hand side makes an $\xrightarrow{1()}$ then so does the right-hand side and vice-versa. If instead the left-hand side makes a communication on 2 then the right-hand side will communicate on 3 then 2. If the the right-hand side communicates on 3 then the left-hand side does nothing. After performing these actions in any given order, the processes will be syntactically equivalent, and therefore trivially bisimilar.

```
1  Lemma app_val_bisim:
2    forall v0 v u r,
3      (Res (Par
4        (In (BN 2) (Out (BN 1) (Out (BN 2) (Out (BN 3) (Par
5          (Link (BN 2) (BN (9 + u)))
6          (Par
7            (Link (BN 1) (BN (9 + r)))
8            (Link (BN 0) (BN 3))
9            )
10       )))))
11       (Par
12         ($ v0; [] $)
13         (Out (BN 2) ($ v; [] $ [[lift_subst S]]))
14         )
15     )) ~~
16     (Par
17       (In (BN 3) (In (BN 2) (Out (BN 1) (Out (BN 2) (Out (BN 3) (Par
18         (Link (BN 2) (BN (9 + u)))
19         (Par
20           (Link (BN 1) (BN (9 + r)))
21           (Link (BN 0) (BN 3))
22           )
23       (Par
24         (Out (BN 3) ($ v0; [] $))
25         (Out (BN 1) ($ v; [] $))
26       )))))))
27     ).
```

In `app_bind_bisim` if the left-hand side makes an action then the right-hand side will perform a communication on 1 and then the action. If the right-hand side makes a communication on 1 then the left-hand side will do nothing. After any of these actions the processes will be syntactically equivalent.

```
1   Lemma app_bind_bisim:
2     forall v0,
3       (Res (Par
4         ($ v0; [] $)
5         (Out (BN 0) (Out (BN 0) (Out (BN 1) (Par
6           (Link (BN 1) (BN 9))
7           (Link (BN 0) (BN 8))
8         ))))
9       )) ~~
10      (Par
11        (Out (BN 1) ($ v0; [] $))
12        (In (BN 1) (Out (BN 0) (Out (BN 0) (Out (BN 1) (Par
13          (Link (BN 1) (BN 9))
14          (Link (BN 0) (BN 8))
15        )))))
16      ).
```

`nil_proc_1` holds as the left-hand side can make no actions, because the action $\xrightarrow{1()}$ available through any of the links, is not possible due to the restrictions.

```
1   Lemma nil_proc_1:
2     forall u r s,
3       Res
4         (Res
5           (Par (Res (Link (BN 0) (BN (S (S (S u)))))))
6             (Res (Par (Link (BN 0) (FN s)) (Link (BN 0) (BN (S (S (S r))))))))))))
6           ~~
7       Nil.
```

`nil_proc_2` holds as the right-hand side can only make $\xrightarrow{\tau}$ actions, after which it can take no actions for the same reasons as in the previous lemma.

```
1   Lemma nil_proc_2:
2     forall u r s,
3       Nil ~~
4       Res
5         (Res
6           (Par
7             (Out (BN 1)
8               (Rep (In (BN 0) (In (BN 0) (In (BN 1) (Link (BN 0) (FN s)))))))
9             (In (BN 1)
10              (Out (BN 0)
11                (Out (BN 0)
```

```
12                        (Out (BN 1)
13                      (Par (Link (BN 1) (BN (S (S (S (S (S (S u)))))))))
14                        (Link (BN 0) (BN (S (S (S (S (S (S r))))))))))))))))))).
```

### 4.6.2 Lemmas of Well-formed terms

We have a couple of lemmas about how well-formed terms behave. The first being the somewhat trivial property that if a term is well-formed for some $n$, then it is also well-formed for the successor of $n$.

The second being that if a term is well-formed with respect to 1, and one substitutes in a value that is well-formed with respect to 0, then the resulting term is well-formed with respect to 0. Take the term $(\lambda s)v$, by assumption if $\lambda s$ and $v$ are well-formed with respect to 0 then we know that $s$ must be well-formed with respect to 1. The resulting substitution $s\{v \triangleright \mathrm{id}\}$, will be a term where all the references to the binder $\lambda$ are replaced with a value that is well-formed with respect to 0, as such the whole term must now also be well-formed with respect to 0.

```
1  Lemma wf_term_extend:
2    forall n s,
3      wf_term n s -> wf_term (S n) s.
4
5  Lemma wf_term_subst:
6    forall s v,
7      wf_term 1 s ->
8      wf_value 0 v ->
9      wf_term 0 (s {{v {}> Var <<< BV}}).
```

### 4.6.3 De Bruijn index substitutions

Most of our lemmas about de Bruijn index substitutions are about the simplification and removal of redundant substitutions. By this we mean $p\sigma = p$. In general the problem with proving these in a formal setting is that induction on the structure of processes becomes difficult due to lifting. As soon as you traverse a binder the substitution is lifted. However, one's inductive hypothesis only works on a non-lifted substitution.

The two lemmas about substitutions on encodings of values and terms, follow from an inspection of the encoding rules. One can see that the encoding of a term never reference any index higher than $u$ or $r$. Therefore it follows from this that any substitution lifted higher than those indices must therefore be redundant. In the same manner, one can see that the encoding of a value never references an index higher than 0, it therefore follows that any lifted substitution is redundant.

```
1  Lemma redundant_subst_term:
2    forall s u r refs n subst,
3      wf_term 0 s ->
4      n > u ->
5      n > r ->
6      $ s ; u ; r ; refs $ [[Nat.iter n lift_subst subst]] =
7      $ s ; u ; r ; refs $.
8
```

```
9    Lemma redundant_subst_value:
10     forall v refs subst,
11       wf_value 0 v ->
12       $ v ; refs $ [[lift_subst subst]] =
13       $ v ; refs $.
```

The following lemmas states that a shift followed by an extend of id by 0 is redundant. One can argue for the correctness of this pictorially as seen on Figure 3.

```
1   Lemma shift_extend_proc:
2     forall p,
3       (p [[S]] [[0 []> id]]) = p.
```

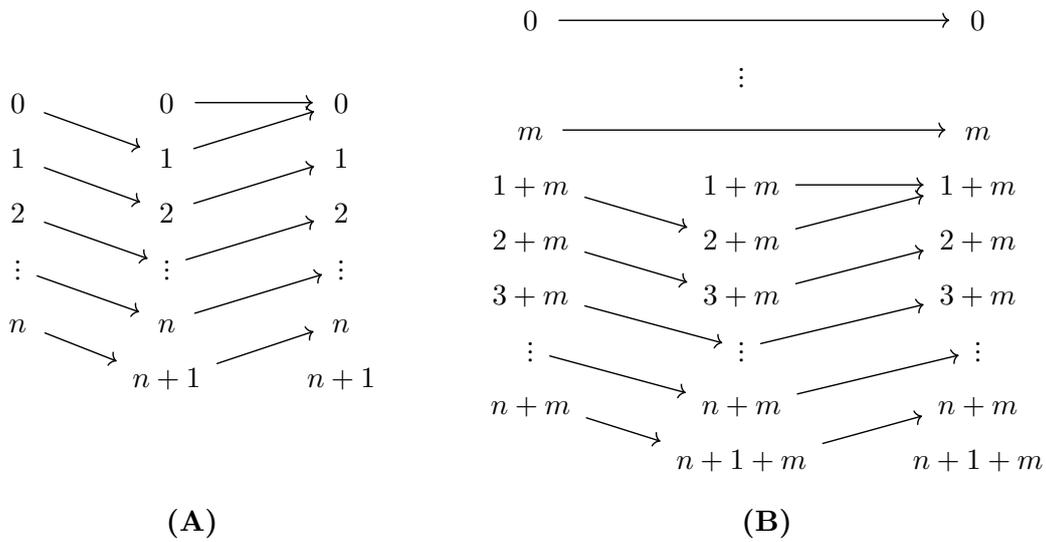

**(A)**                    **(B)**

Figure 3: Pictoral depiction of a shift followed be an extend of id by 0. **(A)** base case. **(B)** lifted one or more times.

The last lemma about de Bruijn index substitutions states that shifted process will never make a reference to the index 0. This follows trivially from the definition of shift and lifting.

```
1   Lemma ref_n_in_proc_shift:
2     forall p,
3       ref_n_in_proc 0 (p[[S]]) = false.
```

### 4.6.4 Prefix

We have a few lemmas regarding the immediate action that an encoded term can make. By inspection of the rules one can see that an encoded term never makes an input, and never does an output on anything other than $u$ and $r$. These lemmas are often used to prove the falsehood of premises, stating that an encoded term can not take an input or an output.

```
1   Lemma encode_no_input:
2     forall s u r refs p n,
3       ($ s ; u ; r ; refs $) -( a_in n )> p -> False.
```

```
1   Lemma encode_u_r_bn_output:
```

```
2    forall s u r refs p n,
3      n <> u ->
4      n <> r ->
5      ($ s ; u ; r ; refs $) -( a_out (BN n) )> p -> False.
```

```
1  Lemma encode_no_fn_output:
2    forall s u r refs p n,
3      ($ s ; u ; r ; refs $) -( a_out (FN n) )> p -> False.
```

### 4.6.5 From hand-written proof

In the formalization we have used some of the same lemmas as in the hand-written proof. We have not formally proven any of them yet, but we will argue here as to why they have been translated correctly, thus they should be correct under the assumption that the hand-written proofs are correct. From Lemma 6 we have clause (2) and (4) that are translated to `link_lift` and `link_handlers` respectively, and Lemma 5 which is translated to `subtitution`.

From Lemma 6 clause (2) the restriction on $y$ becomes 0, and $x$ becomes $n$. The substitution $\{y/x\}$ is replaced by the de Bruijn index substitution $\{\Uparrow^n S\}$ which means that no $n$'s can appear in the process. We see that $s$ is well-formed with respect to 1. This is because the lemma is used when $s$ is a sub-term of an abstraction, as such $s$ may not be well-formed with respect to 0. Therefore, the lemma states that if an bound occurrence does not appear in a process, but one can it access trough a link, one can remove the link and access it without the use of a link.

An adjustment that should be added to this lemma is that $n$ should be different from 0, $u$, and $r$.

$$(\nu y)(y \to x \mid \mathcal{J}[\![M]\!]_u^r \{y/x\}) \approx_E \mathcal{J}[\![M]\!]_u^r$$

```
1  Lemma link_lift:
2    forall n s u r refs,
3      wf_term 1 s ->
4      Res (Par (Link (BN 0) (BN n)) ($ s ; S u ; S r ; refs $ [[Nat.iter n
       lift_subst S]])) ~~
5      $ s ; u ; r ; refs $.
```

We now look at Lemma 6 clause (4):

$$(\nu u' r')\big(u' \to u \mid r' \to r \mid \mathcal{J}[\![M]\!]_{u'}^{r'}\big) \approx_E \mathcal{J}[\![M]\!]_u^r$$

From Lemma 6 clause (4) $u'$ becomes a 1 and $r'$ becomes a 0, and then the handles are decremented as two restrictions are removed. Again there needs to be made adjustments that keeps $u$, $r$, $n$ and $m$ different.

```
1  Lemma link_handlers: forall s n m refs,
2    wf_term 1 s ->
3    (Res (Res (Par
4      (encode s 1 0 refs)
5      (Par
6        (Link (BN 1) (BN (S (S n))))
```

```
7         (Link (BN 0) (BN (S (S m)))) 
8       ) 
9     ))) ~~ 
10    (encode s n m refs).
```

One of the most elaborate lemmas, in this report is the substitution lemma from Lemma 5. Given the detailed by-hand proof, and time restraint, formalizing it in Coq remains a promising direction for future work.

$$\llbracket M\{^V/_x\} \rrbracket_u^r \approx_E (\nu y)(\llbracket M \rrbracket_u^r \{^y/_x\} \mid [y := V])$$

```
1  Lemma substitution: 
2    forall s v, 
3      wf_term 1 s -> 
4      wf_value 0 v -> 
5      forall u r, 
6        Res (Par ($ v ; [] $) ($ s ; S u ; S r ; [(0,0)] $)) ~~ 
7        $ (s {{ extend_subst_lam v (Var <<< BV) }}) ; u ; r ; []$.
```

### 4.6.6 Encode reach

An important lemma used in the formalization, which does not appear in the unformalized proof is that of `encode_reach`. This lemma captures the idea, that the only binders referenced within an encoding of a term, are those that bind $u$ and $r$. By inspecting the rules of the encoding, one can see that the only names which refer outside the immediate surrounding context are $u$ and $r$. Here we also see an advantage of the locally nameless approach. Had we simply used de Bruijn indices then corresponding names to any free variable in $s$ would also point out of the encoding. The lemma is particularly useful when moving around restrictions, without having to unfold the encoding.

```
1  Lemma encode_reach: 
2    forall n s u r, 
3      wf_term 0 s -> 
4      n <> u -> 
5      n <> r -> 
6      ref_n_in_proc n ($ s ; u ; r ; [] $) = false.
```

### 4.6.7 Weak tau transition in context

The following lemma has been made since during the formalization, it was often the case that we had the premise $P \overset{\tau}{\Rightarrow} Q$ and the goal $C[P] \overset{\tau}{\Rightarrow} C[Q]$. However due to the nature of the implementation, the way to prove the goal would be to decompose the goal into $C[P] \overset{\tau}{\Rightarrow} C[P'] \overset{\tau}{\Rightarrow} C[Q]$, and then $C[P] \overset{\tau}{\rightarrow} C[P'] \overset{\tau}{\Rightarrow} C[Q]$. We could then traverse the process until we needed to prove that $P$ could make a single step transition. However, the premise says nothing about a single step transition, it only states that $P$ can make a weak transition. As such we make the following lemma, which states that if the context hole is not under a prefix (`hup c`), then if a process in the context hole can make a weak tau transition then so can the context with the process.

```
1  Lemma wt_tau_context:
2    forall c p q,
3      hup c = false ->
4      p =()> q ->
5      plug c p =()> plug c q.
```

### 4.6.8 Restricted encoding

The following lemma says something about unnecessary restrictions surrounding an encoding of a term. The intuition behind this lemma is that we know that the only actions performed outwards would be inputs and outputs on the handles $u$ and $r$. On the left hand side this will be the case because any actions will have an $S$ attached to them due to the handles being $S\ u$ and $S\ r$. When this action traverses the restriction, the $S$ will be removed, a swap substitution will then occur to account for the swapping of binders. However, as previously mentioned, this lemma is not quite correct, as the references would need to be updated, and most likely $u$ and $r$ would need to be different.

```
1  Lemma res_encoding:
2    forall s,
3      wf_term 0 s ->
4      forall u r refs,
5        (Res (encode s (S u) (S r) refs)) ~~ (encode s u r refs).
```

# 5 Conclusion

In this chapter, we will summarize the results of our report, as well as future work that could follow the results presented.

## 5.1 Results

We have in this report shown our encoding of a simplified version of CBPV, that had the necessary constructs to encode CBN and CBV, in the monadic $\pi$-calculus, and proved that this encoding is both sound and complete by proving Theorem 1. Our encoding also satisfies the properties presented in [1] for a good encoding. This means that our encoding closely mimics the behaviour of CBPV, with both divergence and termination capabilities.

Our encoding has a close correspondence between encoding CBN and CBV directly in the $\pi$-calculus using the encodings introduced in "Functions as Processes"[14] by Milner, and if we instead encode CBV or CBN to CBPV, and then use our encoding to encode in the $\pi$-calculus.

We show how to create a encoding in the polyadic $\pi$-calculus, that allowed for more concise correspondence between the original encodings in [14] and our encoding of CBPV.

We show an encoding of CBPV in the $\pi I$-calculus that is more similar to Sangiorgi's output based encodings. We then proved that our encoding to CBPV are sound and complete. Using this proof, we also showed that this encoding satisfies the properties of [1].

We began formalizing the soundness and completeness proof of our encoding in the $\pi I$-calculus using a formalization in Coq with de Bruijn index and the locally nameless approach. We have not completed the formalized proof, but we have reached a state where the main proof of soundness and a weaker version of completeness is proven, given a number of lemmas and bisimilarities hold, which we did not have time to complete for this project. However, we believe it is easy to see that the lemmas and bisimilarities are true, and could be proven by hand without larger issues.

## 5.2 Future work

The most pressing future work would be to fully complete our formal proofs of soundness and completeness in Coq. This would help solidify the correctness of our encoding, and also hopefully give a framework for formally proving the other theorems and conjectures presented in this paper.

We did, for example, not prove the conjecture of the correspondence between Milners original encodings in [14], and our encoding in the polyadic $\pi$-calculus. This proof would also be a nice result, and could show a way to combine CBN and CBV paradigms in parallel programming languages, since each paradigm would have its own barbs, that could be converted to correspond with the other paradigm. This means, that in case you had a function, and you were unsure if it would be most efficient to execute it as CBN or CBV (for example, if it would only terminate in CBN), this could give a method to execute them in parallel, in the $\pi$-calculus, and then only use the result from the process that terminated first.

Another approach to extend the work in this report could be to extend our encoding, to also include more of the original constructs in Levy's CBPV. These constructs allow more advanced CBN and CBV $\lambda$-calculus to be encoded in the CBPV, and then with an extended encoding of the CBPV, a unified encoding of these extended paradigms could then be made to the $\pi$-calculus. But since we wanted to show similarities between our encoding and Milners original encodings, we leave this extension as future work.

As mentioned, it would also be possible to specify a more specific correctness property for property 5 from [1]. This would take into account all the other terminal processes that CBPV could reduce to. We deem this to be a lesser result, but it could be interesting to look into and prove. But it would require that terms like **force** $x$ is seen as a goal-state, which is not the usual case when looking at CBPV.

There is also the potential to modify our encoding. We did some preliminary work in showing an encoding of CBPV in the asynchronous $\pi$-calculus, as well as the local $\pi$-calculus (L$\pi$), which is an asynchronous $\pi$-calculus that only sends "output capabilities" on channels. Our encoding in L$\pi$ and the asynchronous $\pi$-calculus can be found in Appendix D. We have not proven any soundness or completeness properties for these encodings.

One could also look into typing our encoding. We suspect that there is a possible correspondence between the type system of CBPV, and a receptiveness type system of our $\pi$-calculus encoding, such as the one presented in [43], as well as a polyadic type system. We believe this, since the places where there is multiple outputs on the same channel, such as when encoding $[\![V\ V']\!]$, it would also be ill-typed in CBPV. However, we deemed that defining a relevant type-system, and proving that our encoding preserves well-typedness, is outside the scope of this project.

Lastly, a nice property of an encoding is also that it preserves equivalence. In [4], they present a formal equational theory for CBPV. We strongly believe that if two CBPV term $M, N$ are bisimilar, then $[\![M]\!]^r_u \approx_E [\![N]\!]^r_u$. However, we leave it as future work to actually show that this holds.

# A) Soundness and Completeness Proof of CBPV encoding in $\pi$-calculus

In this appendix we will prove that our encoding of CBPV in the $\pi$-calculus is both sound and complete.

Before we can prove Theorem 1, we need Lemma 5 (Substitution lemma). This lemma is necessary with regards to substitution of values, as we in the $\pi$-calculus cannot substitute whole processes, as we would in the CBPV using thunks, but instead rely on pointers.

To prove Lemma 5 we need a few other lemmas. Lemma 2 states that we are able to move a sequence of inputs or outputs out in front of a restriction. Lemma 3 states that we may create an extra pointer to a value. Lemma 4 states that we can move a replication outside of a parallel process that has a restriction.

Lemma 2 and Lemma 4 are identical to that of lemma 3.3 and 3.2 by Sangiorgi in [33].

We will throughout the proof take note of how many reductions in our encoding is equivalent to a reduction in CBPV, as this will be used later for a theorem about divergence reflection.

## A.1) Restriction prefix lemma

Lemma 2 is identical to that of lemma 3.3 in [33].

**Lemma 2.** (Restriction prefix) Let $\mu$ be a sequence of inputs and output, then:

$$(\nu y)(\mu.P \mid [\![ y := V ]\!]) \approx_E \mu.(\nu y)(P \mid [\![ y := V ]\!]) \quad \begin{cases} y \notin \mathrm{n}(\mu) \\ \mathcal{V}(\mathrm{bn}(\mu)) \cap \mathrm{fv}(V) = \emptyset \end{cases}$$

*Proof of Lemma 2.*

We construct the bisimulation relation $R$:

$$R' = \{((\nu y)(\mu.P \mid [\![ y := V ]\!]), \mu.(\nu y)(P \mid [\![ y := V ]\!])),$$
$$((\nu y)(P \mid [\![ y := V ]\!]), (\nu y)(P \mid [\![ y := V ]\!]))\}$$
$$R = R' \cup \{(Q, P) \mid (P, Q) \in R'\} \cup \{(P, P) \mid P \in \mathbf{Proc}\}$$

We have that:

$$\text{if and only if } (\nu y)(\mu.P \mid [\![ y := V ]\!]) \xrightarrow{\mu} (\nu y)(P \mid [\![ y := V ]\!])$$
$$\text{then} \qquad \mu.(\nu y)(P \mid [\![ y := V ]\!]) \xrightarrow{\mu} (\nu y)(P \mid [\![ y := V ]\!])$$

Based on the restriction $y \notin \mathrm{n}(\mu)$, and that the encoding of $[\![ y := V ]\!]$ always starts with an input on the channel $y$, this is also the only action that can be taken. Because the pair of resulting processes are in $R$, and the rest of the pairs have 2 identical processes, $R$ must be a bisimulation, and therefore the lemma must hold.

$\square$

## A.2) Pointer lemma

**Lemma 3.** (Pointer)

$$(\nu y)(P\{^y/_x\} \mid Q\{^y/_x\} \mid [\![y := V]\!])$$
$$\approx_E (\nu y)(\nu z)(P\{^y/_x\} \mid Q\{^z/_x\} \mid [\![y := V]\!] \mid [\![z := V]\!])$$
$$\text{where } x \notin \mathrm{sub}_{\mathrm{in}}(P) \cup \mathrm{sub}_{\mathrm{in}}(Q) \cup \mathrm{obj}_{\mathrm{out}}(P) \cup \mathrm{obj}_{\mathrm{out}}(Q)$$
$$\text{and } z \notin \mathrm{fn}(P, Q) \cup \mathcal{V}^{-1}(\mathrm{fv}(V))$$

This lemma is needed to show that we can create a new pointer, that points to a new location with the same value as in the original process, and it will not introduce any new behaviour to the processes. By inspection of the rules of our encoding, it is clear that the requirements set for $x$ and $z$ are always satisfied.

*Proof of Lemma 3.*

We construct the relation $R'$ and $R^*$:

$$R' = \{((\nu y)(P\{^y/_x\} \mid Q\{^y/_x\} \mid [\![y := V]\!]),$$
$$(\nu y)(\nu z)(P\{^y/_x\} \mid Q\{^z/_x\} \mid [\![y := V]\!] \mid [\![z := V]\!]))\}$$

$$R^* = \Bigg\{ \left( (\nu y)\left( P' \mid Q'_y \mid [\![y := V]\!] \mid \prod_{i=0}^{k} V' \right), \right.$$
$$\left. (\nu y)(\nu z)\left( P' \mid Q'_z \mid [\![y := V]\!] \mid [\![z := V]\!] \mid \prod_{i=0}^{k} V' \right) \right)$$

$$\mid P\{^y/_x\} \overset{\mu'}{\Rightarrow} P' \ \lor \ P' = P\{^y/_x\} \ ,$$
$$Q\{^y/_x\} \overset{\mu\{^y/_x\}}{\Rightarrow} Q'_y \ \lor \ Q'_y = Q\{^y/_x\} \ ,$$
$$Q\{^z/_x\} \overset{\mu\{^z/_x\}}{\Rightarrow} Q'_z \ \lor \ Q'_z = Q\{^z/_x\} \ ,$$
$$[\![y := V]\!] \overset{y(a)}{\to} [\![y := V]\!] \mid V' \ , \ [\![z := V]\!] \overset{z(a)}{\to} [\![z := V]\!] \mid V' \Bigg\}$$

From the encoding we know that $[\![y := V]\!] \overset{y(a)}{\to} [\![y := V]\!] \mid V'$ is always the case, since all encodings on the form $[\![y := V]\!]$ begins with a replicated input.

We write $\overset{\mu}{\Rightarrow}$ where $\mu$ is a sequence of inputs and outputs eg. $\mu = \mu_1, ... \mu_n$, then $P \overset{\mu}{\Rightarrow} P' = P \overset{\tau}{\to} ... \overset{\tau}{\to} P_1 \overset{\mu_1}{\to} P_2 \overset{\tau}{\to} ... \overset{\tau}{\to} P_n \overset{\mu_n}{\to} P_{n+1} \overset{\tau}{\to} ... \overset{\tau}{\to} P'$. We close $R^*$ with regards to the behaviour of $V'$ to obtain $R'^*$. What we mean by this is, that for each pair, where each process contain a $V'$, we add the pairs where that $V'$ has been exchanged with $V''$ and we have that $V' \overset{\mu}{\Rightarrow} V''$. We then have the relation:

$$R = R' \cup R'^* \cup \{(Q, P) \mid (P, Q) \in R' \cup R'^*\}$$

We will now show that $R$ is an early weak bisimulation.

We start with the pair in $R'$. From looking at the full processes we know that $[\![y := V]\!]$ and $[\![z := V]\!]$ cannot make the entire process do an input, because of the restrictions. Therefore, we look at $P'$ and $Q'$ instead. We inspect the case where $P \overset{\overline{x}a}{\to} P'$

We know that: $\llbracket y := V \rrbracket \overset{y(a)}{\to} \llbracket y := V \rrbracket \mid V'$ We then have:

$$(\nu y)(P\{^y/_x\} \mid Q\{^y/_x\} \mid \llbracket y := V \rrbracket) \overset{\tau}{\to} (\nu y)(P'\{^y/_x\} \mid Q\{^y/_x\} \mid \llbracket y := V \rrbracket \mid V')$$

and

$$(\nu y)(\nu z)(P\{^y/_x\} \mid Q\{^z/_x\} \mid \llbracket y := V \rrbracket \mid \llbracket z := V \rrbracket) \overset{\tau}{\to}$$
$$(\nu y)(\nu z)(P'\{^y/_x\} \mid Q\{^z/_x\} \mid \llbracket y := V \rrbracket \mid V' \mid \llbracket z := V \rrbracket)$$

We see the resulting pair is still in $R$. Now, if instead we have that $Q \overset{\overline{x}a}{\to} Q'$, we also have that:

$$\llbracket z := V \rrbracket \overset{z(a)}{\to} \llbracket z := V \rrbracket \mid V'$$

we then have:

$$(\nu y)(P\{^y/_x\} \mid Q\{^y/_x\} \mid \llbracket y := V \rrbracket) \overset{\tau}{\to} (\nu y)(P\{^y/_x\} \mid Q'\{^y/_x\} \mid \llbracket y := V \rrbracket \mid V')$$

and

$$(\nu y)(\nu z)(P\{^y/_x\} \mid Q\{^z/_x\} \mid \llbracket y := V \rrbracket \mid \llbracket z := V \rrbracket) \overset{\tau}{\to}$$
$$(\nu y)(\nu z)(P\{^y/_x\} \mid Q'\{^z/_x\} \mid \llbracket y := V \rrbracket \mid \llbracket z := V \rrbracket \mid V')$$

In this case the resulting pair is also in the relation. In case that $P$ or $Q$ can output on another channel than $x$, then the processes can still easily match the other, as they both have a $P$ and $Q$, and their behaviour only differ when communicating with $y$ and $z$ to get the value $V$.

For the pairs in $R'^*$, we see from the structure of each pair, that a process can always match the others transitions. The only interesting case is if $Q'_y$ can take an output on the channel $y$, but since there is a restriction on $y$, it can only do so to communicate with $\llbracket y := V \rrbracket$ and $P$ does not have $y$ as subject of an input (and can never get it, as neither $P$ nor $Q$ can have $y$ as the object of an output). The same logic applies to $Q'_z$ for the channel $z$.

We can clearly see that each process match the others transitions. In all cases, if $P', Q', V' \overset{\mu}{\to}$ where $y, z \notin (\mu)$, then both process are able to do so and since we, by definition of $R$ regardless of what actions the sub-processes take, still end up being in $R$, $R$ must be an early weak bisimulation. And since we have that:

$$((\nu y)(P\{^y/_x\} \mid Q\{^y/_x\} \mid \llbracket y := V \rrbracket), (\nu y)(\nu z)(P\{^y/_x\} \mid Q\{^z/_x\} \mid \llbracket y := V \rrbracket \mid \llbracket z := V \rrbracket)) \in R$$

It must be the case that:

$$(\nu y)(P\{^y/_x\} \mid Q\{^y/_x\} \mid \llbracket y := V \rrbracket) \approx_E (\nu y)(\nu z)(P\{^y/_x\} \mid Q\{^z/_x\} \mid \llbracket y := V \rrbracket \mid \llbracket z := V \rrbracket)$$

$\square$

## A.3) Restriction replication lemma

This lemma is identical to that of lemma 3.2 in [33].

**Lemma 4.** (Restriction replication)

$$(\nu y)(!P\{^y/_x\} \mid \llbracket y := V \rrbracket) \approx_E \, !((\nu y)(P\{^y/_x\} \mid \llbracket y := V \rrbracket))$$
$$\text{where } y \notin \text{sub}_{\text{in}}(P) \cup \text{obj}_{\text{out}}(P)$$

*Proof of Lemma 4.*

We use up-to-bisimulation, which means we can use our bisimulation laws to rewrite. We have that:

$$(\nu y)(!P\{^y/_x\} \mid [\![y := V]\!]) \approx_E (\nu y)(!P\{^y/_x\} \mid P\{^y/_x\} \mid [\![y := V]\!])$$

Using Lemma 3 we get:

$$(\nu y)(!P\{^y/_x\} \mid P\{^y/_x\} \mid [\![y := V]\!]) \approx_E (\nu y)(\nu z)(!P\{^z/_x\} \mid [\![z := V]\!] \mid P\{^y/_x\} \mid [\![y := V]\!])$$
$$\approx_E (\nu z)(!P\{^z/_x\} \mid [\![z := V]\!]) \mid (\nu y)(P\{^y/_x\} \mid [\![y := V]\!])$$

We also have that:

$$!((\nu y)(P\{^y/_x\} \mid [\![y := V]\!])) \approx_E !((\nu y)(P\{^y/_x\} \mid [\![y := V]\!])) \mid (\nu y)(P\{^y/_x\} \mid [\![y := V]\!])$$
$$\approx_E !((\nu z)(P\{^z/_x\} \mid [\![z := V]\!])) \mid (\nu y)(P\{^y/_x\} \mid [\![y := V]\!])$$

We see that when we unfold both processes, we obtain exactly the same context. Since this is the only behaviour of both processes, to create arbitrarily many $(\nu y)(P\{^y/_x\} \mid [\![y := V]\!])$, we can conclude that:

$$(\nu y)(!P\{^y/_x\} \mid [\![y := V]\!]) \approx_E !((\nu y)(P\{^y/_x\} \mid [\![y := V]\!]))$$

$\square$

## A.4) Substitution lemma

**Lemma 5.** (Substitution lemma)
$$[\![M\{^V/_x\}]\!]_u^r \approx_E (\nu y)([\![M]\!]_u^r\{^y/_x\} \mid [y := V])$$

*Proof of Lemma 5.*

We use structural induction on $M$, and prove it using up-to-bisimilarity.

**Abstraction**: For the encoding of abstraction:

$$[\![\lambda z.M]\!]_u^r = u(s).s(u).s(r).s(z).[\![M]\!]_u^r$$

We need to prove that:

$$[\![\lambda z.M\{^V/_x\}]\!]_u^r \approx_E (\nu y)([\![\lambda z.M]\!]_u^r\{^y/_x\} \mid [\![y := V]\!])$$

In (3.1) we unfold using the definition of the encoding. For (3.2) to (3.3) we move the restriction over the inputs using Lemma 2. In case there would be a name clash, we can easily alpha convert the object of the input to avoid it. From (3.3) to (3.4) we use the inductive hypothesis. And as a last step we use the definition of the encoding to fold (3.5).

$$(\nu y)([\![\lambda z.M]\!]_u^r\{^y/_x\} \mid [\![y := V]\!]) = (\nu y)(u(s).s(u).s(r).s(z).[\![M]\!]_v^r\{^y/_x\} \mid [y := V]) \quad (3.1)$$
$$\approx_E u(s).(\nu y)(s(u).s(r).s(z).[\![M]\!]_v^r\{^y/_x\} \mid [y := V]) \quad (3.2)$$
$$\approx_E u(s).s(v).(\nu y)(s(r).s(z).[\![M]\!]_v^r\{^y/_x\} \mid [y := V])$$
$$\approx_E u(s).s(v).s(r).(\nu y)(s(z).[\![M]\!]_v^r\{^y/_x\} \mid [y := V])$$
$$\approx_E u(s).s(v).s(r).s(z).(\nu y)([\![M]\!]_v^r\{^y/_x\} \mid [y := V]) \quad (3.3)$$

IV

$$\approx_E u(s).s(v).s(r).s(z).[\![M\{^V/_x\}]\!]_v^r \tag{3.4}$$

$$= [\![\lambda z.M\{^V/_x\}]\!]_u^r \tag{3.5}$$

**Application**: For the encoding of application:

$$[\![M\ V']\!]_u^r = (\nu p)(\nu q)\big([\![M\ V']\!]_a^b \mid (\nu s)\overline{p}(s).\overline{s}u.\overline{s}r.[\![V']\!]_s^r\big)$$

We need to prove that:

$$[\![M\ V'\{^V/_x\}]\!]_u^r \approx_E (\nu y)\big([\![M\ V']\!]_u^r\{^y/_x\} \mid [\![y := V]\!]\big)$$

First, in (4.1) we unfold the definition of the encoding. In (4.2) we use Lemma 3 to split $[\![y :=V]\!]$ up into 2. In (4.3) and (4.4) we use our bisimilarity laws from Proposition 1 to split up the process. In (4.5) we use Lemma 2 to move the restriction $(\nu z)$ under the outputs. In (4.6) we use the inductive hypothesis, and end up with exactly the encoding of $[\![M\ V'\{^V/_x\}]\!]$.

$$(\nu y)\big([\![M\ V']\!]_u^r\{^y/_x\} \mid [\![y := V]\!]\big)$$

$$= (\nu y)\big((\nu p)(\nu q)\big([\![M]\!]_p^q\{^y/_x\} \mid (\nu s)\overline{p}s.\overline{s}u.\overline{s}r.[\![V']\!]_s^r\{^y/_x\}\big) \mid [\![y := V]\!]\big) \tag{4.1}$$

$$\approx_E (\nu y)(\nu z)\big((\nu p)(\nu q)\big([\![M]\!]_p^q\{^y/_x\} \mid (\nu s)\overline{p}s.\overline{s}u.\overline{s}r.[\![V']\!]_s^r\{^z/_x\}\big) \mid [\![y := V]\!] \mid [\![z := V]\!]\big) \tag{4.2}$$

$$\approx_E (\nu p)(\nu q)(\nu y)(\nu z)\big([\![M]\!]_p^r\{^y/_x\} \mid (\nu s)\overline{p}s.\overline{s}u.\overline{s}r.[\![V']\!]_s^r\{^z/_x\} \mid [\![y := V]\!] \mid [\![z := V]\!]\big) \tag{4.3}$$

$$\approx_E (\nu p)(\nu q)\big((\nu y)\big([\![M]\!]_p^r\{^y/_x\} \mid [\![y := V]\!]\big) \mid (\nu z)\big((\nu s)\overline{p}s.\overline{s}u.\overline{s}r.[\![V']\!]_s^r\{^z/_x\} \mid [\![z := V]\!]\big)\big) \tag{4.4}$$

$$\approx_E (\nu p)(\nu q)\big((\nu y)\big([\![M]\!]_p^r\{^y/_x\} \mid [\![y := V]\!]\big) \mid (\nu s)\overline{p}s.\overline{s}u.\overline{s}r.(\nu z)\big([\![V']\!]_s^r\{^z/_x\} \mid [\![z := V]\!]\big)\big) \tag{4.5}$$

$$\approx_E (\nu p)(\nu q)\big([\![M\{^V/_x\}]\!]_p^r \mid (\nu s)\overline{p}s.\overline{s}u.\overline{s}r.[\![V'\{^V/_x\}]\!]_s^r\big) \tag{4.6}$$

$$= [\![M\ V'\{^V/_x\}]\!]_u^r.$$

**Force**: For the encoding of force:

$$[\![\mathbf{force}\ V]\!]_u^r = (\nu p)\big([\![V]\!]_p^r \mid p(y).(\nu s)\overline{y}s.\overline{s}u.\overline{s}r\big) \quad \mathcal{V}(p) \notin \mathrm{fv}(V)$$

We need to prove that:

$$[\![(\mathbf{force}\ V')\{^V/_x\}]\!]_u^r \approx_E (\nu y)([\![\mathbf{force}\ V]\!]_u^r\{^y/_x\} \mid [\![y := V']\!])$$

First, in (5.1) we unfold the definition of the encoding. Until (5.4) we use the bisimilarity laws to move $[\![y := V]\!]$ and $(\nu y)$ around, and we then use the inductive hypothesis to reach $[\![(\mathbf{force}\ V_1)\{^V/_x\}]\!]_u^r.$

$$(\nu y)\big([\![\mathbf{force}\ V']\!]_u^r\{^y/_x\} \mid [\![y := V]\!]\big)$$

$$= (\nu y)\big((\nu p)\big([\![V']\!]_p^r\{^y/_x\} \mid p(z).(\nu s)\overline{z}s.\overline{s}u.\overline{s}r\big) \mid [\![y := V]\!]\big) \tag{5.1}$$

$$\approx_E (\nu y)\big((\nu p)\big([\![V']\!]_p^r\{^y/_x\} \mid [\![y := V]\!] \mid p(z).(\nu s)\overline{z}s.\overline{s}u.\overline{s}r\big)\big) \tag{5.2}$$

$$\approx_E (\nu p)\big((\nu y)\big([\![V']\!]_p^r\{^y/_x\} \mid [\![y := V]\!]\big) \mid p(z).(\nu s)\overline{z}s.\overline{s}u.\overline{s}r\big) \tag{5.3}$$

$$\approx_E (\nu p)\big([\![V'\{^V/_x\}]\!]_p^r \mid p(z).(\nu s)\overline{z}s.\overline{s}u.\overline{s}r\big) \tag{5.4}$$

V

$$= [\![(\textbf{force } V')\{^V/_x\}]\!]_u^r$$

**Return**: For the encoding of return:

$$[\![\textbf{return } V]\!]_u^r = [\![V]\!]_r^r$$

We need to prove that:

$$[\![(\textbf{return } V')\{^V/_x\}]\!]_u^r \approx_E (\nu y)\big([\![\textbf{return } V']\!]_u^r\{^y/_x\} \mid [\![y := V]\!]\big)$$

We unfold the encoding:

$$(\nu y)\big([\![\textbf{return } V']\!]_u^r\{^y/_x\} \mid [\![y := V]\!]\big) = (\nu y)\big([\![V']\!]_r^r\{^y/_x\} \mid [y := V]\!]\big)$$

We use the inductive hypothesis, and then fold the encoding again. Since the argument handle $u$ is not present in the encoding of **return** $V_1$, when we fold the encoding, we can choose whatever handle we want, and since it will not be present in the encoding, they will still be equal to the unfolding:

$$(\nu y)\big([\![V']\!]_r^r\{^y/_x\} \mid [\![y := V]\!]\big) \approx_E [\![V'\{^V/_x\}]\!]_r^r$$
$$= [\![(\textbf{return } V')\{^V/_x\}]\!]_u^r$$

**Binding**: For the encoding of bind:

$$[\![M \ggg= \lambda x.N]\!]_u^r = (\nu z)((\nu u)[\![M]\!]_u^z \mid z(x).[\![N]\!]_u^r) \quad \mathcal{V}(z) \notin \text{fv}(M, N)$$

We need to prove that:

$$[\![(M \ggg= \lambda x.N)\{^V/_x\}]\!]_u^r \approx_E (\nu y)([\![M \ggg= \lambda x.N]\!]_u^r\{^y/_x\} \mid [\![y := V]\!])$$

Firstly, we unfold the encoding in (6.1) using its definition. To reach (6.2) we split $[\![y := V]\!]$ up into two with Lemma 3, such that $y$ and $a$ point to two distinct locations, both of which contain an encoding of the same value. To reach (6.4) we move the restriction behind the input using Lemma 2, and in (6.5) we use the inductive hypothesis to end up with the encoding of $[\![(M \ggg= \lambda x.N)\{^V/_x\}]\!]_u^r$.

$$(\nu y)([\![M \ggg= \lambda x.N]\!]_u^r\{^y/_x\} \mid [\![y := V]\!])$$

$$= (\nu y)((\nu z)((\nu u)[\![M]\!]_u^z\{^y/_x\} \mid z(q).[\![N]\!]_u^r\{^y/_x\}) \mid [\![y := V]\!]) \tag{6.1}$$

$$\approx_E (\nu y)(\nu a)((\nu z)((\nu u)[\![M]\!]_u^z\{^y/_x\} \mid z(q).[\![N]\!]_u^r\{^a/_x\}) \mid [\![y := V]\!] \mid [\![a := V]\!]) \tag{6.2}$$

$$\approx_E (\nu z)((\nu y)((\nu u)[\![M]\!]_u^z\{^y/_x\} \mid [\![y := V]\!]) \mid (\nu a)(z(q).[\![N]\!]_u^r\{^a/_x\} \mid [\![a := V]\!])) \tag{6.3}$$

$$\approx_E (\nu z)((\nu u)(\nu y)([\![M]\!]_u^z\{^y/_x\} \mid [\![y := V]\!]) \mid z(q).(\nu a)([\![N]\!]_u^r\{^a/_x\} \mid [\![a := V]\!])) \tag{6.4}$$

$$\approx_E (\nu z)\big((\nu u)[\![M\{^V/_x\}]\!]_u^z \mid z(q).[\![N\{^V/_x\}]\!]_u^r\big) \tag{6.5}$$

$$= [\![(M \ggg= \lambda x.N)\{^V/_x\}]\!]_u^r$$

**Thunk**: For the encoding of thunk:

$$[\![V]\!]_u^r = (\nu y)(\overline{u}y.[\![y := V]\!]) \qquad \mathcal{V}(y) \notin \text{fv}(V)$$
$$[\![y := \textbf{thunk } M]\!] = !y(s).s(u).s(r).[\![M]\!]_u^r \quad \mathcal{V}(s) \notin \text{fv}(M)$$

We need to prove that:

$$\llbracket (\textbf{thunk } M)\{^{V}/_{x}\} \rrbracket^{r}_{u} \approx_{E} (\nu y)(\llbracket \textbf{thunk } M \rrbracket^{r}_{u}\{^{y}/_{x}\} \mid \llbracket y := V \rrbracket)$$

In (7.1) we unfold using the definition of a binding. In (7.2) we use Lemma 2 to move the restriction and $\llbracket y := V \rrbracket$ behind the prefix, and move the restriction on $z$ out. In (7.3) we use Lemma 4 to move the restriction behind the replication. In (7.4) to reach (7.5) we use Lemma 2. To reach (7.6) we use the inductive hypothesis. And finally, to reach (7.7) we fold using the definition of the encoding of a thunk.

$$(\nu y)(\llbracket \textbf{thunk } M \rrbracket^{r}_{u}\{^{y}/_{x}\} \mid \llbracket y := V \rrbracket) \tag{7.1}$$

$$= (\nu y)((\nu z)(\overline{u}z.!z(s).s(u).s(r).\llbracket M \rrbracket^{r}_{u}\{^{y}/_{x}\}) \mid \llbracket y := V \rrbracket) \tag{7.2}$$

$$\approx_{E} (\nu z)(\overline{u}z.(\nu y)(!z(s).s(u).s(r).\llbracket M \rrbracket^{r}_{u}\{^{y}/_{x}\} \mid \llbracket y := V \rrbracket)) \tag{7.3}$$

$$\approx_{E} (\nu z)(\overline{u}z.!z(s).(\nu y)(s(u).s(r).\llbracket M \rrbracket^{r}_{u}\{^{y}/_{x}\} \mid \llbracket y := V \rrbracket)) \tag{7.4}$$

$$\approx_{E} (\nu z)(\overline{u}z.!z(s).s(u).s(r).(\nu y)(\llbracket M \rrbracket^{r}_{u}\{^{y}/_{x}\} \mid \llbracket y := V \rrbracket)) \tag{7.5}$$

$$\approx_{E} (\nu z)\left(\overline{u}z.!z(s).s(u).s(r).\llbracket M\{^{V}/_{x}\} \rrbracket^{r}_{u}\right) \tag{7.6}$$

$$= \llbracket (\textbf{thunk } M)\{^{V}/_{x}\} \rrbracket^{r}_{u} \tag{7.7}$$

**Variable**: For the encoding of a variable:

$$\llbracket V \rrbracket^{r}_{u} = (\nu y)(\overline{u}y.\llbracket y := V \rrbracket) \quad \mathcal{V}(y) \notin \text{fv}(V)$$

$$\llbracket y := x \rrbracket = !y(w).\overline{x}w$$

We need to prove that:

$$\llbracket z\{^{V}/_{x}\} \rrbracket^{r}_{u} \approx_{E} (\nu y)(\llbracket z \rrbracket\{^{y}/_{x}\} \mid \llbracket y := V \rrbracket)$$

We split this into two cases, first when $z \neq x$, and then the interesting case when $z = x$.

$z \neq x$:

In the case that $z$ is another variable than $x$, we need to prove that:

$$\llbracket z\{^{V}/_{x}\} \rrbracket^{r}_{u} = \llbracket z \rrbracket^{r}_{u} \approx_{E} (\nu y)(\llbracket z \rrbracket^{r}_{u}\{^{y}/_{x}\} \mid \llbracket y := V \rrbracket)$$

We have that in (8.1) the substitution changes nothing, since $x$ is not free in $\llbracket z \rrbracket^{r}_{u}$. In (8.2) we can move the restriction since $y$ is not free in $\llbracket z \rrbracket^{r}_{u}$, and we can then garbage collect the restricted replication in $\llbracket y := V \rrbracket$ to reach (8.3).

$$(\nu y)(\llbracket z \rrbracket^{r}_{u}\{^{y}/_{x}\} \mid \llbracket y := V \rrbracket) = (\nu y)(\llbracket z \rrbracket \mid \llbracket y := V \rrbracket) \tag{8.1}$$

$$\approx_{E} \llbracket z \rrbracket^{r}_{u} \mid (\nu y)\llbracket y := V \rrbracket \tag{8.2}$$

$$\approx_{E} \llbracket z \rrbracket^{r}_{u} \tag{8.3}$$

Therefore, it holds when $z \neq x$.

$z = x$: In this case we need to prove that:

$$\llbracket x\{^{V}/_{x}\} \rrbracket^{r}_{u} = \llbracket V \rrbracket^{r}_{u}$$

$$\approx_{E} (\nu y)(\llbracket x \rrbracket^{r}_{u}\{^{y}/_{x}\} \mid \llbracket y := V \rrbracket)$$

We have that:

$$(\nu y)(\llbracket x \rrbracket^{r}_{u}\{^{y}/_{x}\} \mid \llbracket y := V \rrbracket) = (\nu y)(\llbracket x \rrbracket\{^{y}/_{x}\} \mid \llbracket y := V \rrbracket)$$

$$= (\nu y)((\nu z)(\overline{u}z.!z(w).\overline{x}w)\{^y/_x\} \mid [\![y := V]\!])$$
$$= (\nu y)((\nu z)(\overline{u}z.!z(w).\overline{y}w) \mid [\![y := V]\!])$$

We now split this into the two cases of $V$: The first being when $V$ is a variable, the second when $V$ is a thunk.

**$V = q$:**

We now look at the case where $V$ is a variable, meaning $V = q$

We then have:

$$(\nu y)((\nu z)(\overline{u}z.!z(w).\overline{y}w) \mid [\![y := q]\!]) = (\nu y)((\nu z)(\overline{u}z.!z(w).\overline{y}w) \mid !y(w).\overline{q}w)$$

We also have that:

$$[\![q]\!]_u^r = (\nu z)\overline{u}z.!z(w).\overline{q}w$$

We now need to show that:

$$(\nu z)\overline{u}z.!z(w).\overline{q}w \approx_E (\nu y)((\nu z)(\overline{u}z.!z(w).\overline{y}w) \mid !y(w).\overline{q}w)$$

We therefore want to create an early weak bisimulation relation $R$. It will be created from the following two relations:

$$R' = \{((\nu z)\overline{u}z.!z(w).\overline{q}w, (\nu y)((\nu z)(\overline{u}z.!z(w).\overline{y}w) \mid !y(w).\overline{q}w))\}$$

$$R^* = \left\{ \left( !z(w).\overline{q}w \mid \prod_{i=0}^{m} \overline{q}w_i, (\nu y)\left( (!z(w).\overline{y}w) \mid \prod_{i=0}^{j} \overline{y}w_i \mid \prod_{k=j+1}^{m} \overline{q}w_k \mid !y(w).\overline{q}w \right) \right) \right\}$$

We then have that:

$$R = R' \cup R^*$$

We now need to show that we can match all actions, and the resulting processes are still in the relation $R$. We will prove this using up-to-early-weak-bisimulation: We begin with the pair in $R'$. By inspection we see that in both cases, the only action they can take is a bound output:

$$(\nu z)\overline{u}z.!z(w).\overline{q}w \xrightarrow{\overline{u}(z)} !z(w).\overline{q}w$$

and

$$(\nu y)((\nu z)(\overline{u}z.!z(w).\overline{y}w) \mid !y(w).\overline{q}w) \xrightarrow{\overline{u}(z)} (\nu y)(!z(w).\overline{y}w) \mid !y(w).\overline{q}w$$

The pair of the resulting processes are in $R^*$, and therefore, they must also be in $R$. We now need to show that regardless of what transition a process in $R^*$ takes, it stays within $R$, and the corresponding process can match it. We begin with the case that $m = 0$. Here we see that we can only take an input on the channel $z$:

$$!z(w).\overline{q}w \xrightarrow{z(w)} !z(w).\overline{q}w \mid \overline{q}w$$

and

$$(\nu y)(!z(w).\overline{y}w) \mid !y(w).\overline{q}w \xrightarrow{z(w)} (\nu y)(!z(w).\overline{y}w) \mid \overline{y}w \mid !y(w).\overline{q}w$$

VIII

Now, the resulting processes are still in $R$. Now we look at the case where $m = 1$. Here the first process in the pair can do an output:

$$!z(w).\overline{q}w \mid \overline{q}w \xrightarrow{\overline{q}w} \; !z(w).\overline{q}w$$

If we have that $j = 0$, the second process can also do an output on the channel $q$. However, if $j = 1$, the process can do an internal communication, and then it can also make an output on channel $q$.

$$(\nu y)(!z(w).\overline{y}w) \mid \overline{y}w \mid !y(w).\overline{q}w \xrightarrow{\tau} (\nu y)(!z(w).\overline{y}w) \mid !y(w).\overline{q}w \mid \overline{q}w \xrightarrow{\overline{q}w} (\nu y)(!z(w).\overline{y}w) \mid !y(w).\overline{q}w$$

It can easily be seen, that for any $m$ larger than 1, the behaviour is still the same. Since $R$ satisfies the criteria for being a early weak bisimulation, and we have that:

$$([\![q]\!]^r_u, (\nu y)([\![x]\!]\{^y/_x\} \mid [\![y := q]\!]) \in R$$

it must be the case that:

$$[\![q]\!]^r_u \approx_E (\nu y)([\![x]\!]\{^y/_x\} \mid [\![y := q]\!]$$

**$V = $ thunk $M$**: We now look at the case where $V = $ **thunk** $V'$. The proof follows a very similar approach.

$$(\nu y)((\nu z)(\overline{u}z.!z(w).\overline{y}w) \mid [\![y := \textbf{thunk } V']\!]) = (\nu y)\Big((\nu z)(\overline{u}z.!z(s).\overline{y}w) \mid !y(s).s(u).s(r).[\![M']\!]^r_u\Big)$$

We also have that:

$$[\![\textbf{thunk } M']\!]^r_u = (\nu z)\overline{u}z.!z(s).s(u).s(r).[\![M']\!]^r_u)$$

We will create the relation

$$R' = \{\Big((\nu z)\overline{u}z.!z(s).s(u).s(r).[\![M']\!]^r_u, (\nu y)\big((\nu z)(\overline{u}z.!z(s).\overline{y}s) \mid !y(s).s(u).s(r).[\![M']\!]^r_u\big)\Big),$$

$$R^* = \Bigg\{ \Bigg(!z(s).s(u).s(r).[\![M']\!]^r_u \mid \prod_{i=0}^{l} s_i(u).s_i(r).[\![M']\!]^r_u \mid \prod_{j=l+1}^{m} s_j(r).[\![M']\!]^r_{u_j} \mid \prod_{k=m+1}^{n} [\![M']\!]^{r_k}_{u_k},$$

$$(\nu y)\Bigg(!z(s).\overline{y}s \mid !y(s).s(u).s(r).[\![M']\!]^r_u \mid \prod_{i'=0}^{w} \overline{y}s_{i'} \mid \prod_{i=w+1}^{l} s_i(u).s_i(r).[\![M']\!]^r_u$$

$$\mid \prod_{j=l+1}^{m} s_j(r).[\![M']\!]^r_{u_j} \mid \prod_{k=m+1}^{n} [\![M']\!]^{r_k}_{u_k}\Bigg)\Bigg)\Bigg\}$$

We then have that:

$$R = R' \cup R^* \cup \{(Q,P) \mid (P,Q) \in R' \cup R^*\}$$

We will now show that $R$ is an early weak bisimulation. This means that we need to show that we stay inside $R$, regardless of the transitions taken.

We begin with the pair in $R'$. We have that:

$$(\nu z)\overline{u}z.!z(s).s(u).s(r).[\![M']\!]^r_u \xrightarrow{\overline{u}(z)} \; !z(s).s(u).s(r).[\![M']\!]^r_u$$

and

$$(\nu y)\Big((\nu z)(\overline{u}z.!z(s).\overline{y}s) \mid !y(s).s(u).s(r).[\![M']\!]^r_u\Big) \xrightarrow{\overline{u}(z)} (\nu y)\Big((!z(s).\overline{y}s) \mid !y(s).s(u).s(r).[\![M']\!]^r_u\Big)$$

The resulting pair of processes are still in the relation, and it is the only transition possible for the pair. We now look at the case where $w, l, m, n = 0$. Here we have that both processes can make an input on the channel $z$:

$$!z(s).s(u).s(r).[\![M']\!]^r_u \xrightarrow{z(s)} !z(s).s(u).s(r).[\![M']\!]^r_u \mid s(u).s(r).[\![M']\!]^r_u$$

and

$$(\nu y)\Big((!z(s).\overline{y}s) \mid !y(s).s(u).s(r).[\![M']\!]^r_u\Big) \xrightarrow{z(s)} (\nu y)\Big((!z(s).\overline{y}s) \mid \overline{y}s \mid !y(s).s(u).s(r).[\![M']\!]^r_u)\Big)$$

Both processes are still in the relation, and it was the only action we could take. We now look at the case where $l = 1$. Here the only transition is an input on some channel $s$:

$$!z(s).s(u).s(r).[\![M']\!]^r_u \mid s(u).s(r).[\![M']\!]^r_u \xrightarrow{s(u)} !z(s).s(u).s(r).[\![M']\!]^r_u \mid s(r).[\![M']\!]^r_u$$

However, in case $w = 1$, we need to first make an internal communication, before we can match it. In the case where $w = 0$, then it can just match it.

$$(\nu y)\Big((!z(s).\overline{y}s) \mid \overline{y}s \mid !y(s).s(u).s(r).[\![M']\!]^r_u)\Big)$$
$$\xrightarrow{s(u)} (\nu y)\Big((!z(s).\overline{y}s) \mid !y(s).s(u).s(r).[\![M']\!]^r_u \mid s(r).[\![M']\!]^r_u\Big)$$

The resulting processes are still in the relation. We now look at the case where $m = 1$. Here both processes can take an input on some channel $s$:

$$!z(s).s(u).s(r).[\![M']\!]^r_u \mid s(r).[\![M']\!]^r_u \xrightarrow{s(r)} !z(s).s(u).s(r).[\![M']\!]^r_u \mid [\![M']\!]^r_u$$

and

$$(\nu y)\Big((!z(s).\overline{y}s) \mid !y(s).s(u).s(r).[\![M']\!]^r_u \mid s(r).[\![M']\!]^r_u\Big) \xrightarrow{s(r)} (\nu y)\Big((!z(s).\overline{y}s) \mid !y(s).s(u).s(r).[\![M']\!]^r_u \mid [\![M']\!]^r_u\Big)$$

These are also still in the relation. But when we look at the case where $n = 1$, if $[\![M']\!]^r_u$ is able to do a transition, we will end up outside of $R$. Therefore, we need a greater relation, called $R_c$, which is the closure of behaviour of $[\![M']\!]^r_u$. What this means is that for each pair where both processes contain a $[\![M']\!]^r_u$, we also add the pair, that contains the processes where that $[\![M']\!]^r_u$ has been substituted with $M''$ and $[\![M']\!]^r_u \Rightarrow M''$ or $[\![M']\!]^r_u \xrightarrow{\mu} \Rightarrow M''$ However, since we know that $y \notin \operatorname{fn}(M')$, we also know that both processes can match the behaviour of the other, and stay in the relation. It is also easily seen, that for all $w, l, m, n > 1$, they will stay inside the relation. Since $R_c$ satisfies the requirements for being a early weak bisimulation, and we have that:

$$\Big((\nu z)\overline{u}z.!z(s).s(u).s(r).[\![M']\!]^r_u, (\nu y)\big((\nu z)(\overline{u}z.!z(s).\overline{y}s) \mid !y(s).s(u).s(r).[\![M']\!]^r_u\big)\Big) \in R_c$$

then it must be the case that:

$$(\nu z)\overline{u}z.!z(s).s(u).s(r).[\![M']\!]^r_u \approx_E (\nu y)\big((\nu z)(\overline{u}z.!z(s).\overline{y}s) \mid !y(s).s(u).s(r).[\![M']\!]^r_u\big)$$

With this, we have proven that for all $M$ the lemma holds.

$\square$

## A.5) Proof of sound and complete encoding

*Proof of Theorem 1.*

We will split the proof into two. First, we will prove soundness (1). We will also take note of the amount of reductions in the $\pi$-calculus it is necessary to do, to match a reduction in CBPV:

(1)   if $M \to N$ then $\exists P. [\![M]\!]_u^r \Rightarrow P$ and $P \approx_E [\![N]\!]_u^r$.

We will prove this using induction on the reduction $M \to N$, and up-to-bisimulation.

**(Force-thunk)**: If the reduction was derived using the (force-thunk) rule, which is:

$$(\text{force-thunk}) \; \frac{}{\textbf{force } (\textbf{thunk } M) \to M}$$

We then need to show that $[\![\textbf{force } (\textbf{thunk } M)]\!]_u^r \Rightarrow P$ and $P \approx_E [\![M]\!]_u^r$.

We begin by expanding the encodings:

$$[\![\textbf{force } (\textbf{thunk } M)]\!]_u^r = (\nu p)((\nu y)\overline{p}y.!y(s).s(u).s(r).[\![M]\!]_u^r \mid p(y).(\nu s)\overline{y}s.\overline{s}u.\overline{s}r)$$

We can reduce this using the following derivation:

$$(\nu p)((\nu y)\overline{p}y.!y(s).s(u).s(r).[\![M]\!]_u^r \mid p(y).(\nu s)\overline{y}s.\overline{s}u.\overline{s}r)$$
$$\xrightarrow{\tau} (\nu p)(\nu y)(!y(s).s(u).s(r).[\![M]\!]_u^r \mid (\nu s)\overline{y}s.\overline{s}u.\overline{s}r)$$
$$\xrightarrow{\tau} (\nu p)(\nu y)(\nu s)(!y(s).s(u).s(r).[\![M]\!]_u^r \mid s(u).s(r).[\![M]\!]_u^r \mid \overline{s}u.\overline{s}r)$$
$$\xrightarrow{\tau} (\nu p)(\nu y)(\nu s)(!y(s).s(u).s(r).[\![M]\!]_u^r \mid s(r).[\![M]\!]_u^r \mid \overline{s}r)$$
$$\xrightarrow{\tau} (\nu p)(\nu y)(\nu s)(!y(s).s(u).s(r).[\![M]\!]_u^r \mid [\![M]\!]_u^r)$$

Since there is a restriction on $y$, and the replication has an input prefix on the channel $y$, a communication can not happen with the replication, since $p$ and $y$ are not present in $[\![M]\!]_u^r$. We therefore have:

$$(\nu p)(\nu y)(\nu s)(!y(s).s(u).s(r).[\![M]\!]_u^r \mid [\![M]\!]_u^r) \approx_E ((\nu p)(\nu y)(\nu s)!y(s).s(u).s(r).[\![M]\!]_u^r) \mid [\![M]\!]_u^r$$
$$\approx_E [\![M]\!]_u^r$$

Therefore, we can see that (1) holds for (force-thunk). We furthermore saw that we used 4 reductions in the $\pi$-calculus to match the reduction in CBPV. This is because the place where a value is substituted in would be instead of a variable, which of course when encoded would have been a link. So in the case of the value being a thunk, we once again remove a link.

**(Application)**: We now look at the (Application-base)-rule.

$$(\text{Application-base}) \; \frac{}{(\lambda x.M) \; V \to M\{^V/_x\}}$$

For (1) to hold, it must be the case that $[\![(\lambda x.M) \; V]\!]_u^r$ can reduce to a process that is early bisimilar with $[\![M\{^V/_x\}]\!]_u^r$.

We begin with the encoding:

$$[\![(\lambda x.M) \; V]\!]_u^r = (\nu p)(\nu q)(p(s).s(u).s(r).s(x).[\![M]\!]_u^r \mid (\nu s)\overline{p}s.\overline{s}u.\overline{s}r.((\nu q')\overline{s}q'.[\![q' := V]\!]))$$

We will now reduce this:

$$(\nu p)(\nu q)(p(s).s(u).s(r).s(x).[\![M]\!]_u^r \mid (\nu s)\overline{p}s.\overline{s}u.\overline{s}r.((\nu q')\overline{s}q'.[\![q' := V]\!]))$$

$$\xrightarrow{\tau} (\nu p)(\nu q)(\nu s)(s(u).s(r).s(x).[\![M]\!]_u^r \mid \overline{s}u.\overline{s}r.((\nu q')\overline{s}q'.[\![q' := V]\!]))$$

$$\xrightarrow{\tau} (\nu p)(\nu q)(\nu s)(s(r).s(x).[\![M]\!]_u^r \mid \overline{s}r.((\nu q')\overline{s}q'.[\![q' := V]\!]))$$

$$\xrightarrow{\tau} (\nu p)(\nu q)(\nu s)(s(x).[\![M]\!]_u^r \mid ((\nu q')\overline{s}q'.[\![q' := V]\!]))$$

$$\xrightarrow{\tau} (\nu p)(\nu q)(\nu s)(\nu q')\left([\![M]\!]_u^r\{q'/_x\} \mid [\![q' := V]\!]\right)$$

We will now show that:

$$(\nu p)(\nu q)(\nu q')(\nu s)\left([\![M]\!]_u^r\{q'/_x\} \mid [\![q := V]\!]\right) \approx_E [\![M\{V/_x\}]\!]_u^r$$

First of all, we can ignore the restriction $(\nu p)$, $(\nu q)$ and $(\nu s)$ as there is not any instances of $p, q$ or $s$ anymore, and $(\nu q')\left([\![M]\!]_u^r\{q'/_x\} \mid [\![q' := V]\!]\right) \approx_E [\![M\{V/_x\}]\!]_u^r$ can be concluded using Lemma 5. With this, we showed we can match the reduction in CBPV with 4 reductions in the $\pi$-calculus. We also notice, that because we use Lemma 5, we "remove" a link, that might increase the number of reductions to match future reductions by 1, since if $V = y$, we have that $[\![q' = y]\!] = !q'(a).\overline{y}a$ because if:

$$[\![M]\!]_u^r\{y/_x\} \xrightarrow{y(a)}$$

Then without using Lemma 5 we would have:

$$(\nu q')\left([\![M]\!]_u^r\{q'/_x\} \mid !q'(a).\overline{y}a\right) \xrightarrow{\tau} (\nu q')\left([\![M]\!]_u^r\{q'/_x\} \mid !q'(a).\overline{y}a \mid \overline{y}a\right) \xrightarrow{y(a)}$$

Here we see the extra $\tau$-reduction needed for future reductions. In case $V = \mathbf{thunk}\ N$, we would also at most add 1 extra reduction, as then when we for example had $M = \mathbf{force}\ x$, then

$$[\![\mathbf{force}\ x\{\mathbf{thunk}\ N/_x\}]\!]_u^r = [\![\mathbf{force}\ \mathbf{thunk}\ N]\!]_u^r$$

Which we have previously shown the reduction of. On the other hand, if we do not apply Lemma 5, we have:

$$(\nu q')\left([\![\mathbf{force}\ q']\!]_u^r \mid !q'(s).s(u).s(r).[\![N]\!]_u^r\right)$$

$$= (\nu q')\left((\nu a)(\overline{a}z.!z(b).\overline{q'}b \mid a(z).(\nu s)\overline{z}s.\overline{s}u.\overline{s}r) \mid !q'(s).s(u).s(r).[\![N]\!]_u^r\right)$$

$$\xrightarrow{\tau} (\nu q')\left((\nu a)(!z(b).\overline{q'}b \mid (\nu s)\overline{z}s.\overline{s}u.\overline{s}r) \mid !q'(s).s(u).s(r).[\![N]\!]_u^r\right)$$

$$\xrightarrow{\tau} (\nu q')\left((\nu a)(\nu s)(!z(b).\overline{q'}b \mid \overline{q'}s \mid \overline{s}u.\overline{s}r) \mid !q'(s).s(u).s(r).[\![N]\!]_u^r\right)$$

$$\xrightarrow{\tau} (\nu q')(\nu s)\left((\nu a)(!z(b).\overline{q'}b \mid \overline{s}u.\overline{s}r) \mid !q'(s).s(u).s(r).[\![N]\!]_u^r \mid s(u).s(r).[\![N]\!]_u^r\right)$$

$$\xrightarrow{\tau} (\nu q')(\nu s)\left((\nu a)(!z(b).\overline{q'}b \mid \overline{s}r) \mid !q'(s).s(u).s(r).[\![N]\!]_u^r \mid s(r).[\![N]\!]_u^r\right)$$

$$\xrightarrow{\tau} (\nu q')(\nu s)\left((\nu a)(!z(b).\overline{q'}b) \mid !q'(s).s(u).s(r).[\![N]\!]_u^r \mid [\![N]\!]_u^r\right)$$

We see that this reduction now needed 5 $\tau$-reductions, instead of 4, once again showing that we increment future reductions with 1, when we use Lemma 5.

**(Binding-base)**: We now look at the case where the reduction is concluded using the rule (Binding-base)

$$\text{(Binding-base)} \ \frac{}{(\mathbf{return}\ V) \gg= \lambda x.N \rightarrow N\{V/_x\}}$$

We need to show that $[\![(\mathbf{return}\ V) \ggeq \lambda x.N]\!]_u^r$ can reduce to a process that is early bisimilar with $[\![N\{^V/_x\}]\!]_u^r$.

First we will encode $(\mathbf{return}\ V) \ggeq \lambda x.N$:

$$[\![(\mathbf{return}\ V) \ggeq \lambda x.N]\!]_u^r = (\nu z)((\nu u)(\nu y)\overline{z}y.[\![y := V]\!] \mid z(x).[\![N]\!]_u^r)$$

We will now reduce this:

$$(\nu z)((\nu u)(\nu y)\overline{z}y.[\![y := V]\!] \mid z(x).[\![N]\!]_u^r) \xrightarrow{\tau} (\nu z)(\nu y)((\nu u)[\![y := V]\!] \mid [\![N]\!]_u^r\{^y/_x\})$$

With this, we can use Lemma 5 to show that:

$$[\![N\{^V/_x\}]\!]_u^r \approx_E (\nu y)([\![y := V]\!] \mid [\![N]\!]_u^r\{^y/_x\})$$

We can remove the restriction $(\nu u)$ since $u$ does not appear in $[\![y := V]\!]$. Therefore, (1) must hold for (Binding-base). We also see we only use one $\tau$-reduction, and introduce 1 link.

**(Application-evolve)**:

We now look at the case when the reduction is concluded using the (Application-evolve)

$$\text{(Application-evolve)}\ \frac{M \to M'}{M\ V \to M'\ V}$$

We need to show that $[\![M\ V]\!]_u^r$ can reduce to a process that is early bisimilar to $[\![M'\ V]\!]_u^r$ given that we know that $[\![M]\!]_p^r \Rightarrow [\![M']\!]_p^r$ (or a process that is early bisimilar to $[\![M']\!]_p^r$).

We will begin with the encoding:

$$[\![M\ V]\!]_u^r = (\nu p)\big([\![M]\!]_p^r \mid (\nu s)\overline{p}s.\overline{s}u.\overline{s}r.[\![V]\!]_p^r\big)$$

We will also show the encoding of $M'\ V$

$$[\![M'\ V]\!]_u^r = (\nu p)\big([\![M']\!]_p^r \mid (\nu s)\overline{p}s.\overline{s}u.\overline{s}r.[\![V]\!]_p^r\big)$$

We then derive the following reduction tree. We remind the reader that since $\Rightarrow$ is the transitive and reflexive closure of $\xrightarrow{\tau}$, the $\xrightarrow{\tau}$-rules of the labeled semantics will also apply to the $\Rightarrow$ relation:

$$\text{(Par)}\ \frac{[\![M]\!]_p^q \Rightarrow [\![M']\!]_p^q}{[\![M]\!]_p^q \mid (\nu s)\overline{p}s.\overline{s}u.\overline{s}r.[\![V]\!]_s^r) \Rightarrow [\![M']\!]_p^r \mid (\nu s)\overline{p}s.\overline{s}u.\overline{s}r.[\![V]\!]_s^r)}$$

$$\text{(Res)}\ \frac{}{(\nu q)([\![M]\!]_p^q \mid (\nu s)\overline{p}s.\overline{s}u.\overline{s}r.[\![V]\!]_s^r) \Rightarrow (\nu q)\big([\![M']\!]_p^q \mid (\nu s)\overline{p}s.\overline{s}u.\overline{s}r.[\![V]\!]_s^r\big)}$$

$$\text{(Res)}\ \frac{}{(\nu p)(\nu q)([\![M]\!]_p^q \mid (\nu s)\overline{p}s.\overline{s}u.\overline{s}r.[\![V]\!]_s^r) \Rightarrow (\nu p)(\nu q)\big([\![M']\!]_p^q \mid (\nu s)\overline{p}s.\overline{s}u.\overline{s}r.[\![V]\!]_s^r\big)}$$

Since this is exactly the encoding of $(M'\ V)$, the theorem holds for (Application-evolve). With regards to the amount of reductions needed in the $\pi$-calculus to match a reduction in CBPV, we see that reduction uses the same amount of reductions as $[\![M]\!]_p^r \Rightarrow [\![M']\!]_p^r$.

**(Binding-evolve)**: We now look at the case when the reduction is concluded using (Binding-evolve)

$$\text{(Binding-evolve)} \ \frac{M \to M'}{M \ggeq \lambda x.N \to M' \ggeq \lambda x.N}$$

We need to show that $[\![M \ggeq \lambda x.N]\!]_u^r$ can reduce to a process that is early bisimilar to $[\![M' \ggeq \lambda x.N]\!]_u^r$ given that we know that $[\![M]\!]_u^p \Rightarrow [\![M']\!]_u^p$ (or a process that is early bisimilar to $[\![M']\!]_u^p$)

We will begin with the encoding:

$$[\![M \ggeq \lambda x.N]\!]_u^r = (\nu p)((\nu u)[\![M]\!]_u^p \mid p(x).[\![N]\!]_u^r)$$

We will also show the encoding of $M' \ggeq \lambda x.N$:

$$[\![M' \ggeq \lambda x.N]\!]_u^r = (\nu p)\Big((\nu u)[\![M']\!]_u^p \mid p(x).[\![N]\!]_u^r\Big)$$

We will now reduce $[\![M \ggeq \lambda x.N]\!]_u^r$ :

$$\text{(Res)} \ \frac{\text{(Par)} \ \dfrac{[\![M]\!]_u^p \Rightarrow [\![M']\!]_u^p}{(\nu u)[\![M]\!]_u^p \mid p(x).[\![N]\!]_u^r \Rightarrow (\nu u)[\![M']\!]_u^p \mid p(x).[\![N]\!]_u^r}}{(\nu p)((\nu u)[\![M]\!]_u^p \mid p(x).[\![N]\!]_u^r) \Rightarrow (\nu p)\Big((\nu u)[\![M']\!]_u^p \mid p(x).[\![N]\!]_u^r\Big)}$$

With this derivation we can conclude that (1) also holds for the rule (Binding-evolve). We can also say that this reduction uses no more reductions in the $\pi$-calculus than the reduction $[\![M]\!]_p^r \Rightarrow [\![M']\!]_p^r$. With this, we have proven that (1) holds for all semantic rules.

We will now prove that (2) holds for all possible encodings:

(2) $\forall P$. If $[\![M]\!]_u^r \Rightarrow P$, then $\exists N, P'.P \Rightarrow P'$ and $P' \approx_E [\![N]\!]_u^r$, and $M \mapsto N$. Furthermore, if $P \neq P'$ then $\forall a, b.P \overset{\overline{a}b}{\nrightarrow}$ and $P \overset{\overline{a}(b)}{\nrightarrow}$ and $P \overset{a(b)}{\nrightarrow}$.

We will use structural induction on the term $M$. This means that we assume that (2) will hold for sub-expressions $M', M'', N'$ and $V'$, and then look at each case separately. We will only show, until we reach a process that is early weak bisimilar with the encoding of a process. From induction, we can then conclude that the following reductions will also uphold (2), meaning we show that we assume (2) holds for $N$, if $M \to N$.

**Abstraction**:

When $M = \lambda x.M'$, we have that:

$$[\![\lambda x.M']\!]_u^r = u(s).s(u).s(r).s(x).[\![M']\!]_u^r$$

The abstraction encoding cannot reduce, as it is behind input prefix. This means, that the only possible $P$ is $[\![\lambda x.M]\!]_u^r$. We can choose $P' = P$, and $N = M$, and then (2) holds. Therefore (2) holds when $M$ is an abstraction. We can use this argument whenever an encoded term is behind prefix.

**Application**:

We look at the encoding of an application, where

$$M = M' \ V'$$

We will be using structural induction on $M$, and our inductive hypothesis will be the following: If $\llbracket M' \rrbracket_a^b \xrightarrow{\tau} Q$, then $Q \Rightarrow Q'$ and $Q' \approx_E \llbracket N \rrbracket_a^b$, without being able to communicate with its environment until it reaches Q. We also use induction on $\Rightarrow$, which means we will assume that forall Q, where $\llbracket M\, V \rrbracket_u^r \Rightarrow Q$ and $Q \approx_E \llbracket N' \rrbracket_u^r$, then if $N' \neq M$, (2) holds for $N'$. This essentially mean we will prove (2) until we reach a process that is early weak bisimilar with a new term.

We have that the encoding of this will be:

$$\llbracket M'\, V' \rrbracket_u^r = (\nu a)(\nu b)\big(\llbracket M' \rrbracket_a^b \mid (\nu s)\overline{a}s.\overline{s}u.\overline{s}r.\llbracket V' \rrbracket_u^r\big)$$
$$= (\nu a)(\nu b)\big(\llbracket M' \rrbracket_a^b \mid (\nu s)\overline{a}s.\overline{s}u.\overline{s}r.(\nu y).\overline{s}y.\llbracket y := V' \rrbracket\big)$$

For this to reduce, there are two cases; either $\llbracket M' \rrbracket_a^b \xrightarrow{\tau}$, or $\llbracket M' \rrbracket_a^b$ communicates with the output $\overline{a}s$. In case of $\llbracket M' \rrbracket_a^b \xrightarrow{\tau}$, where we have internal communication, we have from the inductive hypothesis know, that it will further reduce to a process that is early weak bisimilar to the encoding of a CBPV expression, without having any outgoing outputs or inputs, until it reaches a process weak bisimilar with an encoding. Therefore, without loss of generality, we can ignore the internal communications in $\llbracket M' \rrbracket_a^b$, as it will eventually end up as a process that is early weak bisimilar to an encoding of a CBPV expression again.

In the other case, if it is able to reduce, it must be the case that $\llbracket M' \rrbracket_a^b \xrightarrow{a(s)} Q$. By inspection of the encoding rules, for this to be the case, $M'$ must be of the form $\lambda x.M''$, as in all other encoding rules, either their sub-expressions are restricted with regards to the argument handle, or behind output prefixes. We therefore have:

$$(\nu a)(\nu b)\big(\llbracket \lambda x.M'' \rrbracket_a^b \mid (\nu s)\overline{a}s.\overline{s}u.\overline{s}r.(\nu y).\overline{s}y.\llbracket y := V' \rrbracket\big)$$
$$= (\nu a)(\nu b)\big(a(s).s(u).s(r).s(x).\llbracket M'' \rrbracket_u^r \mid (\nu s)\overline{a}s.\overline{s}u.\overline{s}r.(\nu y).\overline{s}y.\llbracket y := V' \rrbracket\big)$$
$$\rightarrow (\nu a)(\nu b)(\nu s)\big(s(u).s(r).s(x).\llbracket M'' \rrbracket_u^r \mid \overline{s}u.\overline{s}r.(\nu y).\overline{s}y.\llbracket y := V' \rrbracket\big)$$

Now we need to show that there exists a $P'$ and a $N$, where we can reduce to $P'$, and $P' \approx_E \llbracket N \rrbracket_u^r$, and we have that $M \rightarrow N$. We begin by further reducing, until we reach a suitable process. We notice that in all intermediate processes, the process as a whole has no barbs, meaning that it can only do internal communications, and not output or input actions. This is the case since all communication happens on the channel $s$, that is restricted:

$$(\nu a)(\nu b)(\nu s)\big(s(u).s(r).s(x).\llbracket M' \rrbracket_u^r \mid \overline{s}u.\overline{s}r.(\nu y).\overline{s}y.\llbracket y := V' \rrbracket\big)$$
$$\Rightarrow (\nu a)(\nu b)(\nu s)(\nu y)\big(\llbracket M' \rrbracket_u^r\{^y/_x\} \mid \llbracket y := V' \rrbracket\big)$$

This will be our $P'$. We then see that if $N = M'\big\{^{V'}/_x\big\}$, we have that $(\lambda x.M')V' \rightarrow M'\{^{V'}/_x\}$, and we can then show that $P' \approx_E \llbracket M'\big\{^{V'}/_x\big\} \rrbracket_u^r$. We note that for each intermediate process until here, we can choose the same $P'$ and $N$, and they will not have had any outgoing behaviour, as the channels for communication have been restricted until now. From Lemma 1 and our bisimilarity laws we have the following:

$$(\nu a)(\nu b)(\nu s)(\nu y)\big(\llbracket M' \rrbracket_u^r\{^y/_x\} \mid \llbracket y := V' \rrbracket\big) \approx_E (\nu y)\big(\llbracket M' \rrbracket_u^r\{^y/_x\} \mid \llbracket y := V' \rrbracket\big)$$
$$\approx_E \llbracket M'\big\{^{V'}/_x\big\} \rrbracket_u^r$$

We have reached a process that is early weak bisimilar to an encoding of a $\lambda$-expression, so by the inductive hypothesis, this will hold for future reductions. With this, we can say that (2) holds when $M$ is an application. We of course also noticed that this is the same reduction as the reduction in the proof of (1) for application.

**Force**:

If $M$ is a force, the encoding will look in the following way:

$$\llbracket \textbf{force } V \rrbracket_u^r = (\nu p)\big(\llbracket V \rrbracket_p^r \mid p(y).(\nu s)\overline{y}s.\overline{s}u.\overline{y}r\big)$$
$$= (\nu p)((\nu y)\overline{p}y.\llbracket y := V \rrbracket \mid p(y).(\nu s)\overline{y}s.\overline{s}u.\overline{y}r)$$

We now look at each case of $V$. First we look at the case that $V$ is a thunk. We then have:

$$(\nu p((\nu y)\overline{p}y.\llbracket y := \textbf{thunk } M' \rrbracket \mid p(y).(\nu s)\overline{y}s.\overline{s}u.\overline{y}r) = (\nu p\big((\nu y)\overline{p}y.!y(s).s(u).s(r).\llbracket M' \rrbracket_u^r \mid p(y).(\nu s)\overline{y}s.\overline{s}u.\overline{y}r\big)$$
$$\rightarrow (\nu y)\big(!y(s).s(u).s(r).\llbracket M' \rrbracket_u^r \mid (\nu s)\overline{y}s.\overline{s}u.\overline{y}r\big)$$

Since we have a reduction, we need to find a $P'$. We reduce further:

$$(\nu y)\big(!y(s).s(u).s(r).\llbracket M' \rrbracket_u^r \mid (\nu s)\overline{y}s.\overline{s}u.\overline{y}r\big) \Rightarrow (\nu y)(\nu s)\big(!y(s).s(u).s(r).\llbracket M' \rrbracket_u^r \mid \llbracket M' \rrbracket_u^r\big)$$

This will be our $P'$. We then choose our $N$ to be our $M'$, since we know that $(\textbf{force } (\textbf{thunk } M')) \rightarrow M'$ . We then need to show that $P' \approx_E \llbracket M' \rrbracket_u^r$. This is easy, since $y$ is not present in $\llbracket M' \rrbracket_u^r$, so the replication can be garbage collected:

$$(\nu y)(\nu s)\big(!y(s).s(u).s(r).\llbracket M' \rrbracket_u^r \mid \llbracket M' \rrbracket_u^r\big) \approx_E \llbracket M' \rrbracket_u^r$$

Therefore, (2) holds when $M$ is a force, and the value is a thunk. We note that it here is possible to unfold the replication to gain an infinite amount of processes in paralel, however, because they need to take an input on y, it amounts to an infinite amount of 0. In general, we will ignore processes that could be reached by unfolding replications multiple times, as they will always be restricted.

We now look at the case where $V$ is a variable $z$. The encoding is then:

$$(\nu p)((\nu y)\overline{p}y.\llbracket y := z \rrbracket \mid p(y).(\nu s)\overline{y}s.\overline{s}u.\overline{y}r) = (\nu p)((\nu y)\overline{p}y.!y(w).\overline{z}w \mid p(y).(\nu s)\overline{y}s.\overline{s}u.\overline{s}r)$$
$$\rightarrow (\nu p)(\nu y)(!y(w).\overline{z}w \mid (\nu s)\overline{y}s.\overline{s}u.\overline{s}r)$$

Since we have a reduction, we need to find a $P'$. We call this current intermediate process $P^*$. We reduce this further:

$$(\nu p)(\nu y)(!y(w).\overline{z}w \mid (\nu s)\overline{y}s.\overline{s}u.\overline{s}r) \rightarrow (\nu p)(\nu y)(\nu s)(!y(w).\overline{z}w \mid \overline{z}w \mid \overline{s}u.\overline{s}r)$$

This will be our $P'$.

We create a early weak bisimilarity relation $R$:

$$R' = \{(\llbracket \textbf{force } z \rrbracket_u^r, P'), (\llbracket \textbf{force } z \rrbracket_u^r, P^*), (P', P'), (P^*, P^*)\} \cup \{(P, P) \mid \forall P\}$$

We say that $R$ is the symmetric closure of $R'$. We need to show that $R$ is a bisimulation relation. We easily see that whichever process you choose to do a $\tau$-reduction with, which is the only possible action, the other process can either do a $\tau$-reduction to become exactly the same process, or just stay the same process, as the other one will have reduced to the exact same process. This is enough to show that $R'$ is a bisimulation relation.

We note here, that the process $\llbracket \mathbf{force}\ x \rrbracket_u^r$ is able to do two $\tau$-reductions, before not being able to do an internal reduction, even though the term $\mathbf{force}\ x$ is terminal in CBPV, and cannot reduce.

**Return**:

In the case that $M = \mathbf{return}\ V$ we have that:

$$\llbracket \mathbf{return}\ V \rrbracket_u^r = (\nu y)\bar{r}y.\llbracket y := V \rrbracket$$

The return encoding can not reduce, as it is behind output prefix. Therefore (2) holds when $M$ is a return.

**Binding**:

In the case that $M$ is a binding, we have the following encoding:

$$\llbracket M \ggg \lambda x.N \rrbracket_u^r = (\nu z)(\nu u)\Big(\llbracket M' \rrbracket_u^z \mid z(x).\llbracket N' \rrbracket_u^r\Big)$$

For this to be able to reduce, either $\llbracket M' \rrbracket_u^z \overset{\tau}{\rightarrow}$, or it communicates with the input on the channel $z$. If it is the case it does an internal reduction, then from the inductive hypothesis, we know it will eventually reach a process that is early weak bisimilar with an encoded term, without being able to interact with the input $z(x)$ until it has. In the other case, $\llbracket M' \rrbracket_u^z$ must output on the channel $z$. The only case where this can happen, is if $M'$ is a return:

$$(\nu z)\Big((\nu u)\llbracket \mathbf{return}\ V' \rrbracket_u^z \mid z(x).\llbracket N' \rrbracket_u^r\Big) = (\nu z)(\nu u)\Big(\llbracket V' \rrbracket_z^z \mid z(x).\llbracket N' \rrbracket_u^r\Big)$$

$$= (\nu z)(\nu u)\Big((\nu y).\bar{z}y.\llbracket y := V \rrbracket \mid z(x).\llbracket N' \rrbracket_u^r\Big)$$

$$\rightarrow (\nu z)(\nu u)(\nu y)\Big(\llbracket y := V \rrbracket \mid \llbracket N' \rrbracket_u^r\{^y/_x\}\Big)$$

We now see that if $N = \llbracket N'\{^V/_x\} \rrbracket_u^r$ we can show that $P' \approx_E \llbracket N'\{^V/_x\} \rrbracket_u^r$ with the bisimilarity laws and Lemma 5.

$$(\nu z)(\nu u)(\nu y)\Big(\llbracket y := V \rrbracket \mid \llbracket N' \rrbracket_u^r\{^y/_x\}\Big) \approx_E (\nu y)\Big(\llbracket N' \rrbracket_u^r\{^y/_x\} \mid \llbracket y := V \rrbracket\Big) \approx_E \llbracket N' \rrbracket_u^r$$

With this, we have shown that if $M$ is a binding, (2) holds.

**Value**:

In the case of $M$ being a value, we have that:

$$\llbracket V \rrbracket_u^r = (\nu y)\bar{u}y.\llbracket y := V \rrbracket$$

A reduction can not happen as it is behind an output prefix.

We have now shown that regardless of how $M$ is constructed, (2) holds, and therefore it must be the case that (2) always holds.

Since both (1) and (2) hold in all cases, we have now proven Theorem 1.

$\square$

# B) Proof of soundness and completeness for $\pi I$-calculus

This is the proof of soundness and completeness for the following encoding:

$$\mathcal{I}[\![\lambda x.M]\!]_u^r = \overline{u}(e).e(u,r,x).\mathcal{I}[\![M]\!]_u^r \qquad\qquad e \notin \text{fn}(M)$$

$$\mathcal{I}[\![M\ V]\!]_u^r = (\nu abcd)\big(a(e).c(x).\overline{e}(u',r',x').(u' \to u \mid r' \to r \mid x' \to x)$$

$$\mid \mathcal{I}[\![M]\!]_a^b \mid \mathcal{I}[\![V]\!]_c^d\big) \qquad\qquad a,b,c,d \notin \text{fn}(M) \cup \text{fn}(V)$$

$$\mathcal{I}[\![\textbf{force}\ V]\!]_u^r = (\nu ab)\big(\mathcal{I}[\![V]\!]_a^b \mid a(y).\overline{y}(u',r').(u' \to u \mid r' \to r)\big) \qquad\qquad a,b \notin \text{fn}(V)$$

$$\mathcal{I}[\![\textbf{return}\ V]\!]_u^r = \mathcal{I}[\![V]\!]_r^r$$

$$\mathcal{I}[\![M \ggg \lambda x.N]\!]_u^r = (\nu ab)\big(\mathcal{I}[\![M]\!]_a^b \mid b(x).\mathcal{I}[\![N]\!]_u^r\big) \qquad\qquad a,b \notin \text{fn}(M) \cup \text{fn}(N)$$

$$\mathcal{I}[\![V]\!]_u^r = (\overline{u}(y).\mathcal{I}[\![y := V]\!]) \qquad\qquad y \notin \text{fn}(V)$$

$$\mathcal{I}[\![y := x]\!] = !y(u,r).\overline{x}(u',r').(u' \to u \mid r' \to r)$$

$$\mathcal{I}[\![y := \textbf{thunk}\ M]\!] = !y(u,r).\mathcal{I}[\![M]\!]_u^r$$

We assume that $u, r \notin \text{fn}(M)$, and that introduced names from the side conditions are distinct. We note that since the encoding is the polyadic encoding, the proof will also be true for the monadic encoding, as the polyadic encoding can easily be encoded to the monadic $\pi$-calculus.

We will prove the following theorem:

**Theorem 2.** (Encoding)

For some term $M$, $N$, and two distinct names $u, r \notin \text{fn}(M)$ we have:

(1.) If $M \to N$ then $\exists P.\mathcal{I}[\![M]\!]_u^r \Rightarrow P$, and $P \approx_E \mathcal{I}[\![N]\!]_u^r$.

(2.) If $\mathcal{I}[\![M]\!]_u^r \Rightarrow P$ then $\exists P', N.P \Rightarrow P'$, $P' \approx_E \mathcal{I}[\![N]\!]_u^r$ and $M \Rightarrow N$, and if $P \neq P'$ then $\forall a,b.P \overset{\overline{a}(b)}{\not\to}$ and $P \overset{a(b)}{\not\to}$

To do this, we need the following lemmas:

**Lemma 6.** (Encoding and name bisimilarity)

For some term $M$, $N$, and distinct names $u, r, y \notin \text{fn}(M)$, and a name $x$ we have:

(1). $(\nu y)(\mathcal{I}[\![y := V]\!] \mid P\{^y/_x\} \mid Q\{^y/_x\}) \approx_E$
$\qquad (\nu y)(\mathcal{I}[\![y := V]\!] \mid P\{^y/_x\}) \mid (\nu z)(Q\{^z/_x\} \mid \mathcal{I}[\![z := V]\!]))$

$\qquad$ where $z, y \notin \text{fn}(V) \cup \text{fn}(P) \cup \text{fn}(Q)$

(2). $(\nu y)(y \to x \mid \mathcal{I}[\![M]\!]_u^r\{^y/_x\}) \approx_E \mathcal{I}[\![M]\!]_u^r$

(3). $(\nu y)(\mathcal{I}[\![y := V]\!] \mid \mathcal{I}[\![M]\!]_u^r\{^y/_x\}) \approx_E \mathcal{I}[\![M\{^V/_x\}]\!]_u^r$

(4). $(\nu u'r')\big(u' \to u \mid r' \to r \mid \mathcal{I}[\![M]\!]_{u'}^{r'}\big) \approx_E \mathcal{I}[\![M]\!]_u^r$

(5). $(\nu y)(!\mathcal{I}[\![M]\!]_u^r\{^y/_x\} \mid \mathcal{I}[\![y := V]\!]) \approx_E$
$\qquad !(\nu y)(\mathcal{I}[\![M]\!]_u^r\{^y/_x\} \mid \mathcal{I}[\![y := V]\!])$

We will use a lemma proven by Sangiorgi in [11] that states:

**Lemma 7.** (Combine links) if $a, b$ and $c$ are distinct names, then:

$$(\nu b)(a \to b \mid b \to c) \approx_E a \to c$$

*Proof of Lemma 6.*

For the proofs of the lemmas, we will use the notation that $\mathcal{I}[\![y := x]\!]_u^r = y \to x$. This is a single folding of a definition of a link, that is allowed since in all cases a variable only receives 2 arguments in our encoding, and notationally it is much easier to read than $!y(u,r).\overline{x}(u',r').(u' \to u \mid r' \to r)$. We do this, even though it is not strictly correct, as in a single case $\mathcal{I}[\![V\ V']\!]$ This might lead to unintended behaviour, (we will note in the proof when this might be an issue). However, if anyone should ever be able to read this proof, we thought it was necessary to make the notation more compact. Furthermore, since we will be doing the proof in the polyadic encoding, again, for readability sake, we will note that the semantics of a link is slightly different from those presented in Definition 45. Namely, a link in the polyadic $\pi$-calculus might first be used for one arity of names, and then later for another arity of names, depending on what communicates with them. The polyadic semantic of links can be presented as:

$$\text{(Link-Polyadic)} \frac{x \to y \mid x(\mu).\overline{y}(\mu').(\mu' \to \mu) \xrightarrow{x(\mu)} P'}{x \to y \xrightarrow{x(\mu)} P'}$$

Where $\mu$ can be a list of names. We will also say that $(\nu x, y, z) = (\nu x)(\nu y)(\nu z)$. We also have the following law:

$$x \to y \approx_E x \to y \mid x(\mu).\overline{y}(\mu').(\mu' \to \mu)$$

We also use the following bisimilarity law from [25]:

$$a(b).(\nu c)(\overline{c}(x).Q \mid c(x).P) \approx_E a(b).(\nu c)(\nu x)(Q \mid P)$$

**(1)**: We begin with the proof of (1). This proof is the same as with the proof of Lemma 3, as in this case, $\mathcal{I}[\![y := V]\!]$ is also always behind prefix of replication and a input, that unlocks a version of $V$.

**(2)**: We then look at (2). We have that since $y \to x \approx_E y \to x \mid y(a).\overline{x}(a').a' \to a$, if $\mathcal{I}[\![M]\!]_u^r$ ever outputs on the channel $x$, then $(\nu y)(\mathcal{I}[\![M]\!]_u^r\{y/_x\} \mid y \to x)$ would be able to output on the channel $y$, and then a pointer would be send on the channel $x$. This essentially reduces to the base case of a variable, where we need to show the following:

$$\mathcal{I}[\![x]\!]_u^r \approx_E (\nu y)(y \to x \mid \mathcal{I}[\![x]\!]_u^r\{y/_x\})$$

We unfold the encoding, and move behind prefix:

$$(\nu y)(y \to x \mid \mathcal{I}[\![x]\!]_u^r\{y/_x\}) = (\nu y)(y \to x \mid \overline{u}(y').y' \to y)$$
$$\approx_E \overline{u}(y').(\nu y)(y \to x \mid y' \to y)$$

We use Sangiorgi's lemma:

$$\overline{u}(y').(\nu y)(y \to x \mid y' \to y) \approx_E \overline{u}(y').y' \to x$$
$$= \mathcal{I}[\![x]\!]_u^r$$

In all other encodings, it is just a question of moving the link and restriction further inside the encoding, and use (1) to split it up in the case of an application or binding (similarly to the proof of Lemma 5).

**(3)**: We now look at (3). It comes down to the base case where $M = x$, as in all other rules, we can move it further inside the encoding using the other clauses, and laws of bisimilarity, and use (1) to split it up in case of the application and binding. We then need to show that:

$$(\nu y)(\mathcal{J}[\![x]\!]_u^r\{^y/_x\} \mid \mathcal{J}[\![y := V]\!]) \approx_E \mathcal{J}[\![y]\!]_u^r$$

We have that:

$$(\nu y)(\mathcal{J}[\![x]\!]_u^r\{^y/_x\} \mid \mathcal{J}[\![y := V]\!]) = (\nu y)(\overline{u}(z).z \to x\{^y/_x\} \mid \mathcal{J}[\![y := V]\!])$$
$$= (\nu y)(\overline{u}(z).z \to y \mid \mathcal{J}[\![y := V]\!])$$

We now look at each case of $V$.

We first look at $V = a$:

$$(\nu y)(\overline{u}(z).z \to y \mid \mathcal{J}[\![y := a]\!]) = (\nu y)(\overline{u}(z).z \to y \mid y \to a)$$
$$\approx_E \overline{u}(z).(\nu y)(z \to y \mid y \to a)$$
$$\approx_E \overline{u}(z).(z \to a)$$

Since this is exactly the encoding of $a$, (3) holds when $V = a$.

We now look at when $V = \textbf{thunk } N$: We then have:

$$(\nu y)(\overline{u}(z).z \to y \mid \mathcal{J}[\![y := \textbf{thunk } N]\!]) = (\nu y)(\overline{u}(z).z \to y \mid !y(u,r).\mathcal{J}[\![N]\!]_u^r)$$
$$\approx_E (\nu y)(\overline{u}(z).(!y(u,r).\mathcal{J}[\![N]\!]_u^r \mid z \to y))$$

If we look at the encoding of $\textbf{thunk } N$, and use (2) to replace behind output:
$\mathcal{J}[\![\textbf{thunk } N]\!]_u^r = \overline{u}(z).!z(u,r).\mathcal{J}[\![N]\!]_u$
$$\approx_E \overline{u}(z).(\nu y)(!y(u,r).\mathcal{J}[\![N]\!]_u^r \mid z \to y)$$
$$\approx_E (\nu y)\overline{u}(z).(!y(u,r).\mathcal{J}[\![N]\!]_u^r \mid z \to y)$$

We see we reach the exact same process. Therefore, (3) holds for all cases.

**(4)**: We now look at (4). It states: $(\nu u'r')\big(u' \to u \mid r' \to r \mid \mathcal{J}[\![M]\!]_{u'}^{r'}\big) \approx_E \mathcal{J}[\![M]\!]_u^r$

This we prove using structural induction on $M$ and up-to-bisimulation.

**Abstraction**: We have that $M = \lambda x''.M'$

We need to show that:

$$(\nu u'r')\big(u' \to u \mid r' \to r \mid \overline{u'}(e).e(u'',r'',x'').\mathcal{J}[\![M']\!]_{u''}^{r''}\big) \approx_E \overline{u}(e).e(u'',r'',x'').\mathcal{J}[\![M']\!]_{u''}^{r''}$$

From the inductive hypothesis, we have that:

$$\mathcal{J}[\![M']\!]_{u''}^{r''} \approx_E (\nu u'r')\big(u' \to u'' \mid r' \to r'' \mid \mathcal{J}[\![M']\!]_{u'}^{r'}\big)$$

Since early weak bisimulation allows $\tau$-reductions, we will do a single reduction step:

$$(\nu u' r')\Big( u' \to u \mid r' \to r \mid \overline{u'}(e).e(u'', r'', x'').\mathcal{J}[\![M']\!]_{u''}^{r''}\Big)$$

$$\xrightarrow{\tau} (\nu u' r' e)\Big( \overline{u}(e').(e' \to e) \mid u' \to u \mid r' \to r \mid e(u'', r'', x'').\mathcal{J}[\![M']\!]_{u''}^{r''}\Big)$$

$$\approx_E \overline{u}(e').(\nu u' r' e)\Big( (e' \to e) \mid r' \to r \mid e(u'', r'', x'').\mathcal{J}[\![M']\!]_{u''}^{r''}\Big)$$

We now unfold the link once.

$$\approx_E \overline{u}(e').(\nu u' r' e)\Big( (e'(u''', r''', x''').\overline{e}(u'', r'', x'').(u'' \to u''' \mid r'' \to r''' \mid x'' \to x'''))$$

$$\mid r' \to r \mid e(u'', r'', x'').\mathcal{J}[\![M']\!]_{u''}^{r''} \mid e' \to e\Big)$$

We can then move the input outside, because of everything else being restricted.

$$\approx_E \overline{u}(e').e'(u''', r''', x''').(\nu u', r', e)\Big( (\overline{e}(u'', r'', x'').(u'' \to u''' \mid r'' \to r''' \mid x'' \to x'''))$$

$$\mid r' \to r \mid e(u'', r'', x'').\mathcal{J}[\![M']\!]_{u''}^{r''} \mid e' \to e\Big)$$

$$\approx_E \overline{u}(e').e'(u''', r''', x''').(\nu u', r', e)(\nu u'', r'', x'')\Big( (u'' \to u''' \mid r'' \to r''' \mid x'' \to x''')$$

$$\mid r' \to r \mid \mathcal{J}[\![M']\!]_{u''}^{r''} \mid e' \to e\Big)$$

We can remove $r' \to r$, as neither $r$ nor $r'$ are anywhere, and $r'$ is restricted.

$$\approx_E \overline{u}(e').e'(u''', r''', x''').(\nu u', e)(\nu u'', r'', x'')\Big( (u'' \to u''' \mid r'' \to r''' \mid x'' \to x''') \mid \mathcal{J}[\![M']\!]_{u''}^{r''} \mid e' \to e\Big)$$

$$\approx_E \overline{u}(e').e'(u''', r''', x''').(\nu u' r' e)(\nu u'', r'', x'')\Big( (u'' \to u''' \mid r'' \to r''') \mid \mathcal{J}[\![M']\!]_{u''}^{r''} \mid e' \to e \mid x'' \to x'''\Big)$$

We use the inductive hypothesis:

$$\overline{u}(e').e'(u''', r''', x''').(\nu u' r' e)(\nu u'', r'', x'')\Big( (u'' \to u''' \mid r'' \to r''') \mid \mathcal{J}[\![M']\!]_{u''}^{r''} \mid x'' \to x'''\Big)$$

$$\approx_E \overline{u}(e').e'(u''', r''', x''').(\nu u' r' e)(\nu x'')\Big( (\nu u'', r'')\big( (u'' \to u''' \mid r'' \to r''') \mid \mathcal{J}[\![M']\!]_{u''}^{r''}\big) \mid x'' \to x'''\Big)$$

$$\approx_E \overline{u}(e').e'(u''', r''', x''').(\nu u' r' e)(\nu x'')\Big( \mathcal{J}[\![M']\!]_{u'''}^{r'''} \mid x'' \to x'''\Big)$$

$$\approx_E \overline{u}(e').e'(u''', r''', x''').(\nu x'')\Big( \mathcal{J}[\![M']\!]_{u'''}^{r'''} \mid x'' \to x'''\Big)$$

We can now use (2) to prove that

$$\overline{u}(e').e'(u''', r''', x''').(\nu x'')\Big( \mathcal{J}[\![M']\!]_{u'''}^{r'''} \mid x'' \to x'''\Big) \approx_E \overline{u}(e').e'(u''', r''', x''').\Big( \mathcal{J}[\![M']\!]_{u'''}^{r'''}\big\{ x'''/x'' \big\}\Big)$$

$$\approx_E \overline{u}(e').e'(u''', r''', x''').\mathcal{J}[\![M']\!]_{u'''}^{r'''}$$

$$= \mathcal{J}[\![\lambda x''.M']\!]_{u}^{r}$$

With this, we have now shown that (4) holds when $M$ is an abstraction.

**Application**: We need to show that:

$(\nu u'r')\big(u' \to u \mid r' \to r \mid \big((\nu abcd)\big(a(e).c(x).\overline{e}(u'',r'',x').(u'' \to u' \mid r'' \to r' \mid x' \to x)$

$\mid \mathcal{J}[\![M]\!]_a^b \mid \mathcal{J}[\![V]\!]_c^d\big)\big)\big)$

$\approx_E (\nu abcd)\big(a(e).c(x).\overline{e}(u'',r'',x').(u'' \to u \mid r'' \to r \mid x' \to x) \mid \mathcal{J}[\![M]\!]_a^b \mid \mathcal{J}[\![V]\!]_c^d\big)$

We can move the links and restriction down under the prefix, due to the fact that $u'$ and $r'$ are not present in $\mathcal{J}[\![M]\!]_a^b$ and $\mathcal{J}[\![M]\!]_c^d$.

$(\nu u'r')\big(u' \to u \mid r' \to r \mid \big((\nu abcd)\big(a(e).c(x).\overline{e}(u'',r'',x').(u'' \to u' \mid r'' \to r' \mid x' \to x) \mid \mathcal{J}[\![M]\!]_a^b \mid \mathcal{J}[\![V]\!]_c^d\big)\big)\big)$

$\approx_E (\nu abcd)\big(a(e).c(x).\overline{e}(u'',r'',x').(\nu u'r')(u' \to u \mid r' \to r \mid u'' \to u' \mid r'' \to r' \mid x' \to x) \mid \mathcal{J}[\![M]\!]_a^b \mid \mathcal{J}[\![V]\!]_c^d\big)$

We then use Sangiorgi's Lemma 7, that we can combine the links:

$(\nu abcd)\big(a(e).c(x).\overline{e}(u'',r'',x').(\nu u'r')(u' \to u \mid r' \to r \mid u'' \to u' \mid r'' \to r' \mid x' \to x) \mid \mathcal{J}[\![M]\!]_a^b \mid \mathcal{J}[\![V]\!]_c^d\big)$

$\approx_E (\nu abcd)\big(a(e).c(x).\overline{e}(u'',r'',x').(u'' \to u \mid r'' \to r \mid x' \to x) \mid \mathcal{J}[\![M]\!]_a^b \mid \mathcal{J}[\![V]\!]_c^d\big)$

With this, we have shown (4) holds for application.

**Force**: We need to show that:

$(\nu u'r')\big(u' \to u \mid r' \to r \mid \big((\nu ab)\big(\mathcal{J}[\![V]\!]_a^b \mid a(y).\overline{y}(u'',r'').(u'' \to u' \mid r'' \to r')\big)\big)\big)$

$\approx_E (\nu ab)\big(\mathcal{J}[\![V]\!]_a^b \mid a(y).\overline{y}(u'',r'').(u'' \to u \mid r'' \to r)\big)$

We can again move these behind prefix, as they are not present in $\mathcal{J}[\![V]\!]_a^b$.

$(\nu u'r')\big(u' \to u \mid r' \to r \mid \big((\nu ab)\big(\mathcal{J}[\![V]\!]_a^b \mid a(y).\overline{y}(u'',r'').(u'' \to u' \mid r'' \to r')\big)\big)\big)$

$\approx_E (\nu ab)\big(\mathcal{J}[\![V]\!]_a^b \mid a(y).\overline{y}(u'',r'').(\nu u'r')(u' \to u \mid r' \to r \mid u'' \to u' \mid r'' \to r')\big)$

We can use Lemma 7 to combine the links:

$(\nu ab)\big(\mathcal{J}[\![V]\!]_a^b \mid a(y).\overline{y}(u'',r'').(\nu u'r')(u' \to u \mid r' \to r \mid u'' \to u' \mid r'' \to r')\big)$

$\approx_E (\nu ab)\big(\mathcal{J}[\![V]\!]_a^b \mid a(y).\overline{y}(u'',r'').(u'' \to u \mid r'' \to r)\big)$

With this, we have shown that (4) holds for **force**.

**Binding**: We need to show that:

$(\nu u'r')\big(u' \to u \mid r' \to r \mid (\nu ab)\big(\mathcal{J}[\![M]\!]_a^b \mid b(x).\mathcal{J}[\![N]\!]_{u'}^{r'}\big)\big)$

$\approx_E (\nu ab)\big(\mathcal{J}[\![M]\!]_a^b \mid b(x).\mathcal{J}[\![N]\!]_u^r\big)$

We move everything behind the input on $b$. We can do this, since $u'$ and $r'$ are not present in $\mathcal{J}[\![M]\!]_a^b$, and we can with alpha conversion make sure they are different from a,b and x.

$(\nu u'r')\big(u' \to u \mid r' \to r \mid (\nu ab)\big(\mathcal{J}[\![M]\!]_a^b \mid b(x).\mathcal{J}[\![N]\!]_{u'}^{r'}\big)\big)$

$\approx_E (\nu ab)\big(\mathcal{J}[\![M]\!]_a^b \mid b(x).\big((\nu u'r')\big(u' \to u \mid r' \to r \mid \mathcal{J}[\![N]\!]_{u'}^{r'}\big)\big)\big)$

With this, we can use the inductive hypothesis:

$(\nu ab)\big(\mathcal{J}[\![M]\!]_a^b \mid b(x).\big((\nu u'r')\big(u' \to u \mid r' \to r \mid \mathcal{J}[\![N]\!]_{u'}^{r'}\big)\big)\big) \approx_E (\nu ab)\big(\mathcal{J}[\![M]\!]_a^b \mid b(x).\mathcal{J}[\![N]\!]_u^r\big)$

With this, we have shown (4) holds for Binding.

**Value**: We need to show that:

$$(\nu u' r')\big(u' \to u \mid r' \to r \mid (\overline{u'}(y).\mathcal{J}[\![y := V]\!])\big) \approx_E \overline{u}(y).\mathcal{J}[\![y := V]\!]$$

This follows from the case of the encoding of a value. First of all we can disregard $r'$, as it is not present in $V$:

$$(\nu u' r')\big(u' \to u \mid r' \to r \mid (\overline{u'}(y).\mathcal{J}[\![y := V]\!])\big) \approx_E (\nu u')\big(u' \to u \mid (\overline{u'}(y).\mathcal{J}[\![y := V]\!])\big)$$

We use a single internal communication, as it is the only thing that can happen, given the restriction on $u'$:

$$(\nu u')\big(u' \to u \mid (\overline{u'}(y).\mathcal{J}[\![y := V]\!])\big) \xrightarrow{\tau} (\nu u' y)(u' \to u \mid \mathcal{J}[\![y := V]\!] \mid \overline{u}(z).z \to y)$$

Now we move the prefix out:

$$(\nu u' y)(u' \to u \mid \mathcal{J}[\![y := V]\!] \mid \overline{u}(z).z \to y) \approx_E \overline{u}(z).(\nu u' y)(u' \to u \mid \mathcal{J}[\![y := V]\!] \mid z \to y)$$

We now look at each case of $V$:

**Variable:** In the case $V$ is a variable, we have:

$$\overline{u}(z).(\nu u' y)(u' \to u \mid \mathcal{J}[\![y := x]\!] \mid z \to y) = \overline{u}(z).(\nu u' y)(u' \to u \mid y \to x \mid z \to y)$$

We can remove the link $(u' \to u)$, as $u'$ does not exist anymore.

$$\overline{u}(z).(\nu u' y)(u' \to u \mid \mathcal{J}[\![y := x]\!] \mid z \to y) \approx_E \overline{u}(z).(\nu y)(\mathcal{J}[\![y := x]\!] \mid z \to y)$$

We use Lemma 7:

$$\overline{u}(z).(\nu y)(\mathcal{J}[\![y := x]\!] \mid z \to y) \approx_E \overline{u}(z).z \to x$$
$$\approx_E \mathcal{J}[\![x]\!]_u^r$$

With this, we have shown that (4) holds when $V$ is a variable.

**Thunk:** In the case $V$ is a thunk, we have:

$$\overline{u}(z).(\nu u' y)(u' \to u \mid \mathcal{J}[\![y := N]\!] \mid z \to y) = \overline{u}(z).(\nu u' y)(u' \to u \mid \,!y(u,r).[\![N]\!]_u^r \mid z \to y)$$

We will here use the real definition of the encoding of a variable:

$$\approx_E \overline{u}(z).(\nu u' y)(u' \to u \mid \,!y(u,r).[\![N]\!]_u^r \mid \,!z(u,r).\overline{y}(u',r').(u' \to u \mid r' \to r))$$
$$\approx_E \overline{u}(z).(\nu y)!((y(u,r).[\![N]\!]_u^r \mid z(u,r).\overline{y}(u',r').(u' \to u \mid r' \to r)))$$
$$\approx_E \overline{u}(z).(\nu y)!z(u,r).((y(u,r).[\![N]\!]_u^r \mid \overline{y}(u',r').(u' \to u \mid r' \to r)))$$
$$\approx_E \overline{u}(z).(\nu y)!z(u,r).\Big(\big(y(u',r').[\![N]\!]_{u'}^{r'} \mid \overline{y}(u',r').(u' \to u \mid r' \to r)\big)\Big)$$
$$\approx_E \overline{u}(z).(\nu y)!z(u,r).(\nu u',r')\big([\![N]\!]_{u'}^{r'} \mid .(u' \to u \mid r' \to r)\big)$$
$$\approx_E \overline{u}(z).!z(u,r).([\![N]\!]_u^r))$$

With this, we have shown that (4) holds when $M$ is a thunk.

**Return**: We need to show that:

$$(\nu u' r')\big(u' \to u \mid r' \to r \mid \mathcal{J}[\![V]\!]_{r'}^{r'}\big) \approx_E \mathcal{J}[\![V]\!]_r^r$$

This follows from the case of the encoding of a value.

**(5)**: We now look at the proof of (5). This lemma follows from the proof of Lemma 4, as in this case, $\mathcal{J}[\![y := V]\!]$ is also always behind prefix of replication and a input, that unlocks a version of $V$.

With this we have shown that Lemma 6 holds for all possible encodings.

$\square$

*Proof of Theorem 2.*

We will now prove Theorem 2, where we will be using the lemmas from before. We begin with the proof of (1.). We use induction on $M \to N$ and up-to-bisimulation.

**(Force-thunk)**: If the reduction was derived using the (force-thunk) rule, which is:

$$\text{(force-thunk)} \; \frac{}{\textbf{force } (\textbf{thunk } M) \to M}$$

We then need to show that $\mathcal{J}[\![\textbf{force } (\textbf{thunk } M)]\!]_u^r \Rightarrow P$ and $P \approx_E \mathcal{J}[\![M]\!]_u^r$.

We begin by expanding the encodings:

$\mathcal{J}[\![\textbf{force } (\textbf{thunk } M)]\!]_u^r = (\nu ab)\big(\mathcal{J}[\![\textbf{thunk } M]\!]_a^b \mid a(y).\overline{y}(u',r').(u' \to u \mid r' \to r)\big)$

$\qquad = (\nu ab)\big(\overline{a}(y).!y(u',r').\mathcal{J}[\![M]\!]_{u'}^{r'} \mid a(y).\overline{y}(u',r').(u' \to u \mid r' \to r)\big)$

We can only reduce this using the following derivation (up to multiple unfoldings of the replication:

$$(\nu ab)\big(\overline{a}(y).!y(u',r').\mathcal{J}[\![M]\!]_{u'}^{r'} \mid a(y).\overline{y}(u',r').(u' \to u \mid r' \to r)\big)$$

$$\xrightarrow{\tau} (\nu ab)(\nu y)\big(!y(u',r').\mathcal{J}[\![M]\!]_{u'}^{r'} \mid \overline{y}(u',r').(u' \to u \mid r' \to r)\big)$$

$$\xrightarrow{\tau} (\nu ab)(\nu y)(\nu u'r')\big(!y(u',r').\mathcal{J}[\![M]\!]_{u'}^{r'} \mid \mathcal{J}[\![M]\!]_{u'}^{r'} \mid (u' \to u \mid r' \to r)\big)$$

This is the final process $P$. We now have to show that $P' \approx_E \mathcal{J}[\![M]\!]_u^r$. We prove this using up-to-bisimilarity. We note that until here, there have not been any outgoing barbs in the intermediate processes, because of the restrictions. From the side conditions of the encoding we have that $a, b, y \notin \text{fn}(M)$. Therefore we have that:

$$(\nu ab)(\nu y)(\nu u'r')\big(!y(u',r').\mathcal{J}[\![M]\!]_{u'}^{r'} \mid \mathcal{J}[\![M]\!]_{u'}^{r'} \mid (u' \to u \mid r' \to r)\big)$$

$$\approx_E (\nu ab)(\nu u'r')\big(\big((\nu y)\big(!y(u',r').\mathcal{J}[\![M]\!]_{u'}^{r'}\big)\big) \mid \mathcal{J}[\![M]\!]_{u'}^{r'} \mid (u' \to u \mid r' \to r)\big)$$

$$\approx_E (\nu ab)(\nu u'r')\big(\mathcal{J}[\![M]\!]_{u'}^{r'} \mid (u' \to u \mid r' \to r)\big)$$

$$\approx_E (\nu u'r')\big(\mathcal{J}[\![M]\!]_{u'}^{r'} \mid (u' \to u \mid r' \to r)\big)$$

We can use (3) from Lemma 6 to conclude that: $(\nu u'r')\big(\mathcal{J}[\![M]\!]_{u'}^{r'} \mid (u' \to u \mid r' \to r)\big) \approx_E \mathcal{J}[\![M]\!]_u^r$ Therefore, (1) holds for (Force-thunk).

**(Application)**: We now look at the (Application)-rule.

$$\text{(Application-base)} \; \frac{}{(\lambda x.M) \; V \to M\{^V/_x\}}$$

For (1.) to hold, it must be the case that $\mathcal{J}[\![(\lambda x.M) \; V]\!]_u^r$ can reduce to a process that is that is bisimilar with $\mathcal{J}[\![M\{^V/_x\}]\!]_u^r$.

We begin with the encoding:

$$\mathcal{I}[\![(\lambda x.M)\ V]\!]_u^r = (\nu abcd)(a(e).c(y).\overline{e}(u',r',y').(u' \to u \mid r' \to r \mid y' \to y) \mid \mathcal{I}[\![\lambda x.M]\!]_a^b \mid \mathcal{I}[\![V]\!]_c^d)$$

$$= (\nu abcd)(a(e).c(x).\overline{e}(u',r',x').(u' \to u \mid r' \to r \mid y' \to y) \mid \overline{a}(e).e(u',r',x).\mathcal{I}[\![M]\!]_{u'}^{r'} \mid (\overline{c}(y).\mathcal{I}[\![y := V]\!])$$

We will now reduce this. The only reduction possible is the following:

$$(\nu abcd)(a(e).c(y).\overline{e}(u',r',x').(u' \to u \mid r' \to r \mid y' \to y) \mid \overline{a}(e).e(u',r',x).\mathcal{I}[\![M]\!]_{u'}^{r'} \mid (\overline{c}(y).\mathcal{I}[\![y := V]\!])$$

$$\xrightarrow{\tau} (\nu abcd)((\nu e)(c(y).\overline{e}(u',r',y').(u' \to u \mid r' \to r \mid y' \to y) \mid e(u',r',x).\mathcal{I}[\![M]\!]_{u'}^{r'}) \mid (\overline{c}(y).\mathcal{I}[\![y := V]\!]))$$

$$\xrightarrow{\tau} (\nu abcdy)\big((\nu e)(\overline{e}(u',r',y').(u' \to u \mid r' \to r \mid y' \to y) \mid e(u',r',x).\mathcal{I}[\![M]\!]_{u'}^{r'}) \mid \mathcal{I}[\![y := V]\!]\big)$$

$$\xrightarrow{\tau} (\nu abcdy)\big((\nu e)(\nu u'r'y')\big((u' \to u \mid r' \to r \mid y' \to y) \mid \mathcal{I}[\![M]\!]_{u'}^{r'}\{^y/_x\}\big) \mid \mathcal{I}[\![y := V]\!]\big)$$

This will be the final process. We now need to show that this is early weak bisimilar to $\mathcal{I}[\![M]\!]_{u'}^{r'}\{^V/_x\}$

From (2) and (4) from Lemma 6, and our bisimilarity rules we have that:

$$(\nu abcdy)\big((\nu e)(\nu u',r',y')\big((u' \to u \mid r' \to r \mid y' \to y) \mid \mathcal{I}[\![M]\!]_{u'}^{r'}\{^y/_x\}\big) \mid \mathcal{I}[\![y := V]\!]\big)$$

$$\approx_E (\nu y)(\mathcal{I}[\![M]\!]_u^r\{^y/_x\} \mid \mathcal{I}[\![y := V]\!])$$

This matches (3) from Lemma 6:

$$(\nu y)(\mathcal{I}[\![M]\!]_u^r\{^y/_x\} \mid \mathcal{I}[\![y := V]\!]) \approx_E \mathcal{I}[\![M\{^V/_x\}]\!]_u^r$$

Therefore, (1.) holds for (Application).

**(Binding-base)**: We now look at the case where the reduction is concluded using the rule (Binding-base)

$$\text{(Binding-base)} \ \frac{}{(\mathbf{return}\ V) \ggg= \lambda x.N \to N\{^V/_x\}}$$

We need to show that $\mathcal{I}[\![(\mathbf{return}\ V) \ggg= \lambda x.N]\!]_u^r$ can reduce to a process that is weak bisimilar with $\mathcal{I}[\![N\{^V/_x\}]\!]_u^r$.

First we will encode $(\mathbf{return}\ V) \ggg= \lambda x.N$:

$$\mathcal{I}[\![(\mathbf{return}\ V) \ggg= \lambda x.N]\!]_u^r = (\nu ab)\big(\mathcal{I}[\![\mathbf{return}\ V]\!]_a^b \mid b(x).\mathcal{I}[\![N]\!]_u^r\big)$$

$$= (\nu ab)\big(\mathcal{I}[\![V]\!]_b^b \mid b(x).\mathcal{I}[\![N]\!]_u^r\big)$$

$$= (\nu ab)\big((\overline{b}(y).\mathcal{I}[\![y := V]\!]) \mid b(x).\mathcal{I}[\![N]\!]_u^r\big)$$

We will now reduce this:

$$(\nu ab)\big((\overline{b}(y).\mathcal{I}[\![y := V]\!]) \mid b(x).\mathcal{I}[\![N]\!]_u^r\big) \xrightarrow{\tau} (\nu aby)(\mathcal{I}[\![y := V]\!] \mid \mathcal{I}[\![N]\!]_u^r\{^y/_x\})$$

This is our final process, and this is also the only reduction possible. We now need to prove that this is early weak bisimilar to $\mathcal{I}[\![N\{^V/_x\}]\!]_u^r$.

$$(\nu aby)(\mathcal{I}[\![y := V]\!] \mid \mathcal{I}[\![N]\!]_u^r\{^y/_x\})$$

$$\approx_E (\nu y)(\mathcal{I}[\![y := V]\!] \mid \mathcal{I}[\![N]\!]_u^r\{^y/_x\})$$

This matches (3) from Lemma 6:

$$(\nu y)(\mathcal{J}[\![y := V]\!] \mid \mathcal{J}[\![N]\!]_u^r\{^y/_x\}) \approx_E \mathcal{J}[\![N\{^V/_x\}]\!]_u^r$$

Therefore, (1.) holds for (Binding-base).

**(Application-evolve)**:

We now look at the case when the reduction is concluded using the (Application-evolve)

$$\text{(Application-evolve)} \; \frac{M \to M'}{M \; V \to M' \; V}$$

We need to show that $\mathcal{J}[\![M \; V]\!]_u^r$ can reduce to a process that is early bisimilar to $\mathcal{J}[\![M' \; V]\!]_u^r$ given that we know that $\mathcal{J}[\![M]\!]_a^b \Rightarrow \mathcal{J}[\![M']\!]_a^b$ (or a process that is early bisimilar to $\mathcal{J}[\![M']\!]_a^b$) for some $a$ and $b$.

We will begin with the encoding:

$$\mathcal{J}[\![M \; V]\!]_u^r = (\nu abcd)\big(a(e).c(x).\overline{e}(u',r',x').(u' \to u \mid r' \to r \mid x' \to x) \mid \mathcal{J}[\![M]\!]_a^b \mid \mathcal{J}[\![V]\!]_c^d\big)$$

Then we have that since $\mathcal{J}[\![M]\!]_a^b$ is not behind prefix, we have the following reduction using the inductive hypothesis, where $\mathcal{J}[\![M]\!]_a^b \Rightarrow P'$ and that $P' \approx_E \mathcal{J}[\![M']\!]_a^b$:

$$(\nu abcd)\big(a(e).c(x).\overline{e}(u',r',x').(u' \to u \mid r' \to r \mid x' \to x) \mid \mathcal{J}[\![M]\!]_a^b \mid \mathcal{J}[\![V]\!]_c^d\big)$$
$$\Rightarrow (\nu abcd)\big(a(e).c(x).\overline{e}(u',r',x').(u' \to u \mid r' \to r \mid x' \to x) \mid P \mid \mathcal{J}[\![V]\!]_c^d\big)$$
$$\approx_E (\nu abcd)\big(a(e).c(x).\overline{e}(u',r',x').(u' \to u \mid r' \to r \mid x' \to x) \mid \mathcal{J}[\![M']\!]_a^b \mid \mathcal{J}[\![V]\!]_c^d\big)$$

This is exactly the encoding of $\mathcal{J}[\![M' \; V]\!]_u^r$, and therefore (1.) holds for (Application-evolve)

**(Binding-evolve)**: We now look at the case when the reduction is concluded using (Binding-evolve)

$$\text{(Binding-evolve)} \; \frac{M \to M'}{M \gg= \lambda x.N \to M' \gg= \lambda x.N}$$

We need to show that $\mathcal{J}[\![M \gg= \lambda x.N]\!]_u^r$ can reduce to a process that is early bisimilar to $\mathcal{J}[\![M' \gg= \lambda x.N]\!]$ given that we know that $\mathcal{J}[\![M]\!]_a^b \Rightarrow P$ and $P \approx_E \mathcal{J}[\![M']\!]_a^b$.

We will begin with the encoding:

$$\mathcal{J}[\![M \gg= \lambda x.N]\!]_u^r = (\nu ab)\big(\mathcal{J}[\![M]\!]_a^b \mid b(x).\mathcal{J}[\![N]\!]_u^r\big)$$

We have from the inductive hypothesis that $\mathcal{J}[\![M]\!]_a^b \Rightarrow P$ and that $P \approx_E \mathcal{J}[\![M']\!]_a^b$. Since $\mathcal{J}[\![M]\!]_a^b$ is not behind prefix, we have the following derivation:

$$(\nu ab)\big(\mathcal{J}[\![M]\!]_a^b \mid b(x).\mathcal{J}[\![N]\!]_u^r\big)$$
$$\Rightarrow (\nu ab)(P \mid b(x).\mathcal{J}[\![N]\!]_u^r)$$
$$\approx_E (\nu ab)\big(\mathcal{J}[\![M']\!]_a^b \mid b(x).\mathcal{J}[\![N]\!]_u^r\big)$$

This is exactly the encoding of $\mathcal{J}[\![M' \gg= \lambda x.N]\!]_u^r$, and therefore (1.) holds for (Binding-evolve)

With this, we have proven (1.) holds in all cases. We will now prove (2.) using structural induction on the term $M$. This means that we assume that (2.) will hold for sub-expressions $M', M'', N'$ and $V'$, and then look at each case separately. We will also notice in this proof, that all intermediate processes do not have any barbs, meaning they cannot do an output or input because of restrictions. We will use this as part of our inductive hypothesis. We will be proving bisimilarity using up-to-bisimilarity, meaning we can use our bisimilarity rules.

**Abstraction**:

when $M = \lambda x.M'$, we have that $\mathcal{I}[\![\lambda x.M']\!]_u^r = \overline{u}(e).e(u, r, x).\mathcal{I}[\![M]\!]_u^r$

The abstraction encoding cannot reduce, as it is behind input prefix. Therefore (2.) holds when $M$ is an abstraction.

**Application**:

We look at the encoding of an application, where

$$M = M' \; V'$$

We have that the encoding of this will be:

$$\mathcal{I}[\![M' \; V']\!]_u^r = (\nu abcd)\big(a(e).c(x).\overline{e}(u', r', x').(u' \to u \mid r' \to r \mid x' \to x) \mid \mathcal{I}[\![M']\!]_a^b \mid \mathcal{I}[\![V']\!]_c^d\big)$$

$$= (\nu abcd)\big(a(e).c(x).\overline{e}(u', r', x').(u' \to u \mid r' \to r \mid x' \to x) \mid \mathcal{I}[\![M']\!]_a^b \mid (\overline{c}(y).\mathcal{I}[\![y := V']\!])\big)$$

For this to reduce, there are two cases; either $\mathcal{I}[\![M']\!]_a^b \xrightarrow{\tau}$, or $\mathcal{I}[\![M']\!]_a^b$ communicates with the input $a(e)$. In case $\mathcal{I}[\![M']\!]_a^b \xrightarrow{\tau}$, from the inductive hypothesis, we have that it will eventually end up as a process that is weak bisimilar to $\mathcal{I}[\![M'']\!]_a^b$, and $M' \mapsto M''$. Since $c$ is not present in $M'$ from the side condition of the encoding, we now look at what $M$ could be, to be able to communicate with $a(e)$. The two cases where this is possible is $M' = \lambda x.M''$ and the other is when $M' = V''$. We look at the case where $M' = \lambda x.M''$ first. This reduction is described in the earlier proof, where, since we have that $M = (\lambda x.M'') \; V$, and it is shown that the only possible reduction (up to multiple unfoldings of a replication) is to a process that is weak bisimilar to $\mathcal{I}[\![M''\{^V/_x\}]\!]$.

The other case, where $M' = V''$, we look at the encoding:

$$(\nu abcd)\big(a(e).c(x).\overline{e}(u', r', x').(u' \to u \mid r' \to r \mid x' \to x) \mid \mathcal{I}[\![V'']\!]_a^b \mid (\overline{c}(y).\mathcal{I}[\![y := V']\!])\big)$$

$$= (\nu abcd)(a(e).c(x).\overline{e}(u', r', x').(u' \to u \mid r' \to r \mid x' \to x) \mid (\overline{a}(y').\mathcal{I}[\![y' := V'']\!]) \mid (\overline{c}(y).\mathcal{I}[\![y := V']\!]))$$

We then have the following reduction:

$$(\nu abcd)(a(e).c(x).\overline{e}(u', r', x').(u' \to u \mid r' \to r \mid x' \to x) \mid (\overline{a}(y').\mathcal{I}[\![y' := V'']\!]) \mid (\overline{c}(y).\mathcal{I}[\![y := V']\!]))$$

$$\xrightarrow{\tau} (\nu abcdy')\big(c(x).\overline{y'}(u', r', x').(u' \to u \mid r' \to r \mid x' \to x) \mid (\mathcal{I}[\![y' := V'']\!]) \mid (\overline{c}(y).\mathcal{I}[\![y := V']\!])\big)$$

$$\xrightarrow{\tau} (\nu abcdy'y)(\overline{y'}(u', r', x').(u' \to u \mid r' \to r \mid x' \to y) \mid (\mathcal{I}[\![y' := V'']\!]) \mid (\mathcal{I}[\![y := V']\!]))$$

Now, regardless of what $V''$ is, $\mathcal{I}[\![y := V'']\!]$ always is behind the prefix $!y'(u, r)$ (here we use the real definition of $\mathcal{I}[\![y := V]\!]$, where variables are not encoded as links), and since it does not match the arity of the output $\overline{y'}(u', r', x')$, no more communications can happen, and since everything is restricted, no output or inputs can happen for the whole process, under this derivation. In other words, we have that:

$$\mathcal{I}[\![V'' \; V']\!]_u^r \approx_E \mathbf{0}$$

$$\approx_E (\nu abcdy'y)(\overline{y'}(u', r', x').(u' \to u \mid r' \to r \mid x' \to y) \mid (\mathcal{I}[\![y' := V'']\!]) \mid (\mathcal{I}[\![y := V']\!]))$$

This is also what we would expect, as $V'' \, V'$ in CBPV can never reduce. But this also means that for all possible sub terms $M'$, (2.) holds when $M$ is an application. We also notice that we did 3 $\tau$-reductions, before we stopped, even though $M$ could not do a reduction in CBPV.

**Force**:

If $M$ is a force, the encoding will look in the following way:

$$\mathcal{J}[\![\textbf{force } V]\!]_u^r = (\nu ab)\big(\mathcal{J}[\![V]\!]_a^b \mid a(y).\bar{y}(u',r').(u' \to u \mid r' \to r)\big)$$
$$= (\nu ab)((\bar{u}(y).\mathcal{J}[\![y := V]\!]) \mid a(y).\bar{y}(u',r').(u' \to u \mid r' \to r))$$

We now look at each case of $V$. First we look at the case that $V$ is a thunk. We then have the derivation shown for the proof of (1.), that was the only derivation possible (up to multiple unfoldings of replications). Therefore it holds when $V$ is a thunk. In the case it is a variable $x$, we have the following derivation:

$$(\nu ab)((\bar{u}(y).\mathcal{J}[\![y := x]\!]) \mid a(y).\bar{y}(u',r').(u' \to u \mid r' \to r))$$
$$= (\nu ab)((\bar{u}(y).!y(u,r).\bar{x}(u',r').(u' \to u \mid r' \to r)) \mid a(y).\bar{y}(u',r').(u' \to u \mid r' \to r))$$
$$\xrightarrow{\tau} (\nu aby)((!y(u,r).\bar{x}(u',r').(u' \to u \mid r' \to r)) \mid \bar{y}(u',r').(u' \to u \mid r' \to r))$$
$$\equiv_\alpha (\nu aby)((!y(u',r').\bar{x}(u'',r'').(u'' \to u' \mid r'' \to r')) \mid \bar{y}(u',r').(u' \to u \mid r' \to r))$$
$$\xrightarrow{\tau} (\nu abyu'r')((!y(u',r').\bar{x}(u'',r'').(u'' \to u' \mid r'' \to r')) \mid \bar{x}(u'',r'').(u'' \to u' \mid r'' \to r') \mid (u' \to u \mid r' \to r))$$

This is the final process. With the bisimilarity laws we get:

$$(\nu abyu'r')((!y(u',r').\bar{x}(u'',r'').(u'' \to u' \mid r'' \to r')) \mid \bar{x}(u'',r'').(u'' \to u' \mid r'' \to r') \mid (u' \to u \mid r' \to r))$$
$$\approx_E (\nu yu'r')(\bar{x}(u'',r'').(u'' \to u' \mid r'' \to r') \mid (u' \to u \mid r' \to r))$$

We use Lemma 7 and the bisimilarity laws:

$$(\nu yu'r')(\bar{x}(u'',r'').(u'' \to u' \mid r'' \to r') \mid (u' \to u \mid r' \to r))$$
$$\approx_E \bar{x}(u'',r'').(u'' \to u \mid r'' \to r)$$
$$\approx_E \bar{x}(u',r').(u' \to u \mid r' \to r)$$

This is noticeable, as normally **force** $x$ should not be able to reduce. However, since this is the only way that a reduction can happen, and there only happened $\tau$-reductions, it is still weak bisimilar with $\mathcal{J}[\![\textbf{force } x]\!]_u^r$. However, these small intermediate reductions is just setup for, if a thunk would be substituted in, instead of the variable.

Therefore (2.) holds when $M$ is a **force**.

**Return V**:

In the case that $M = \textbf{return } V$ we have that $\mathcal{J}[\![\textbf{return } V]\!]_u^r = (\bar{r}(y).\mathcal{J}[\![y := V]\!])$.

The return encoding can not reduce, as it is behind output prefix. Therefore (2.) holds when $M$ is a return.

**Binding**:

In the case that $M$ is a binding, we have the following encoding:

$$\mathcal{J}[\![M' \gg= \lambda x.N]\!]_u^r = (\nu ab)\big(\mathcal{J}[\![M']\!]_a^b \mid b(x).\mathcal{J}[\![N]\!]_u^r\big)$$

There are two way this can reduce: If $\mathcal{I}[\![M]\!]_a^b \xrightarrow{\tau} Q$, and it would not be able to communicate with the input on $b$ before reaching the final process, as the intermediate processes do not have barbs then from the inductive hypothesis, we have that $Q \Rightarrow P'$ and $P \approx_E \mathcal{I}[\![M']\!]_a^b$, so this behaviour satisfies (2.). The second case is if $\mathcal{I}[\![M']\!]_a^b$ can do an output on the channel $b$. The only way this can happen is if $M' = \textbf{return } V$, we have already shown this reduction in the proof for (1.).

With this, (2.) holds when $M$ is a binding.

**Value**:

In the case of $M$ being a value, we have that $\mathcal{I}[\![V]\!]_u^r = \overline{u}(y).\mathcal{I}[\![y := V]\!]$. There cannot happen a reduction, as it is behind an output prefix.

Since both (1.) and (2.) holds in all cases, Theorem 2 holds for all possible terms.

□

# C) Gorla's properties for the $\pi I$-calculus encoding

In Definition 54 we present how the 5 properties of [1] would look for the encoding of CBPV in the $\pi I$-calculus.

**Definition 54.** (5 properties of the encoding)

1. (Compositional) An encoding $\mathcal{J}[\![\cdot]\!]_u^r$ is compositional if, for every term constructor *op* of terms in CBPV, there exists a context $C$ and a set of names $N = \{n_1, ... n_m\}$, so that $\mathcal{J}[\![op(M_1, ..., M_k)]\!]_u^r = C\Big(\mathcal{J}[\![M_1]\!]_{n_1}^{n_2}; ... \mathcal{J}[\![M_k]\!]_{n_{m-1}}^{n_m}\Big)$.

2. (Name invariance) An encoding $\mathcal{J}[\![\cdot]\!]$ is name invariant, if, for every term $M$, and every substitution $\sigma$ in CBPV, we have that:

$$\mathcal{J}[\![M\sigma]\!]_u^r \begin{cases} = \mathcal{J}[\![M]\!]_u^r \sigma' & \text{if } \sigma \text{ is injective} \\ \approx_E \mathcal{J}[\![M]\!]_u^r \sigma' & \text{otherwise} \end{cases}$$

   Where $\sigma'$ is a substitution in the $\pi$-calculus, that does not substitute to names $u$ and $r$, but otherwise replaces with equivalent names to $\sigma$.

3. (Operational correspondence) An encoding $\mathcal{J}[\![\cdot]\!]_u^r$ is operational correspondent if it is:
   - *Sound* : For all $M \Rightarrow N$, $\mathcal{J}[\![M]\!]_u^r \Rightarrow P$ and $P \approx_E \mathcal{J}[\![N]\!]_u^r$.
   - *Complete* : $\forall P$ where $\mathcal{J}[\![M]\!]_u^r \Rightarrow P$, there exists an $N$ so that $M \mapsto N'$ and $P \Rightarrow P'$ and $P' \approx_E \mathcal{J}[\![N]\!]_u^r$

4. (Divergence reflection) An encoding $\mathcal{J}[\![\cdot]\!]_u^r$ reflects divergence, if, for every $M$ where $\mathcal{J}[\![M]\!]_u^r \overset{\omega}{\to}$, it holds that $M \overset{\omega}{\mapsto}$

5. (Success sensitiveness) An encoding $\mathcal{J}[\![\cdot]\!]_u^r$ is success sensitive if, for every $M$, it holds that $M \mapsto V$ if and only if $\exists y, P, P'. \mathcal{J}[\![M]\!]_u^r \Rightarrow P$ and $P \overset{\overline{r}(y)}{\to} P'$.

*Proof.*

**Property 1-3**: Just like for the $\pi$-calculus encoding, our encoding easily satisfies property 1 and 2, because the encoding consists of a contexts around encodings of sub-terms, and the variables in CBPV are directly used as names in the $\pi I$-calculus, and all other names are bound.

Property 3 is satisfied through the previous proof, where we proved the soundness theorem, and a stronger completeness theorem.

**Property 4**: For property 4, we will prove it by proving the contra-positive form, which states:

$$\neg\Big(M \overset{\omega}{\mapsto}\Big) \text{ implies } \neg\Big(\mathcal{J}[\![M]\!]_u^r \overset{\omega}{\to}\Big)$$

What this essentially says is that if $M$ reduces to a final state in $n$ reductions. We then need to show that $\mathcal{J}[\![M]\!]_u^r$ also reduces to a final process in a finite amount of reductions. Just as in the proof of the encoding in the $\pi$-calculus, we will argue that as seen in the reductions in the soundness and completeness proof, we never introduce divergent behaviour, unless the terms themselves are divergent. Unlike the encoding in the $\pi$-calculus, after almost every communication, we now introduce multiple new links. Therefore it is harder to specify exactly how many reductions are necessary to reduce to reduce to a final process, but we can see that in application, we need at least 3 communications in the polyadic encoding, (and at least 6 in the monadic), as well as introducing 3 extra links, that might be used in further reductions. This means we can bound the number of reductions the encoding needs to do, to match one reduction in CBPV by $3 + 3n$. We also notice we have the issue with $V\ V'$ making 3 reductions, before stoping (which can also have been incremented by earlier reductions). In this encoding, This means that it will reduce to a final process in at most $(3 + 3n) * n + (3n + 3)$ reductions (simplified to $3n^2 + 6n + 3$ reductions). With this, we have shown the contra-positive form holds, and therefore it can be concluded that property 4 holds for the $\pi I$-calculus encoding.

**Property 5**:

If property 5 is to hold, it must be the case we can have an output on the channel $r$, if and only if we end up with a process that is early weak bisimilar with $\mathcal{I}[\![\mathbf{return}\ V]\!]_u^r$. By inspection of the encoding, this is also the case, as in all other cases, either there is restrictions that stops a subprocesses from outputting on the $r$ handle, or such an output is behind prefix. And from the soundness and completeness proof, we have shown that any encoding only reduces down to a process that is early weak bisimilar with the encoding of a **return**, if and only if the encoded term also would reduce to a **return**.

□

# D) Encodings

## D.1) Encoding CBPV in the monadic $\pi$-calculus

This encoding is also presented in the main body of the report at Definition 24.

$$\llbracket \lambda x.M \rrbracket_u^r = u(s).s(u).s(r).s(x).\llbracket M \rrbracket_u^r \qquad \mathcal{V}(s) \notin \mathrm{fv}(M)$$

$$\llbracket M\ V \rrbracket_u^r = (\nu p)(\nu q)\big(\llbracket M \rrbracket_p^q \mid (\nu s)\overline{p}s.\overline{s}u.\overline{s}r.\llbracket V \rrbracket_s^r\big) \qquad \mathcal{V}(p,q) \notin \mathrm{fv}(M,V)$$

$$\llbracket \mathbf{force}\ V \rrbracket_u^r = (\nu p)\big(\llbracket V \rrbracket_p^r \mid p(y).(\nu s)\overline{y}s.\overline{s}u.\overline{s}r\big) \qquad \mathcal{V}(p) \notin \mathrm{fv}(V)$$

$$\llbracket \mathbf{return}\ V \rrbracket_u^r = \llbracket V \rrbracket_r^r$$

$$\llbracket M \ggg \lambda x.N \rrbracket_u^r = (\nu z)((\nu u)\llbracket M \rrbracket_u^z \mid z(x).\llbracket N \rrbracket_u^r) \qquad \mathcal{V}(z) \notin \mathrm{fv}(M,N)$$

$$\llbracket V \rrbracket_u^r = (\nu y)(\overline{u}y.\llbracket y := V \rrbracket) \qquad \mathcal{V}(y) \notin \mathrm{fv}(V)$$

$$\llbracket y := x \rrbracket = !y(w).\overline{x}w$$

$$\llbracket y := \mathbf{thunk}\ M \rrbracket = !y(s).s(u).s(r).\llbracket M \rrbracket_u^r \qquad \mathcal{V}(s) \notin \mathrm{fv}(M)$$

## D.2) Encoding CBPV in the polyadic $\pi$-calculus

This encoding is also presented in the main body of the report at Definition 27.

$$\mathcal{P}\llbracket \lambda x.M \rrbracket_u^r = u(v,r).u(x).\mathcal{P}\llbracket M \rrbracket_v^r \qquad v \notin \mathrm{fn}(M)$$

$$\mathcal{P}\llbracket M\ V \rrbracket_u^r = (\nu pq)\big(\mathcal{P}\llbracket M \rrbracket_p^q \mid \overline{p}[u,r].\mathcal{P}\llbracket V \rrbracket_p^r\big) \qquad p \notin \mathrm{fn}(M,V)$$

$$\mathcal{P}\llbracket \mathbf{force}\ V \rrbracket_u^r = (\nu p)\big(\mathcal{P}\llbracket V \rrbracket_p^r \mid p(y).\overline{y}[u,r]\big) \qquad p \notin \mathrm{fn}(V)$$

$$\mathcal{P}\llbracket \mathbf{return}\ V \rrbracket_u^r = \mathcal{P}\llbracket V \rrbracket_r^r$$

$$\mathcal{P}\llbracket M \ggg \lambda x.N \rrbracket_u^r = (\nu z)((\nu u)\mathcal{P}\llbracket M \rrbracket_u^z \mid z(x).\mathcal{P}\llbracket N \rrbracket_u^r) \qquad z \notin \mathrm{fn}(M,N)$$

$$\mathcal{P}\llbracket V \rrbracket_u^r = (\nu y)(\overline{u}y.\mathcal{P}\llbracket y := V \rrbracket) \qquad y \notin \mathrm{fn}(V)$$

$$\mathcal{P}\llbracket y := x \rrbracket = !y(u,r).\overline{x}[u,r]$$

$$\mathcal{P}\llbracket y := \mathbf{thunk}\ M \rrbracket = !y(u,r).\mathcal{P}\llbracket M \rrbracket_u^r \qquad s \notin \mathrm{fn}(M)$$

## D.3) Encoding CBPV in the polyadic $\pi I$-calculus

This encoding is also presented in the main body of the report at Definition 46.

$$\mathcal{I}\llbracket \lambda x.M \rrbracket_u^r = \overline{u}(e).e(u,r,x).\mathcal{I}\llbracket M \rrbracket_u^r$$

$$\mathcal{I}\llbracket M\ V \rrbracket_u^r = (\nu abcd)\big(\mathcal{I}\llbracket M \rrbracket_a^b \mid a(e).c(x).\overline{e}(u',r',x').(u' \to u \mid r' \to r \mid x' \to x) \mid \mathcal{I}\llbracket V \rrbracket_c^d\big)$$

$$\mathcal{I}\llbracket \mathbf{force}\ V \rrbracket_u^r = (\nu ab)\big(\mathcal{I}\llbracket V \rrbracket_a^b \mid a(y).\overline{y}(u',r').(u' \to u \mid r' \to r)\big)$$

$$\mathcal{I}\llbracket \mathbf{return}\ V \rrbracket_u^r = \mathcal{I}\llbracket V \rrbracket_r^r$$

$$\mathcal{I}\llbracket M \ggg \lambda x.N \rrbracket_u^r = (\nu ab)\big(\mathcal{I}\llbracket M \rrbracket_a^b \mid b(x).\mathcal{I}\llbracket N \rrbracket_u^r\big)$$

$$\mathcal{I}\llbracket V \rrbracket_u^r = (\overline{u}(y).\mathcal{I}\llbracket y := V \rrbracket)$$

$$\mathcal{I}\llbracket y := x \rrbracket = !y(u,r).\overline{x}(u',r').(u' \to u \mid r' \to r)$$

$$\mathcal{I}\llbracket y := \mathbf{thunk}\ M \rrbracket = !y(u,r).\mathcal{I}\llbracket M \rrbracket_u^r$$

## D.4) Encoding CBPV in the monadic $\pi I$-calculus

Encoding in the monadic $\pi I$-calculus.

$$\mathcal{M}[\![\lambda x.M]\!]_u^r = \overline{u}(e).e(s).s(u).s(r).s(x).\mathcal{M}[\![M]\!]_u^r \qquad\qquad e,s \notin \mathrm{fn}(M)$$

$$\mathcal{M}[\![M\ V]\!]_u^r = (\nu abcd)\big((a(e).c(x).\overline{e}(s).\overline{s}(u').\overline{s}(r').\overline{s}(x').$$
$$(u' \to u \mid r' \to r \mid x' \to x) \mid \mathcal{M}[\![M]\!]_a^b \mid \mathcal{M}[\![V]\!]_c^d)\big) \qquad a,b,c,d \notin \mathrm{fn}(M) \cup \mathrm{fn}(V)$$

$$\mathcal{M}[\![\mathbf{force}\ V]\!]_u^r = (\nu ab)\big(\mathcal{M}[\![V]\!]_a^b \mid a(y).\overline{y}(s).\overline{s}(u').\overline{s}(r').s(s).(u' \to u \mid r' \to r)\big) \qquad a,b \notin \mathrm{fn}(V)$$

$$\mathcal{M}[\![\mathbf{return}\ V]\!]_u^r = \mathcal{M}[\![V]\!]_r^r$$

$$\mathcal{M}[\![M \gg= \lambda x.N]\!]_u^r = (\nu ab)\big(\mathcal{M}[\![M]\!]_a^b \mid b(x).\mathcal{M}[\![N]\!]_u^r\big) \qquad\qquad a,b \notin \mathrm{fn}(M) \cup \mathrm{fn}(N)$$

$$\mathcal{M}[\![V]\!]_u^r = (\overline{u}(y).\mathcal{M}[\![y := V]\!]) \qquad\qquad y \notin \mathrm{fn}(V)$$

$$\mathcal{M}[\![y := x]\!] = y \to x$$

$$\mathcal{M}[\![y := \mathbf{thunk}\ M]\!] = !y(s).s(u).s(r).\overline{s}(s).\mathcal{M}[\![M]\!]_u^r \qquad\qquad s \notin \mathrm{fn}(M)$$

## D.5) Encoding CBPV in the asynchronous polyadic $\pi$-calculus

Encoding in the asynchronous polyadic $\pi$-calculus (not Local like Sangiorgi in [33]):

$$\mathcal{L}[\![\lambda x.M]\!]_u^r = u(v,q,x).\mathcal{L}[\![M]\!]_v^q$$

$$\mathcal{L}[\![M\ V]\!]_u^r = (\nu p)(\nu q)\big(\mathcal{L}[\![M]\!]_p^q \mid (\nu y)(\overline{p}\langle u,r,y\rangle \mid \mathcal{L}[\![y := V]\!])\big)$$

$$\mathcal{L}[\![\mathbf{force}\ V]\!]_u^r = (\nu p)\big(\mathcal{L}[\![V]\!]_p^r \mid p(y).\overline{y}\langle u,r\rangle\big)$$

$$\mathcal{L}[\![\mathbf{return}\ V]\!]_u^r = \mathcal{L}[\![V]\!]_r^r$$

$$\mathcal{L}[\![M \gg= \lambda x.N]\!]_u^r = (\nu z)((\nu u)[\![M]\!]_u^z \mid z(x).\mathcal{L}[\![N]\!]_u^r)$$

$$\mathcal{L}[\![V]\!]_u^r = (\nu y)(\overline{u}y \mid \mathcal{L}[\![y := V]\!])$$

$$\mathcal{L}[\![y := x]\!] = !(y(w).\overline{x}w \mid y(u,r).\overline{x}\langle u,r\rangle)$$

$$\mathcal{L}[\![y := \mathbf{thunk}\ M]\!] = !y(u,r).\mathcal{L}[\![M]\!]_u^r$$

## D.6) Encoding CBPV in the local $\pi$-calculus

A local $\pi$-calculus encoding:

$$\mathcal{L}[\![\lambda x.M]\!]_u^r = (\nu a)(\overline{u}[a] \mid a(u,r,x).\mathcal{L}[\![M]\!]_u^r) \qquad\qquad a \notin \mathrm{fn}(M)$$

$$\mathcal{L}[\![M\ V]\!]_u^r = (\nu abcd)\big(\mathcal{L}[\![M]\!]_a^b \mid a(e).c(x).\overline{e}[u,r,x] \mid \mathcal{L}[\![V]\!]_c^d\big) \qquad a,b,c,d \notin \mathrm{fn}(M) \cup \mathrm{fn}(V)$$

$$\mathcal{L}[\![\mathbf{return}\ V]\!]_u^r = \mathcal{L}[\![V]\!]_r^r$$

$$\mathcal{L}[\![M \gg= \lambda x.N]\!]_u^r = (\nu ab)\big(\mathcal{L}[\![M]\!]_a^b \mid b(x).\mathcal{L}[\![N]\!]_u^r\big) \qquad\qquad a,b \notin \mathrm{fn}(M) \cup \mathrm{fn}(N)$$

$$\mathcal{L}[\![\mathbf{force}\ V]\!]_u^r = (\nu ab)\big(\mathcal{L}[\![V]\!]_a^b \mid a(y).\overline{y}[u,r]\big) \qquad\qquad a,b \notin \mathrm{fn}(V)$$

$$\mathcal{L}[\![V]\!]_u^r = (\nu y)(\overline{u}[y] \mid \mathcal{L}[\![y := V]\!]) \qquad\qquad y \notin \mathrm{fn}(V)$$

$$\mathcal{L}[\![y := \mathbf{thunk}\ M]\!] = !y(u,r).\mathcal{L}[\![M]\!]_u^r$$

$$\mathcal{L}[\![y := x]\!] = !y(u,r).\overline{x}[u,r]$$

It builds upon the same basic idea, with a few exceptions. When doing application, instead of the two encodings directly communicating, they instead communicate with a third parallel process. This is to ensure we stay inside the local $\pi$-calculus, as we otherwise would have issues when encoding $[\![\lambda x.\lambda y.y]\!]_u^r$, as in the monadic encoding from Definition 24, this encoded process has the input $u(s)$–for it to be local, $s$ may now only be used as the subject or object in outputs, but not in inputs. This is also closer to a well-typed encoding, as each name is present only once as output and once as input. We also see that we now only care about the forwarding "force" of values, as that is the behaviour we want from our encoded variables.

# E) Encoding of CBV and CBN in CBPV

In this chapter we will look at a step by step encoding, followed by the reduction, of the $\lambda$-expression $(\lambda x.xx)((\lambda y.\lambda z.y)u)$ into CBPV. We will encode the $\lambda$-expression both as a CBV expression and a CBN expression into CBPV, to demonstrate that the encoding reduces to an expected result. Since the expressions quickly grow very large, we have opted to present the expressions in a similar fashion to code snippets.

**Example 12.** (Encoding a CBV expression) In this example we will encode the following $\lambda$-expression using the CBV encoding, and then show that the terminal values for the encoded and the non-encoded expression correlate. In expressions where a terminal value will never be reached, one would expect that to be the case for both the non-encoded and the encoded expression.

$$(\lambda x.xx)((\lambda y.\lambda z.y)u)$$

```
1  ⟦(λx.xx)((λy.λz.y)u)⟧ᵛ
```
=
```
1  ⟦(λx.xx)⟧ᵛ >>=
2  λf.⟦((λy.λz.y)u)⟧ᵛ >>=
3  λx. force f x
```
=

```
1  ( return thunk λx.
2      ⟦xx⟧ᵛ
3  ) >>=
4  ( λf.
5      ⟦((λy.λz.y)u)⟧ᵛ
6  ) >>=
7  λx.force f x
```
=
```
1  ( return thunk λx.
2      ⟦x⟧ᵛ >>=
3      λf.⟦x⟧ᵛ >>=
4      λx. force f x
5  ) >>=
6  ( λf.
7      ⟦((λy.λz.y)u)⟧ᵛ
8  ) >>=
9  λx.force f x
```
=

```
1  ( return thunk λx.
2      return x >>=
3      λf.return x >>=
4      λx.force f x
5  ) >>=
6  ( λf.
7      ⟦((λy.λz.y)u)⟧ᵛ
8  ) >>=
9  λx.force f x
```
=
```
1   ( return thunk λx.
2       return x >>=
3       λf.return x >>=
4       λx.force f x
5   ) >>=
6   ( λf.
7       ⟦λy.λz.y⟧ᵛ >>=
8       λf.⟦u⟧ᵛ >>=
9       λx.force f x
10  ) >>=
11  λx.force f x
```
=

```
1   ( return thunk λx.
2       return x >>=
3       λf.return x >>=
4       λx.force f x
5   ) >>=
6   ( λf.
7       ( return thunk λy.
8           〚λz.y〛ᵛ
9       ) >>=
10      λf.〚u〛ᵛ >>=
11      λx.force f x
12  ) >>=
13  λx.force f x
```

=

```
1   ( return thunk λx.
2       return x >>=
3       λf.return x >>=
4       λx.force f x
5   ) >>=
6   ( λf.
7       ( return thunk λy.
8           return thunk λz.〚y〛ᵛ
9       ) >>=
10      λf.〚u〛ᵛ >>=
11      λx.force f x
12  ) >>=
13  λx.force f x
```

=

```
1   ( return thunk λx.
2       return x >>=
3       λf.return x >>=
4       λx.force f x
5   ) >>=
6   ( λf.
7       ( return thunk λy.
8           return thunk λz.
9               return y
10      ) >>=
11      λf.return u >>=
12      λx.force f x
13  ) >>=
14  λx.force f x
```

Now that the example have been fully encoded, we then reduce the example, using the CBPV semantics. We begin by using the (Binding-base) rule:

```
1   ( return thunk λy.
2       return thunk λz.
3           return y
4   ) >>=
5   ( λf.return u >>=
6   λx.force f x
7   )>>=
8   λx.force (thunk λx.
9       return x >>=
10      λf.return x >>=
11      λx.force f x) x)
```

```
1   return u >>=
2   ( λx.
```

```
3      (force thunk (λy.return thunk λz.return y))
4      x
5 ) >>=
6 λx.
7   ( force thunk (λx.return x) >>=
8     λf.(return x >>= λx.force f x) )
9   x
```

We reduce the "`return u >>= λx`" using the (Binding-base) rule, and substitute `x` with `u`.

```
1 ( ( force thunk
2         ( lambdly.return thunk (λz.return y) )
3   )
4   u
5 ) >>=
6 λx.
7   ( force thunk (λx.return x) >>=
8     λf.(return x >>= λx.force f x) )
9   x
```

We reduce using the (force-thunk)-rule.

```
1 (λy.return thunk (λz.return y)) u >>=
2 λx.
3   ( force thunk (λx.return x) >>=
4     λf.(return x >>= λx.force f x) )
5   x
```

We reduce using the (Application)-rule, and substitute `y` with `u`

```
1 return thunk (λz.return u) >>=
2 λx.
3   ( force thunk (λx.return x) >>=
4     λf.(return x >>= λx.force f x) )
5   x
```

We reduce using the (Binding-base)-rule, and substitute the `x` with the thunk.

```
1 ( force thunk (λx.return x) >>=
2   λf.(return x >>= λx.force f x)
3 ) (thunk λz.return u)
```

We use the (force-thunk)-rule:

```
1 ( λx.return x >>=
2   λf.(return x >>= λx.force f x) )
3 (thunk λz.return u)
```

We use the (Application) rule, substituting `x` with "`(thunk λz.return u)`"

```

```
1  return (thunk λz.return u) >>=
2  λf.
3    return (thunk λz.return u) >>=
4    λx.force f x
```

We use the (Binding-base) rule, to substitute `f` with the thunk.

```
1  return (thunk λz.return u) >>=
2  λx.force (thunk λz.return u) x
```

We use the (Binding-base) rule to substitute `x` with the thunk.

```
1  force (thunk λz.return u) (thunk λz.return u)
```

We use the (force-thunk) rule:

```
1  (λz.return u) (thunk λz.return u)
```

We use the (Application)-rule, but since there is no `z` in (`return u`) the substitution does nothing.

```
1  return u
```

As we see, it reduces down to `return u`, which is exactly the same as $[\![u]\!]^v$.

**Example 13.** (Encoding a CBN expression) We encode the same expression as in the previous example, $(\lambda x.xx)((\lambda y.\lambda z.y)u)$, but using the CBN encoding instead. Again we will follow this by showing that the terminal value for the encoded and non-encoded expression match.

```
1  [[(λx.xx)((λy.λz.y)u)]]ⁿ
```
=
```
1  [[(λx.xx)]]ⁿ(thunk [[((λy.λz.y)u)]]ⁿ)
```
=

```
1  (λx.[[xx]]ⁿ)(thunk [[((λy.λz.y)u)]]ⁿ)
```
=
```
1  (λx.[[x]]ⁿ(thunk [[x]]ⁿ))
2  (thunk [[((λy.λz.y)u)]]ⁿ)
```
=

```
1  (λx.force x (thunk [[x]]ⁿ))
2  (thunk [[((λy.λz.y)u)]]ⁿ)
```
=
```
1  (λx.force x (thunk force x))
2  (thunk [[((λy.λz.y)u)]]ⁿ)
```
=

```
1  (λx.force x (thunk force x))
2  (thunk ( [[(λy.λz.y)]]ⁿ
3          (thunk [[u]]ⁿ) ) )
```
=
```
1  (λx.force x (thunk force x))
2  (thunk ( (λy.λz.[[y]]ⁿ)
3          (thunk [[u]]ⁿ) ) )
```
=

```
1  (λx.force x (thunk force x))
2  (thunk ( λy.λz.force y
3          (thunk [[u]]ⁿ) ) )
```
=
```
1  (λx.force x (thunk force x))
2  (thunk ( (λy.λz.force y)
3          (thunk force u) ) )
```

Now that the example have been fully encoded, we then reduce the example, using the call-by-push-value semantics.



```
1 (λx.force x (thunk force x))
2 (thunk ( (λy.λz.force y)
3         (thunk force u) ) )
```

We use the (Application)-rule and substitute x with ((thunk (λy.λz.force y)(thunk force u))).

```
1 (force (thunk ( (λy.λz.force y)
2                (thunk force u) ) ) )
3 (thunk force (thunk ( (λy.λz.force y)
4                      (thunk force u) ) ) )
```

We then use the (force-thunk)-rule:

```
1 ( (λy.λz.force y)
2   (thunk force u) )
3 (thunk force (thunk ( (λy.λz.force y)
4                      (thunk force u) ) ) )
```

We use the (Application)-rule, and substitutes y with "(thunk force u)".

```
1 (λz.force (thunk force u))
2 (thunk force (thunk ( (λy.λz.force y)
3                      (thunk force u) ) ) )
```

We use the (Application)-rule, but since there are no occurrences of z in force (thunk force u), we do not perform a substitution and we are left with the expression we where to substitute into:

```
1 force (thunk force u)
```

We use the (force-thunk)-rule, and end up with:

```
1 force u
```

Which is exactly equal to $[\![u]\!]^n$.



# F) Encoding $(\lambda x.\ \textbf{force}\ x)(\textbf{thunk}\ x)$ in the $\pi$-calculus

Here we will show the full encoding of the CBPV process $(\lambda x.\ \textbf{force}\ x)(\textbf{thunk}\ x)$ in the $\pi$-calculus using our encoding. We start by unfolding $M$ as much as possible and $V$ a single time, as we will need the communication it allows.

$= (\nu p)(\nu q)(\llbracket M \rrbracket_p^q \mid (\nu s)\overline{p}s.\overline{p}u.\overline{p}r.\llbracket V \rrbracket_s^r)$

$= (\nu p)(\nu q)\big(p(s).s(u).s(r).s(x).\llbracket M' \rrbracket_u^r \mid (\nu s)\overline{p}s.\overline{p}u.\overline{p}r.\llbracket V \rrbracket_s^r\big)$

$= (\nu p)(\nu q)\big(p(s).s(u).s(r).s(x).(\nu p_1)\big(\llbracket V' \rrbracket_{p_1}^r \mid p_1(y).(\nu s')\overline{y}s'.\overline{s'}u.\overline{s'}r\big) \mid (\nu s)\overline{p}s.\overline{p}u.\overline{p}r.\llbracket V \rrbracket_s^r\big)$

$= (\nu p)(\nu q)(p(s).s(u).s(r).s(x).(\nu p_1)((\nu y)(\overline{p_1}y.\llbracket y := V' \rrbracket) \mid p_1(y).(\nu s')\overline{y}s'.\overline{s'}u.\overline{s'}r) \mid (\nu s)\overline{p}s.\overline{p}u.\overline{p}r.\llbracket V \rrbracket_s^r)$

$= (\nu p)(\nu q)(p(s).s(u).s(r).s(x).(\nu p_1)((\nu y)(\overline{p_1}y.!y(w).\overline{x}w) \mid p_1(y).(\nu s')\overline{y}s'.\overline{s'}u.\overline{s'}r) \mid (\nu s)\overline{p}s.\overline{p}u.\overline{p}r.\llbracket V \rrbracket_s^r)$

$= (\nu p)(\nu q)(p(s).s(u).s(r).s(x).(\nu p_1)((\nu y)(\overline{p_1}y.!y(w).\overline{x}w) \mid p_1(y).(\nu s')\overline{y}s'.\overline{s'}u.\overline{s'}r) \mid (\nu s)\overline{p}s.\overline{p}u.\overline{p}r.(\nu y_1)(\overline{p}y_1.\llbracket y_1 := V \rrbracket))$

We now communicate on $p$ once and then on $s$ three times.

$\rightarrow (\nu p)(\nu q)(\nu s)(s(u).s(r).s(x).(\nu p_1)((\nu y)(\overline{p_1}y.!y(w).\overline{x}w) \mid p_1(y).(\nu s')\overline{y}s'.\overline{s'}u.\overline{s'}r) \mid \overline{p}u.\overline{p}r.(\nu y_1)(\overline{p}y_1.\llbracket y_1 := V \rrbracket))$

$\rightarrow (\nu p)(\nu q)(\nu s)(s(r).s(x).(\nu p_1)((\nu y)(\overline{p_1}y.!y(w).\overline{x}w) \mid p_1(y).(\nu s')\overline{y}s'.\overline{s'}u.\overline{s'}r) \mid \overline{p}r.(\nu y_1)(\overline{p}y_1.\llbracket y_1 := V \rrbracket))$

$\rightarrow (\nu p)(\nu q)(\nu s)(s(x).(\nu p_1)((\nu y)(\overline{p_1}y.!y(w).\overline{x}w) \mid p_1(y).(\nu s')\overline{y}s'.\overline{s'}u.\overline{s'}r) \mid (\nu y_1)(\overline{p}y_1.\llbracket y_1 := V \rrbracket))$

$\rightarrow (\nu p)(\nu q)(\nu s)(\nu y_1)((\nu p_1)((\nu y)(\overline{p_1}y.!y(w).\overline{y_1}w) \mid p_1(y).(\nu s')\overline{y}s'.\overline{s'}u.\overline{s'}r) \mid \llbracket y_1 := V \rrbracket)$

We remove redundant restrictions using the bisimilarity laws and communicate on $p_1$.

We also perform $\alpha$-conversion to get rid of the primes.

$\approx_E (\nu y_1)((\nu p_1)((\nu y)(\overline{p_1}y.!y(w).\overline{y_1}w) \mid p_1(y).(\nu s')\overline{y}s'.\overline{s'}u.\overline{s'}r) \mid \llbracket y_1 := V \rrbracket)$

$\rightarrow (\nu y_1)((\nu p_1)(\nu y)((!y(w).\overline{y_1}w) \mid (\nu s')\overline{y}s'.\overline{s'}u.\overline{s'}r) \mid \llbracket y_1 := V \rrbracket)$

$\approx_E (\nu y_1)((\nu y)(!y(w).\overline{y_1}w) \mid (\nu s)\overline{y}s.\overline{s}u.\overline{s}r \mid \llbracket y_1 := V \rrbracket)$

We communicate on a replication of $!y(w).\overline{y_1}w$, which gets $s$ in its scope.

Remove redundancies and fully unfold $\llbracket y_1 := V \rrbracket$.

$\rightarrow (\nu y_1)((\nu y)(!y(w).\overline{y_1}w) \mid (\nu s)(\overline{y_1}s \mid \overline{s}u.\overline{s}r) \mid \llbracket y_1 := V \rrbracket)$

$\approx_E (\nu y_1)((\nu s)(\overline{y_1}s \mid \overline{s}u.\overline{s}r) \mid \llbracket y_1 := V \rrbracket)$

$= (\nu y_1)\big((\nu s)(\overline{y_1}s \mid \overline{s}u.\overline{s}r) \mid !y_1(s).s(u).s(r).\llbracket M'' \rrbracket_u^r\big)$

$= (\nu y_1)((\nu s)(\overline{y_1}s \mid \overline{s}u.\overline{s}r) \mid !y_1(s).s(u).s(r).(\nu y)\llbracket y := V' \rrbracket)$

$= (\nu y_1)((\nu s)(\overline{y_1}s \mid \overline{s}u.\overline{s}r) \mid !y_1(s).s(u).s(r).(\nu y)(\overline{u}y.!y(w).\overline{x}w))$

We communicate on $y_1$, $s$ and $s$.

$\rightarrow (\nu y_1)(\nu s)(\overline{s}u.\overline{s}r \mid !y_1(s).s(u).s(r).(\nu y)(\overline{u}y.!y(w).\overline{x}w) \mid s(u).s(r).(\nu y)(\overline{u}y.!y(w).\overline{x}w))$

$\rightarrow (\nu y_1)(\nu s)(\overline{s}r \mid !y_1(s).s(u).s(r).(\nu y)(\overline{u}y.!y(w).\overline{x}w) \mid s(r).(\nu y)(\overline{u}y.!y(w).\overline{x}w))$

$\rightarrow (\nu y_1)(\nu s)(!y_1(s).s(u).s(r).(\nu y)(\overline{u}y.!y(w).\overline{x}w) \mid (\nu y)(\overline{u}y.!y(w).\overline{x}w))$

We remove redundancies and is left with:



$$\approx_E (\nu y)(\overline{u}y.!y(w).\overline{x}w))$$

This is exactly the encoding of $[\![x]\!]$.